\documentclass[11pt,a4paper]{article}
\pdfoutput=1
\usepackage{jstylemarg}
\usepackage[utf8]{inputenc}
\usepackage{amsmath}
\usepackage{subfloat}
\usepackage{subcaption}
\usepackage{caption}
\usepackage{empheq}

\usepackage{graphicx}
\usepackage{etoolbox}
\newcommand{\icon}[1]{\includegraphics[height=12pt]{#1}}
\robustify{\icon}

\newcommand{\bi}{\begin{itemize}}
\newcommand{\ei}{\end{itemize}}
\newcommand{\bea}{\begin{align}}
\newcommand{\eea}{\end{align}}
\newcommand{\be}{\begin{equation}}
\newcommand{\ee}{\end{equation}}
\newcommand{\bep}{\begin{picture}}
\newcommand{\eep}{\end{picture}}

\newcommand{\pl}{{\partial}}

\newcommand{\tcb}{\textcolor{blue}}

\makeatletter
\renewcommand*\env@matrix[1][\arraystretch]{%
  \edef\arraystretch{#1}%
  \hskip -\arraycolsep
  \let\@ifnextchar\new@ifnextchar
  \array{*\c@MaxMatrixCols c}}
\makeatother

\author[a]{Simone GIOMBI}

\author[a,b]{\quad Charlotte SLEIGHT}

\author[a,b]{\quad Massimo TARONNA\footnote{Postdoctoral Researcher of the Fund for Scientific Research-FNRS Belgium.}}

\affiliation[a]{Department of Physics, Princeton University, Princeton, NJ 08544}

\affiliation[b]{Universit\'e Libre de Bruxelles
and International Solvay Institutes\\
ULB-Campus Plaine CP231, 1050 Brussels, Belgium}

\emailAdd{sgiombi@princeton.edu, charlotte.sleight@gmail.com, taronnam@gmail.com}

\title{\centering
\LARGE{Spinning AdS Loop Diagrams: Two Point Functions}}

\abstract{We develop a systematic approach to evaluating AdS loop amplitudes with spinning legs based on the spectral (or ``split") representation of bulk-to-bulk propagators, which re-expresses loop diagrams in terms of spectral integrals and higher-point tree diagrams. In this work we mostly focus on 2pt one-loop Witten diagrams involving totally symmetric fields of arbitrary mass and integer spin. As an application of this framework, we  study the contribution to the anomalous dimension of higher-spin currents generated by bubble diagrams in higher-spin gauge theories on AdS.}

\begin{document}

\begin{flushright}    
  \texttt{PUPT-2540}
\end{flushright}

\maketitle

\section{Introduction}

The AdS/CFT correspondence provides a remarkable framework to handle quantum gravity on AdS space. Scattering amplitudes on AdS are identified with correlation functions in the dual CFT picture, through which the perturbative expansion of AdS amplitudes given by the loop expansion of Witten diagrams \cite{Maldacena:1997re,Gubser:1998bc,Witten:1998qj} is mapped to the $1/N$ expansion of CFT correlators. At tree-level in the bulk, this map is rather well understood.\footnote{\label{foo:tree}By now there are numerous techniques available in the literature for evaluating Witten diagrams at tree-level, both in position- \cite{Liu:1998ty,Muck:1998rr,Freedman:1998tz,DHoker:1998ecp,Chalmers:1998wu,DHoker:1999kzh,DHoker:1999mqo,Costa:2014kfa,Bekaert:2014cea,Bekaert:2015tva,Sleight:2016hyl,Sleight:2017fpc}, momentum- \cite{Raju:2010by,Raju:2012zr} and Mellin- \cite{Mack:2009mi,Mack:2009gy,Penedones:2010ue,Paulos:2011ie,Fitzpatrick:2011ia,Nandan:2011wc} space, and also via so-called geodesic diagrams \cite{Hijano:2015zsa,Nishida:2016vds,Castro:2017hpx,Chen:2017yia,Gubser:2017tsi,Tamaoka:2017jce}.}  However, to date the bulk computation of Witten diagrams at loop level has proven rather challenging and unexplored -- with the exception of some preliminary works on the Mellin representation of loop diagrams involving only scalars \cite{Penedones:2010ue,Fitzpatrick:2011hu,Fitzpatrick:2011dm,Cardona:2017tsw} and recent efforts which instead aim to extract predictions for bulk loop-corrections from within the dual CFT picture \cite{Creutzig:2015hta,Hikida:2016wqj,Aharony:2016dwx,Hikida:2017ecj,Alday:2017xua,Aprile:2017bgs}. 

The aim of this work is to develop a systematic framework for the direct bulk computation of loop Witten diagrams, in particular from bulk Lagrangians involving totally symmetric fields of arbitrary integer spin. The approach, which is outlined in more detail below in \S \tcb{\ref{sec::genA}}, is underpinned by the spectral representation of bulk-to-bulk propagators \cite{Costa:2014kfa,Bekaert:2014cea,Sleight:2017cax}, which allows the expression of a given loop diagram in terms of spectral integrals and integrated products of higher-point tree diagrams.
This reduces the loop computation to the evaluation of the aforementioned spectral integrals, as well as conformal integrals arising from the expressions for the tree-diagrams. Evaluating tree-diagrams is comparably straightforward and can be performed systematically with currently available methods (see footnote \ref{foo:tree}), while the subsequent conformal integrals are well-known \cite{Symanzik1972}. The spectral integrals are all of the Mellin-Barnes type, which we demonstrate how to regularise and evaluate - leaving to the future the development of a fully systematic means to do so. This decomposition of AdS loop diagrams is the natural generalisation to AdS of momentum integrals in flat space, with the spectral integrals encoding bulk UV divergences and the conformal integrals encoding the IR divergences.
For simplicity, the focus of the present work is mostly on 2pt one-loop bubble and tadpole diagrams on AdS$_{d+1}$, though our methods allow to deal with the more general loop amplitudes involving arbitrary spinning internal and external legs.

We begin in \S \tcb{\ref{sec::scalardiagrams}} where, for ease of introducing the approach, we consider one-loop diagrams involving only scalar fields. In \S \tcb{\ref{subsec::scalar2ptbub}} we consider the 2pt bubble diagram in $\phi^3$ theory, and 2pt tadpole diagrams generated by quartic scalar self interactions in \S \tcb{\ref{subsec::2pttad}}. This includes $\phi^4$ (\S \tcb{\ref{subsubsec::phi4vertex}}) and the most general dressing with derivatives (\S \tcb{\ref{subsubsec::derivativeint}}). In \S \tcb{\ref{subsec::onepttad}} we also discuss one-point tadpole diagrams with a single off-shell external leg in the bulk. In \S \tcb{\ref{sec::sbd}} we present the extension to bubble diagrams produced by parity even cubic couplings of a generic triplet of totally symmetric fields of arbitrary mass and integer spin. In \S \tcb{\ref{subsec::ssprime0}} we focus on diagrams generated by the cubic coupling of a scalar and two gauge fields of arbitrary spin, and extract the spectral representation of the contributions from such diagrams to the anomalous dimension of higher-spin currents.\footnote{It is worth stressing here that our methods to evaluate loop corrections to 2pt functions can be also applied to the bulk computation of the central charges $C_T$ and $C_J$ for the stress tensor and the spin-1 currents, which do not receive anomalous dimensions. See e.g. \cite{Diab:2016spb,Giombi:2016fct} for some boundary results on these two CFT observables.}

In \S \tcb{\ref{sec::applications}} we turn to some applications in specific theories. In \S \tcb{\ref{subsec::gravloops}} we consider the bubble diagram generated by the minimal coupling of a scalar field to gravity in de Donder gauge. In \S \tcb{\ref{subcsec::criton}} we consider the type A minimal higher-spin gauge theory.

In fact, one of our motivations for considering higher-spin gauge theories is to make progress towards testing higher-spin holography at the quantum level, beyond the one-loop vacuum energy results \cite{Giombi:2013fka,Giombi:2014iua,Giombi:2014yra,Beccaria:2014xda,Basile:2014wua,Giombi:2016pvg,Pang:2016ofv,Bae:2016rgm,Bae:2016hfy,Gunaydin:2016amv,Bae:2017spv,Skvortsov:2017ldz} which only probe the free theory.\footnote{For some loop results in flat space see \cite{Ponomarev:2016jqk}. For some previous investigations of quantum corrections in the context of higher-spin gauge theories on AdS, see \cite{Manvelyan:2004ii,Manvelyan:2008ks}. For some recent work in the AdS$_3$ Chern-Simons formulation using Wilson lines, see \cite{Hikida:2017ehf}.} This endeavour relies on the knowledge of the explicit interacting type-A theory action, which has only recently become available \cite{Bekaert:2015tva,Sleight:2016dba,Sleight:2016xqq,Sleight:2016hyl,Sleight:2017fpc,Sleight:2017pcz,Sleight:2017cax}.\footnote{See \cite{Bekaert:2012ux,Giombi:2012ms,Rahman:2015pzl,Giombi:2016ejx,Sleight:2017krf} for reviews on higher-spin gauge theories and their holographic dualities.}

Such tests are particularly relevant in the context of the higher-spin AdS$_4$/CFT$_3$ duality, which gives striking predictions for the bulk loop expansion. For the $\Delta=1$ boundary condition on the bulk scalar, the type A minimal higher-spin gauge theory is conjectured to be dual to the free scalar $O\left(N\right)$ model in three-dimensions \cite{Sezgin:2002rt}, which suggests that the contribution of bulk loop amplitudes for this boundary condition should vanish identically. In AdS$_4$ the bulk scalar admits a second boundary condition, $\Delta=2$, for which the theory is conjectured to be dual to the critical $O(N)$ model \cite{Klebanov:2002ja}. 
This suggests that the non-trivial contributions to the anomalous dimension of higher-spin currents in the critical $O(N)$ model should arise from loop Witten diagrams appearing in the difference of $\Delta=2$ and $\Delta=1$ boundary conditions for the scalar. While the latter prediction of the duality has been argued to follow from the duality with $\Delta=1$ \cite{Hartman:2006dy,Giombi:2011ya}, to date there has been no \emph{direct} test of the duality for either boundary condition owing to the lack of a full quantum action in the bulk.\footnote{See however \cite{Sleight:2017cax}.} However, in the case of higher-spin gauge theories, considering loop Witten diagrams in the \emph{difference} of $\Delta=2$ and $\Delta=1$ boundary conditions can still teach us a lot about the properties of higher-spin gauge theories, in particular their Witten diagram expansion and how the infinite spectrum/expansion in derivatives should be treated.

Motivated by the above considerations, in \S \tcb{\ref{subsubsec::altquantads4}} we study the contributions to the anomalous dimensions of higher-spin currents from 2pt bubble and $\bep(14,10)\put(0,0){\line(1,0){14}}\put(6,0){\line(0,1){4}}\put(6,7){\circle{6}}\eep$ tadpole diagrams which appear in the difference of $\Delta=2$ and $\Delta=1$ scalar boundary conditions. We leave for the future a complete analysis of the duality in the case of $\Delta=1$ boundary condition, for which all cubic and quartic couplings, as well as the corresponding ghost couplings, must be included. Our analysis allows us to determine the nature of the various types of bulk one-loop contributions to the anomalous dimension of higher-spin currents in the critical $O\left(N\right)$ model. In particular, we find that 2pt bubble diagrams alone are not sufficient to reproduce the anomalous dimensions, and for this $\bep(14,10)\put(0,0){\line(1,0){14}}\put(7,4.25){\circle{8}}\eep$ tadpole diagrams are required. We also point out a puzzle regarding the infinite summation over spin and the Witten diagram expansion.

\subsection{General approach}\label{sec::genA}

We develop a spectral approach to evaluate AdS loop diagrams, a central ingredient for which is the decomposition of bulk-to-bulk propagators $G\left(x_1,x_2\right)$ into bi-tensorial AdS harmonic functions $\Omega\left(x_1,x_2\right)$ \cite{Costa:2014kfa,Bekaert:2014cea}, which we depict as:
\begin{equation}
    \includegraphics[scale=0.45]{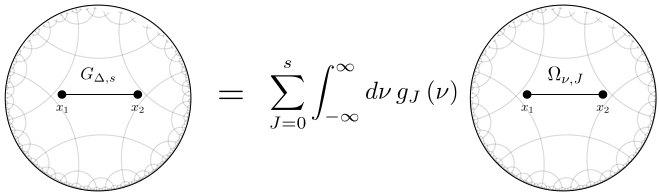}.
\end{equation}
The factorisation of harmonic functions into bulk-to-boundary propagators integrated over the common boundary point \cite{Leonhardt:2003qu}:
\begin{equation}
    \includegraphics[scale=0.45]{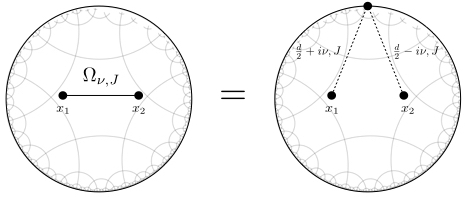},
\end{equation}
leads to the decomposition of loop diagrams into integrated products of higher point tree-level Witten diagrams. Upon evaluating the comparably simple tree-level Witten diagrams, the loop is reduced to the computation of well-known boundary conformal integrals \cite{Symanzik1972} arising from the gluing of the tree-level bulk diagrams, and a spectral integral in the parameters $\nu$.

\begin{figure}[h]
    \centering
    \begin{subfigure}[b]{0.8\textwidth}
        \includegraphics[width=\textwidth]{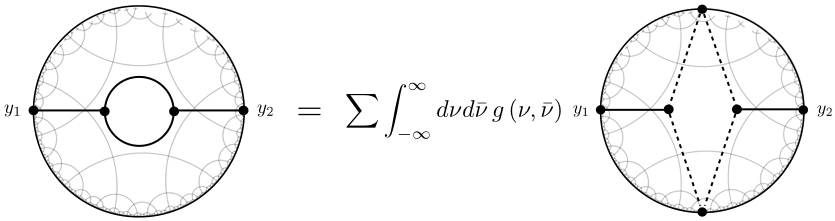}
        \caption{}
        \label{fig::genapp1}
    \end{subfigure}

    \begin{subfigure}[b]{0.8\textwidth}
        \includegraphics[width=\textwidth]{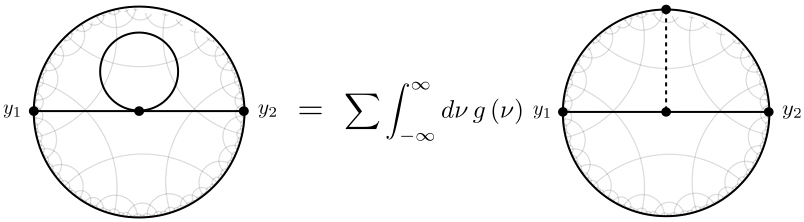}
        \caption{}
        \label{fig::genapp2}
    \end{subfigure}
\caption{}
\label{fig::genapp}
\end{figure}

In this work, we detail this approach for two-point bubble and tadpole diagrams, which induce mass and wave-function renormalisations of the fields which already appear at tree-level. In this case, the task is reduced to the evaluation of tree-level three-point Witten diagrams (illustrated in figures \ref{fig::genapp1} and \ref{fig::genapp2}) which, via the sewing procedure shown in figure \ref{fig::genapp}, give rise to the following three- and, ultimately, two-point conformal integrals:
\begin{subequations}
\begin{align} 
    I_{\text{3pt}}\left(y_1,y_2,y_3\right) & = \int \frac{d^dy}{\left[\left(y_1-y\right)^2\right]^{a_1}\left[\left(y_2-y\right)^2\right]^{a_2}\left[\left(y_3-y\right)^2\right]^{a_3}}, \qquad a_1+a_2+a_3=d, \label{int3ptconf}\\
    \label{int2ptconf}
    I_{\text{2pt}}\left(y_1,y_2\right) & = \int \frac{d^dy}{\left[\left(y_1-y\right)^2\right]^{a_1}\left[\left(y_2-y\right)^2\right]^{a_2}}, \qquad a_1+a_2=d,
\end{align}
\end{subequations}
whose evaluation we give in \S \tcb{\ref{app::confint}}. The two-point integral \eqref{int2ptconf} is divergent, whose regularisation gives rise to the corrections to the wave function and the mass.

For external totally symmetric fields of spin $s$ and tree-level mass $m^2_i R^2=\Delta_i\left(\Delta_i-d\right)-s$, the two-point one-loop diagrams ultimately take the form\footnote{For tadpole diagrams, which have just a single bulk-to-bulk propagator, there is only one spectral integral while for bubble diagram (which instead involve two bulk-to-bulk propagators) there is a double integral as shown above. We emphasise that the presence of the divergent two-point conformal integral on the second line is \emph{universal}. I.e. is generated by any one-loop process, both bubble and tadpole diagrams.} 
\begin{multline}\label{m1loop}
    {\cal M}^{\text{1-loop}}\left(y_1,y_2\right) = \int^\infty_{-\infty}d\nu d{\bar \nu} {\cal F}\left(\nu,{\bar \nu }\right)\\ \times  \frac{{\sf H}^s_{12}}{\left(y^{2}_{12}\right)^{(\tau_1+\tau_2-d)/2}}  \int 
  \frac{d^{d}y}{\left[\left(y_1-y\right)^2\right]^{d/2+(\Delta_1-\Delta_2)/2}\left[\left(y_2-y\right)^2\right]^{d/2-(\Delta_1-\Delta_2)/2}},
\end{multline}
for some spectral function ${\cal F}\left(\nu,{\bar \nu }\right)$. We employ a variant of dimensional regularisation to evaluate the conformal integral on the second line,\footnote{See \S \tcb{\ref{app::two2ptreg}} and \S \tcb{\ref{app::biar}} for a discussion on possible choices of regularisation, including at the level of the bulk harmonic function \eqref{spinharmfact}.} which yields
\begin{align}\label{M12}\nonumber
  I^{\text{1-loop}}_{\text{2pt}}\left(y_1,y_2\right) & = \frac{{\sf H}^s_{12}}{\left(y^{2}_{12}\right)^{(\tau_1+\tau_2-d)/2}}  \int 
  \frac{d^{d+\epsilon}y}{\left[\left(y_1-y\right)^2\right]^{d/2+(\Delta_1-\Delta_2)/2}\left[\left(y_2-y\right)^2\right]^{d/2-(\Delta_1-\Delta_2)/2}}  \\ 
  &=\delta_{\Delta_1,\Delta_2}\ \frac{\pi^{\frac{d+\epsilon}{2}}}{\Gamma(\tfrac{d}{2})^2}\frac{{\sf H}^s_{12}}{\left(y^{2}_{12}\right)^{(\tau_1+\tau_2-\epsilon)/2}}\frac{\Gamma \left(\frac{\epsilon }{2}\right)^2 \Gamma \left(\frac{d-\epsilon }{2}\right)}{\Gamma (\epsilon )}\nonumber\\
  &= \delta_{\Delta_1,\Delta_2}\ \frac{2\pi^{\frac{d}{2}}}{\Gamma(\tfrac{d}{2})}\frac{{\sf H}^s_{12}}{\left(y^{2}_{12}\right)^{(\tau_1+\tau_2)/2}}\left[\frac{2}{\epsilon}+\log(\pi)-\psi\left(\tfrac{d}2\right)+ \log\left(y^{2}_{12}\right)\right]+\mathcal{O}(\epsilon),
\end{align}
where the constant piece generates the wave function renormalisation and the log term the mass correction.\footnote{This can be understood from the expansion of the dual CFT two-point function
\begin{align}\nonumber
    \langle {\cal O}_{\Delta_1,s}\left(y_1\right){\cal O}_{\Delta_2,s}\left(y_2\right) \rangle & = \delta_{\Delta_1 \Delta_2}\,C_{{\cal O}} \frac{{\sf H}^s_{12}}{\left(y^2_{12}\right)^{\tau_1+\gamma}}\\ \nonumber &= \delta_{\Delta_1 \Delta_2}\,C_{{\cal O}} \frac{{\sf H}^s_{12}}{\left(y^2_{12}\right)^{\tau_1}}e^{-\gamma \log\left(y^2_{12}\right) } = \delta_{\Delta_1 \Delta_2}\,C_{{\cal O}} \frac{{\sf H}^s_{12}}{\left(y^2_{12}\right)^{(\tau_1+\tau_2)/2}}\left(1-\gamma \log\left(y^2_{12}\right) + ... \right),
\end{align}
where we see that the anomalous dimension, which is related to the corrected bulk mass via  $m^2R^2=\left(\Delta+\gamma\right)\left(\Delta+\gamma-d\right)-s$, is the coefficient of the log term.} Combining \eqref{M12} with \eqref{m1loop} thus gives the anomalous dimension in the spectral form
\begin{align}
\gamma \sim - \delta_{\Delta_1 \Delta_2} \int^\infty_{-\infty}d\nu d{\bar \nu} \,{\cal F}\left(\nu,{\bar \nu }\right).
\end{align}
The above procedure is not only computationally convenient, but also turns out to disentangle UV and IR bulk divergences. It is indeed easy to see by inspection that the spectral integrals will diverge for large values of the spectral parameter, which therefore should be considered a UV divergence. Such UV divergences translate into divergent anomalous dimensions which require regularisation. While UV finite theories will lead to well-defined predictions for the anomalous dimensions, UV divergent theories will require some subtraction scheme to extract the anomalous dimensions. In the latter case, in this paper we shall use a minimal subtraction scheme. The boundary integrals instead are by construction IR effects, which correspond to short distance singularities from the perspective of the boundary CFT. The fact that it is possible to generate anomalous dimensions even when no UV counter-term is required is a peculiarity of the IR structure of AdS space \cite{Porrati:2001db}.

All of the above spectral integrals will be of the form of Mellin-Barnes integrals, which define generalisations of hypergeometric functions:
\begin{equation}\label{GenHyp}
H^{m,n}_{p,q}\left(z\right) = \int \frac{\prod^m_{j=1}\Gamma\left(b_j-i\nu\right)\prod^n_{j=1}\Gamma\left(1-a_j+i\nu\right)}{\prod^p_{j=n+1}\Gamma\left(a_j-i\nu\right)\prod^q_{j=m+1}\Gamma\left(1-b_j+i\nu\right)}z^{i\nu} d\nu.
\end{equation}
The latter, for $z=\pm 1$ can be expressed in terms of sums of generalised hypergeometric functions of argument $\pm 1$ and can be evaluated by the Gauss hypergeometric formula. Once the anomalous dimension is extracted in terms of a spectral integral the problem of evaluating the loop diagram is drastically simplified and can be solved either analytically (when possible) or numerically. While in this work we focus on some relevant examples, we leave for the future the problem of developing a systematic analytic/numeric method to evaluate the above integrals in general in the case of multiple spectral integrals.

\subsection{Notation, conventions and ambient space}
\label{subsec::ambnotconv}

In this work we consider tensor fields in Euclidean anti-de Sitter (AdS$_{d+1}$) space where, unless specified, the boundary dimension $d$ is taken to be general. We employ an operator notation to package the tensor indices (for a review see e.g. \cite{Sleight:2017krf}, whose conventions we adopt throughout), where a totally symmetric rank-$s$ bulk field $\varphi_{\mu_1...\mu_s}$ represented by the generating function
\begin{equation}\label{genranks}
    \varphi_{\mu_1...\mu_s}\left(x\right)\:\rightarrow\: \varphi_s\left(x;u\right)=\frac{1}{s!} \varphi_{\mu_1...\mu_s}\left(x\right)u^{\mu_1}...u^{\mu_s},
\end{equation}
where we introduced the $\left(d+1\right)$-dimensional constant auxiliary vector $u^{\mu}$. The covariant derivative gets modified when acting on fields expressed in the generating function form \eqref{genranks}:
\begin{equation}\label{modincovder}
    \nabla_{\mu}\;\rightarrow\;\nabla_{\mu}+\omega^{ab}_\mu u_a \frac{\partial}{\partial u^b},
\end{equation}
where $\omega^{ab}_\mu$ is the spin connection and $u^a=e^a_{\mu}\left(x\right)u^\mu$ with vielbein $e^a_{\mu}\left(x\right)$.

One particular virtue of this notation is that tensor operations become an operator calculus, which significantly simplifies manipulations. For instance, the contraction:
\begin{equation}
    \varphi_{\mu_1...\mu_s}\left(x\right)\varphi^{\mu_1...\mu_s}\left(x\right)=s!\,\varphi_s\left(x;\partial_u\right)\varphi\left(x;u\right),
\end{equation}
and the operations:  divergence, symmetrised gradient, box, symmetrised metric, trace and spin are represented by the following operators:
\begin{align}
\textrm{divergence:  } 		& \nabla\cdot\partial_u,&
\textrm{sym.~gradient:  } 	& u\cdot\nabla,	&
	\textrm{box:  } 			& \Box , \\ 
	\textrm{sym.~metric:  } 	& u^2, \nonumber &
	\textrm{trace:  } 			& \partial_u^2,& 
	\textrm{spin:  } 			& u\cdot\partial_u .
\end{align}

Likewise, operators of non-trivial spin living on the conformal boundary of AdS$_{d+1}$ can be expressed in generating function notation. A totally symmetric spin-$s$ operator ${\cal O}_{i_1...i_s}$ at the boundary point $y^i$, $i=1,...,d$, is represented as 
\begin{equation}
    {\cal O}_{i_1...i_s}\left(y\right)\:\rightarrow\: {\cal O}_{s}\left(y;z\right)={\cal O}_{i_1...i_s}\left(y\right)z^{i_1}...z^{i_s},
\end{equation}
with the null auxiliary vector $z^2=0$ enforcing the tracelessness condition. The operator calculus is slightly modified for traceless tensors, since one must instead replace the partial derivative $\partial_z$ with the Thomas derivative \cite{10.2307/84634}:\footnote{In the CFT literature this is sometimes referred to as the Todorov differential operator \cite{Dobrev:1975ru}. The normalisation of the latter is obtained from \eqref{thomasd} by multiplying by the operator $d-2+2z\cdot \partial_z$, and recalling that $z\cdot \partial_z$ gives the spin of the operator being acted on.}
\begin{equation}\label{thomasd}
{\hat \partial}_{z^i} = \partial_{z^i} - \frac{1}{d-2+2 z \cdot \partial_z} z_i \partial^2_z,
\end{equation}
that preserves the condition $z^2=0$. For example,
\begin{equation}
   {\cal O}_{i_1,...,i_s}\left(y\right){\cal O}^{i_1,...,i_s}\left(y\right)=s!\,{\cal O}_{s}(y;{\hat \partial}_{z}){\cal O}_{s}\left(y;z\right).
\end{equation}

\subsubsection*{Ambient space}

The ambient space formalism is an indispensable tool in AdS and CFT, which simplifies computations considerably by making the $SO\left(1,d+1\right)$ symmetry manifest. We employ this formalism throughout, and briefly review the pertinent details here. For further details see e.g. \cite{Grigoriev:2011gp,Joung:2011ww,Taronna:2012gb,Bekaert:2012vt,Sleight:2017krf}.

A perspective first considered by Dirac \cite{Dirac:1936fq}, in the ambient space formalism one regards the AdS$_{d+1}$ space as the co-dimension one hyper-surface
\begin{equation}
    X^2+R^2=0,\label{hyp}
\end{equation}
in an ambient flat space-time parameterised by Cartesian co-ordinates $X^A$ where $A=0,1,..,d+1$ and metric $\eta_{AB}=\text{diag}\left(-++...+\right)$ to describe Euclidean AdS.\footnote{In contrast Lorentzian AdS would require the conformal signature: $\eta_{AB}=\text{diag}\left(-++...+-\right)$.} 

A smooth irreducible $so\left(1,d+1\right)$-tensor field $\varphi_{\mu_1...\mu_s}\left(x\right)$ of mass
\begin{equation}
    m^2 R^2=\Delta\left(\Delta-d\right)-s,
\end{equation}
 is represented uniquely in the ambient space by a field $\varphi_{A_1...A_s}\left(X\right)$ of the same rank subject to the following constraints \cite{Fronsdal:1978vb}:

\begin{itemize}
    \item {\bf Tangentiality} to surfaces of constant $\rho=\sqrt{-X^2}$:
    \begin{equation}
        X^{A_i}\varphi_{A_1...A_i...A_s}=0, \quad i=1,...,s.
    \end{equation}
    Explicitly, one can apply the projection operator:
    \begin{equation}
        {\cal P}_{A}^{B}=\delta^{B}_{A}-\frac{X_A X^B}{X^2},
    \end{equation}
    which acts on ambient tensors as
    \begin{equation}
        \left({\cal P} \varphi \right)_{A_1...A_s}:={\cal P}_{A_1}^{B_1}...{\cal P}_{A_s}^{B_s}\varphi_{B_1...B_s}, \qquad X^{A_i}\left({\cal P} \varphi\right)_{B_1...B_i...B_s}=0
    \end{equation}
    
    \item The {\bf homogeneity} condition:
    \begin{equation}\label{homo}
        \left(X \cdot \partial_X+\mu\right)\varphi_{s}\left(X,U\right)=0, \quad \text{i.e.} \quad \varphi_{s}\left(\lambda X,U\right) = \lambda^{-\mu}\varphi_{s}\left( X,U\right),
    \end{equation}
    where we are free to choose either $\mu=\Delta$ or $\mu=d-\Delta$. In this work we take $\mu=\Delta$. This fixes how the ambient representative extends away from the AdS manifold, in the radial direction $\rho=\sqrt{-X^2}$.
\end{itemize}
The above conditions ensure that the ambient uplift of fields that live on the AdS manifold is well-defined and one-to-one.  

This discussion also extends to differential operators. For instance, the ambient representative of the Levi-Civita connection $\nabla_\mu$ on AdS$_{d+1}$ is given by \cite{Metsaev:1995re,Bekaert:2010hk}:
\begin{equation}
    \nabla_A={\cal P}^B_{A} \frac{\partial}{\partial X^B}, \qquad X \cdot \nabla=0.\label{connongen}
\end{equation}
Crucially, this must act on ambient tensors that are tangent, otherwise extra terms may be introduced which are not killed by the projector acting on the LHS of \eqref{connongen}. The proper action of \eqref{connongen} should thus be regarded as:
\begin{equation}
 \nabla = {\cal P} \circ \partial \circ {\cal P}.
\end{equation}
For example:
\begin{equation}
\nabla_B T_{A_1...A_r} = {\cal P}^{C}_{B}{\cal P}^{C_1}_{A_1}...{\cal P}^{C_r}_{A_r}\frac{\partial}{\partial X^C}\left({\cal P}T\right)_{C_1...C_r},
\end{equation}
for some ambient tensor $ T_{A_1...A_r}\left(X\right)$.

The operator notation for tensor fields introduced in the previous section can also be extended to ambient space. We have:
\begin{equation}
    \varphi_{A_1..A_s}\left(X\right)\:\rightarrow \: \varphi_s\left(X;U\right)=\frac{1}{s!} \varphi_{A_1..A_s}\left(X\right)U^{A_1}...U^{A_s},
\end{equation}
with constant ambient auxiliary vector $U^A$. Like for the intrinsic case \eqref{modincovder}, the covariant derivative \eqref{connongen} also gets modified in the operator formalism \cite{Taronna:2012gb}:
\begin{equation}
    \nabla_A\:\rightarrow \:  \nabla_A-\frac{X^B}{X^2}\Sigma_{AB},
\end{equation}
where 
\begin{equation}
    \Sigma_{AB}=U_A \frac{\partial}{\partial U^B}-U_B \frac{\partial}{\partial U^A}.
\end{equation}\vspace*{0.2cm}

The ambient formalism extends to the boundary of AdS \cite{Dirac:1935zz,Dirac:1936fq,Fronsdal:1978vb,Bars:1998ph,Bekaert:2009fg,Costa:2011mg,Bekaert:2012vt}. Towards the boundary, the hyperboloid \eqref{hyp} asymptotes to the light-cone. This limit does not give rise to a well-defined boundary metric, but a finite limit can be obtained by considering a projective cone of light-rays:
\begin{equation}
    P^A\equiv \epsilon X^A, \quad \epsilon \rightarrow 0.
\end{equation}
Since $X^2$ is fixed, these null co-ordinates satisfy:
\begin{equation}
    P^2=0, \qquad P\cong \lambda P, \qquad \lambda \neq 0,
\end{equation}
and are identified with the AdS boundary. For example, for Euclidean AdS in Poincar\'e co-ordinates $x^\mu=\left(z,y^i\right)$, we have:
\begin{subequations}
\begin{align}\label{poinamb}
    X^0\left(x\right) &= R\frac{z^2+y^2+1}{2z} \\
    X^{d+1}\left(x\right) &=R\frac{1-z^2-y^2}{2z} \\
    X^i\left(x\right) & = \frac{Ry^i}{z}, 
\end{align}
\end{subequations}
and the boundary points are parameterised by the Poincar\'e section:
\begin{equation}\label{sect}
  P^{0}\left(y\right)=\frac{1}{2}\left(1+y^2\right), \quad P^{d+1}\left(y\right)=\frac{1}{2}\left(1-y^2\right), \quad P^i\left(y\right) = y^i.
  \end{equation}

The ambient representative $f_{A_1 ...A_s}\left(P\right)$ of a symmetric spin-$s$ boundary field $f_{i_1...i_s}\left(y\right)$ of scaling dimension $\Delta$ is traceless with respect to the ambient metric\footnote{It is not difficult to see that this follows from the tracelessness of $f_{i_1...i_s}$.}
\begin{equation}
    \eta^{AB}f_{A_1 ...A_s}=0 \label{tlessbound}
\end{equation}
and scales as
\begin{equation}
    f_{A_1 ...A_s}\left(\lambda P\right)=\lambda^{-\Delta}f_{A_1 ...A_s}\left(P\right), \qquad \lambda > 0.
\end{equation}
Like for the ambient description of bulk fields, we require that $f_{A_1 ...A_s}$ is tangent to the light-cone:
\begin{equation}
    P^{A_1}f_{A_1 ...A_s}\left(P\right)=0.\label{transbound}
\end{equation}
However, since $P^2=0$, there is an extra redundancy
\begin{align}\label{extred}
 & \hspace*{3.5cm} f_{A_1...A_s}(P) \rightarrow f_{A_1...A_s}(P) + P_{\left(A_1\right.}\Lambda_{\left. A_2 ... A_s \right)},\\
 & P^{A_1}\Lambda_{ A_1 ... A_{s-1}} = 0, \quad \Lambda_{ A_1 ... A_{s-1}}(\lambda P) = \lambda^{-(\Delta+1)}\Lambda_{ A_1 ... A_{s-1}}(P), \quad \eta^{A_1A_2}\Lambda_{A_1 ... A_{s-1}} = 0,
\end{align}
which, together with \eqref{transbound}, eliminates the extra two degrees of freedom per index of $f_{A_1...A_s}$.

Likewise the operator formalism extends to ambient boundary fields, where we have:
\begin{equation}
    f_{A_1...A_s}\left(P\right)\:\rightarrow\: f_s\left(P;Z\right)=\frac{1}{s!}f_{A_1...A_s}\left(P\right)Z^{A_1}...Z^{A_s}, \quad Z^2=0, \quad P\cdot Z=0,
\end{equation}
where as usual $Z^2=0$ enforces the traceless condition \eqref{tlessbound} and it is useful to impose the new constraint $P\cdot Z=0$ that takes care of tangentiality to the light-cone \eqref{transbound}.

\section{Scalar diagrams}\label{sec::scalardiagrams}

For ease of illustration, we first consider two-point one-loop diagrams involving only scalar fields. We review the basic ingredients below before giving some concrete applications in \S \tcb{\ref{subsec::scalar2ptbub}} and \S \tcb{\ref{subsec::2pttad}}.

Bulk-to-boundary propagators take a very simple form in ambient space. See \S \tcb{\ref{subsec::ambnotconv}} for a review of the ambient space formalism. For a scalar of mass $m^2 R^2 = \Delta\left(\Delta-d\right)$, the bulk-to-boundary propagator\footnote{In the limit we used Poincar\'e co-ordinates \eqref{poinamb} with $x^\mu=\left(z,{\bar y}^i\right)$, where the ${\bar y}^i$ with $i=1,...,d$ parameterise the boundary directions.} 
\begin{equation}
    \left(-\Box+m^2\right)K_{\Delta,0}\left(x;y\right)=0, \qquad \lim_{z \rightarrow 0}\left(z^{\Delta-d}K_{\Delta,0}\left(z,{\bar y};y \right)\right) = \frac{1}{2\Delta-d} \delta^d\left(y-{\bar y}\right),
\end{equation}
is given by the contraction:
\begin{equation}\label{scbubou}
    K_{\Delta,0}\left(X\left(x\right);P\left(y\right)\right) = \frac{C_{\Delta,0}}{\left(-2X \cdot P\right)^{\Delta}},
\end{equation}
with normalisation:
\begin{equation}\label{scbubounorm}
  C_{\Delta,0} = \frac{\Gamma\left(\Delta\right)}{2\pi^{d/2}\Gamma\left(\Delta+1-\tfrac{d}{2}\right)}.
 \end{equation}

We employ the spectral representation of the bulk-to-bulk propagators, which for scalar fields with $\Delta> \frac{d}{2}$ is given by\footnote{The case $\Delta < \frac{d}{2}$ requires a slight modification of the propagator, but the general approach for evaluating loop diagrams is unchanged. This is explained later on in \S \tcb{\ref{subsubsec::altquantads4}}.} 
\begin{align}\label{scalarbubu}
    G_{\Delta,0}\left(x_1;x_2\right) & = \int^\infty_{-\infty}\frac{ d\nu}{ \left[\nu^2+\left(\Delta-\frac{d}{2}\right)^2\right]} \Omega_{\nu,0}\left(x_1,x_2\right),
\end{align}
where $\Omega_{\nu,0}$ is a spin $0$ bi-tensorial harmonic function with equation of motion
\begin{equation}
  \left(\Box_1 +\left(\tfrac{d}{2}\right)^2+\nu^2\right)\Omega_{\nu,0}\left(x_1,x_2\right)=0,
\end{equation}
where the subscript ${}_i$ on differential operators signifies that the derivative is being taken with respect to $x_i$. As is illustrated in figure \ref{fig::genapp}, the factorisation 
\begin{equation}\label{scafacharm}
 \Omega_{\nu,0}\left(x_1,x_2\right) = \frac{\nu^2}{\pi} \int_{\partial \text{AdS}} d^dy\,K_{\frac{d}{2}+i\nu.0}\left(x_1;y\right)K_{\frac{d}{2}-i\nu,0}\left(x_2;y\right),  
\end{equation}
of harmonic functions into bulk-to-boundary propagators \eqref{scbubou} re-expresses two-point one-loop diagrams in terms of conformal integrals of tree-level three-point Witten diagrams. For diagrams involving only scalar fields,  the three-point Witten diagrams are those generated by the basic vertex\footnote{Note that this vertex is the unique cubic vertex of scalars on-shell.}
\begin{equation}\label{basic123cubic}
    {\cal V}^{\left(3\right)} = \phi_1 \phi_2 \phi_3,
\end{equation}
of scalars $\phi_i$ of some mass $m^2_i R^2 = \Delta_i\left(\Delta_i-d\right)$. The tree-level amplitude generated by \eqref{basic123cubic} is well known \cite{Muck:1998rr}, and given in the ambient formalism (see \S \tcb{\ref{subsec::ambnotconv}}) by 
\begin{equation}\label{123scal}
 {\cal M}^{\text{3pt tree}}_{\Delta_1,\Delta_2,\Delta_3}\left(P_1,P_2,P_3\right) 
 =   \frac{{\sf B}\left(\Delta_1,\Delta_2,\Delta_3; 0\right)}{P^{\frac{\Delta_1+\Delta_3-\Delta_2}{2}}_{13}P^{\frac{\Delta_2+\Delta_3-\Delta_1}{2}}_{23}P^{\frac{\Delta_1+\Delta_2-\Delta_3}{2}}_{12}},
\end{equation}
 where $P_{ij}=-2 P_i \cdot P_j$ and 
\begin{multline}
   {\sf B}\left(\Delta_1,\Delta_2,\Delta_3; 0\right)  = \;\frac{1}{2}\pi^{\frac{d}{2}}\Gamma\left(\frac{- d + \sum\nolimits^3_{i=1} \Delta_i}{2}\right) C_{\Delta_1,0}C_{\Delta_2,0}C_{\Delta_3,0} \\ \times \frac{\Gamma\left(\frac{\Delta_1+\Delta_2-\Delta_3}{2}\right)\Gamma\left(\frac{\Delta_1+\Delta_3-\Delta_2}{2}\right)\Gamma\left(\frac{\Delta_2+\Delta_3-\Delta_1}{2}\right)}{\Gamma\left(\Delta_1\right)\Gamma\left(\Delta_2\right)\Gamma\left(\Delta_3\right)}.
\end{multline}
The $C_{\Delta_i,0}$ come from the normalisation \eqref{scbubounorm} of the bulk-to-boundary propagator.

 In \S \tcb{\ref{subsec::scalar2ptbub}} we use this approach to evaluate the two-point one-loop bubble diagram in $\phi^3$ theory. In \S \tcb{\ref{subsec::2pttad}} we move on to tadpole diagrams, showing in \S \tcb{\ref{subsubsec::phi4vertex}} how they are evaluated in $\phi^4$ theory. We extend the latter result to arbitrary derivative quartic self-interactions in \S \tcb{\ref{subsubsec::derivativeint}}.

\subsection{2pt bubble}
\label{subsec::scalar2ptbub}

\begin{figure}[h]
\centering
\includegraphics[scale=0.6]{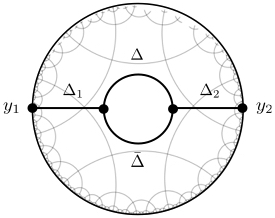}
\caption{Scalar one-loop bubble diagram generated by the cubic couplings \eqref{cusc12}.}
\label{fig::2ptbubscalar}
\end{figure}

We consider the two-point one-loop bubble illustrated in figure \ref{fig::2ptbubscalar}, which is generated by the following cubic couplings:\footnote{In this subsection we drop symmetry factors associated to indistinguishable external legs. In the case of indistinguishable scalar fields, the corresponding symmetry factor is $S=\frac12$.}
\begin{equation}\label{cusc12}
    {\cal V}^{(3)}_1 = g\, \phi_1 \phi {\bar \phi}, \qquad {\cal V}^{(3)}_2 = {\bar g}\, \phi_2 \phi {\bar \phi},
\end{equation}
for arbitrary coupling constants $g$ and ${\bar g}$. The diagram is given by evaluating the bulk integrals
\begin{multline}\label{2ptbubampsca}
{\cal M}^{\text{2pt bubble}}\left(P_1,P_2\right) \\ = g {\bar g} \int_{\text{AdS}} dX_1dX_2\,K_{\Delta_1,0}\left(X_1;P_1\right)G_{\Delta,0}\left(X_1;X_2\right) G_{{\bar \Delta},0}\left(X_1;X_2\right)K_{\Delta_2,0}\left(X_2;P_2\right).
\end{multline}
The spectral representation \eqref{scalarbubu} of the scalar bulk-to-bulk propagators expresses the diagram in terms of two tree-level three-point Witten diagrams \eqref{123scal}, sewn together by their common boundary points (see  figure \ref{fig::genapp1}):
\begin{multline}\label{2ptbubampscafact}
{\cal M}^{\text{2pt bubble}}\left(P_1,P_2\right) = g{\bar g} \int^\infty_{-\infty} \frac{\nu^2 {\bar \nu}^2 d\nu d{\bar \nu}}{\pi^2[\nu^2+(\Delta-\frac{d}{2})^2][{\bar \nu}^2+({\bar \Delta}-\frac{d}{2})^2]}  \\ \times 
\int_{\partial \text{AdS}} dP d{\bar P} \, {\cal M}_{\Delta_1,\,\frac{d}{2}+i\nu,\,\frac{d}{2}+i{\bar \nu}}^{\text{3pt tree}}(P_1,P,{\bar P}) \,{\cal M}_{\Delta_2,\,\frac{d}{2}-i\nu,\,\frac{d}{2}-i{\bar \nu}}^{\text{3pt tree}}(P_2,P,{\bar P}).
\end{multline}

The integrals in $P$ and $\bar P$ are both of the three-point conformal type \eqref{int3ptconf}. Performing first, say, the integration over $\bar P$ leaves the two-point conformal integral \eqref{int2ptconf}:

\begin{multline}
{\cal M}^{\text{2pt bubble}}\left(P_1,P_2\right) = 
\frac{g {\bar g} }{64 \pi ^{\frac{d+8}{2}}} \frac{ C_{\Delta_1,0}C_{\Delta_2,0}}{\Gamma \left(\Delta_1\right)\Gamma \left(\Delta_2\right)} \frac{\Gamma \left(\frac{\Delta_1+\Delta_2-d}{2}\right)}{\Gamma \left(d-\frac{\Delta_1+\Delta_2}{2}\right)}
\int^\infty_{-\infty} 
d\nu d{\bar \nu}\,{\cal F}^{\text{2pt bubble}}\left(\nu,{\bar \nu}\right)\\ \times \underbrace{P_{12}^{\frac{d-\Delta_1-\Delta_2}{2}} \int \frac{dP}{\left(-2 P_1 \cdot P\right)^{\frac{1}{2}\left(d+\Delta_1-\Delta_2\right)} \left(-2 P_2 \cdot P\right)^{\frac{1}{2}\left(d+\Delta_2-\Delta_1\right)}}}_{=I^{\text{1-loop}}_{\text{2pt}}\left(y_1,y_2\right)\ \eqref{M12}}
\end{multline}
where
\begin{multline}\label{phi3specfun}
{\cal F}^{\text{2pt bubble}}\left(\nu,{\bar \nu}\right)=
\frac{\nu {\bar \nu} \sinh (\pi  \nu) \sinh (\pi  {\bar \nu})}{\left[(\tfrac{d}{2}-{\bar \Delta} )^2+{\bar \nu}^2\right] \left[(\tfrac{d}{2}-\Delta)^2+\nu^2\right]}\\
\times \Gamma \left(\tfrac{d-\Delta_1-i (\nu-{\bar \nu})}{2} \right)\Gamma \left(\tfrac{d-\Delta_1+i (\nu+{\bar \nu})}{2}\right)
\Gamma \left( \tfrac{\Delta_1-i (\nu-{\bar \nu})}{2}\right) \Gamma \left(\tfrac{\Delta_1+i (\nu+{\bar \nu})}{2}\right) 
 \\ 
\times  \Gamma \left(\tfrac{d-\Delta_2+i (\nu-{\bar \nu})}{2} \right)\Gamma \left(\tfrac{d-\Delta_2-i (\nu+{\bar \nu})}{2}\right) \Gamma \left(\tfrac{\Delta_2+i (\nu-{\bar \nu})}{2}\right)
\Gamma \left(\tfrac{\Delta_2-i (\nu+{\bar \nu})}{2}\right) \,.
\end{multline}
Focusing on the $\log(y_{12}^2)$ contribution, we can thus extract the leading correction to the anomalous dimension as the following spectral integral: 
\begin{multline}\label{anomscalar}
    \gamma= -g {\bar g}\, \delta_{\Delta_1 \Delta_2}\frac{\Gamma \left(\frac{\Delta_1+\Delta_2-d}{2}\right)}{64 \pi ^{\frac{d+8}{2}} \Gamma \left(\frac{d}{2}\right) \Gamma \left(d-\frac{\Delta_1+\Delta_2}{2}\right)}\frac{1}{ \sqrt{\Gamma (\Delta_1) \Gamma \left(-\frac{d}{2}+\Delta_1+1\right)} \sqrt{\Gamma (\Delta_2) \Gamma \left(-\frac{d}{2}+\Delta_2+1\right)}}\\ \times \int^\infty_{-\infty} d\nu d{\bar \nu}\,{\cal F}^{\text{2pt bubble}}\left(\nu,{\bar \nu}\right)
\end{multline}

In the following sections we first demonstrate how the spectral integrals may be evaluated in some simple examples, and in \S \tcb{\ref{subsec::sumoverresidues}} we detail a general analytic approach based on summing over residues. In \S \tcb{\ref{subsec::ssprime0}} we also discuss the pole structure of the spectral function \eqref{phi3specfun}.

\subsubsection{Conformally coupled scalar ($\Delta=2$) in AdS$_4$}

The simplest case is that of the self-coupling of a conformally coupled scalar in AdS$_4$, i.e.:
\begin{equation}
    {\cal V}^{(3)}_1={\cal V}^{(3)}_2=\frac{g}{3!}\, \phi^3, \label{examphi3}
\end{equation}
with $\Delta=2$. In this section all formulas below will include the corresponding symmetry factor $S=\tfrac12$.

In this case the spectral representation of the anomalous dimension \eqref{anomscalar} is:
\begin{equation}\label{gammaphi324}
    \gamma=-S\,g^2\int_{\mathbb{R}^2}\frac{\nu \bar{\nu} (\nu-\bar{\nu}) (\nu+\bar{\nu}) \sinh (\pi  \nu) \sinh (\pi  \bar{\nu}) \text{csch}(\pi  (\nu-\bar{\nu})) \text{csch}(\pi  (\nu+\bar{\nu}))}{\pi ^2 \left(4 \nu^2+1\right) \left(4 \bar{\nu}^2+1\right)}\,.
\end{equation}
To study the above integral it is convenient to make the following change of variables:
\begin{align}
x&=\nu+{\bar \nu}\,,& y&=\nu-{\bar \nu}\,,
\end{align}
through which the \eqref{gammaphi324} becomes:
\begin{equation}
    \gamma=-\frac{S\,g^2}{2\pi ^2}\int_0^\infty dx\int_0^\infty dy\,\underbrace{\frac{x y \left(x^2-y^2\right) \text{csch}(\pi  x) \text{csch}(\pi  y) \sinh \left(\frac{\pi}{2}(x-y)\right) \sinh \left(\frac{\pi}{2}(x+y)\right)}{ \left((y-x)^2+1\right) \left((x+y)^2+1\right)}}_{I(x,y)}\,,
\end{equation}
where we have used the symmetries of the integral to restrict the region of integration to the first quadrant of the plane.
In the above form it is straightforward to identify the singularity of the integral which arises for $x\to \infty$ or $y\to\infty$ from the asymptotic behavior the integrand:
\begin{align}
I(x,y)&\sim \frac{1}{x}+\mathcal{O}\left(\frac{1}{x^3}\right)\qquad y\ \text{fixed}\,,\\
I(x,y)&\sim \frac{1}{y}+\mathcal{O}\left(\frac{1}{y^3}\right)\qquad x\ \text{fixed}\,.
\end{align}
A standard way to regularise integrals of the above type is to use $\zeta$-function regularisation, which entails introducing a parameter $\mu$:
\begin{equation}\label{gammmusc}
    \gamma\left(\mu\right)=-\frac{S\,g^2}{2\pi ^2}\int_0^\infty dx\int_0^\infty dy\underbrace{\frac{x y \left(x^2-y^2\right) \text{csch}(\pi  x) \text{csch}(\pi  y) \sinh \left(\frac{1}{2} \pi  (x-y)\right) \sinh \left(\frac{1}{2} \pi  (x+y)\right)}{ \left((y-x)^2+1\right)^{1+\mu} \left((x+y)^2+1\right)^{1+\mu}}}_{I^{\mu}(x,y)}\,,
\end{equation}
where, taking a minimal subtraction scheme, the anomalous dimension is given by the finite part as $\mu \rightarrow 0$:
\begin{equation}
    \gamma = \text{finite}\left[\gamma\left(0\right)\right].
\end{equation}

The integral \eqref{gammmusc} is convergent for $\mu$ sufficiently big. For such values of $\mu$ the above integral can be split into two integrals, one of which is convergent for $\mu\to 0$ while the other is divergent:\footnote{This generalises the approach suggested by Camporesi and Higuchi \cite{Camporesi:1994ga}.}
\begin{align}
I^{(\mu)}(x,y)=I_1^{(\mu)}(x,y)+I_2^{(\mu)}(x,y)\,,
\end{align}
with
\begin{align}
I_1^{(\mu)}(x,y)\Big|_{\mu=0}&=\frac{x y}{2} \left[\frac{(y-x) (x+y) \text{csch}(\pi  y)\text{csch}(\pi  x) (\cosh (\pi  x)-\cosh (\pi  y))}{\left((y-x)^2+1\right) \left((x+y)^2+1\right)}+\frac{x^2\text{csch}(\pi  y)}{\left(x^2+1\right)^2}\right.\nonumber\\&\hspace{20pt}\left.+\frac{y^2 \text{csch}(\pi  x)}{\left(y^2+1\right)^2}\right]\,,\\
I_2^{(\mu)}(x,y)&=-\frac12\left[x^3 y \left(x^2+1\right)^{-2 (\mu +1)} \text{csch}(\pi  y)+x y^3 \text{csch}(\pi  x) \left(y^2+1\right)^{-2 (\mu +1)}\right]\,.
\end{align}
The first integral can be evaluated numerically and gives:
\begin{equation}
    \int_0^\infty dx\int_0^\infty dy\, I_1^{(0)}(x,y)=0.0289829\,.
\end{equation}
The second integral diverges, but can be evaluated analytically for arbitrary $\mu$ as:
\begin{equation}
   \int_0^\infty dx\int_0^\infty dy\, I_2^{(\mu)}(x,y) = -\frac{1}{32 \mu ^2+16 \mu }\sim -\frac{1}{16 \mu}+\frac{1}{8}+O\left(\mu\right)
\end{equation}
The final result for the anomalous dimension can thus be given numerically as:
\begin{equation}
    \gamma=0.0156017\; \times \; S\,g^2.
\end{equation}

\subsubsection{$\Delta=3/2$ in AdS$_3$}\label{Delta32}

Another simple case that we can study in detail is that of the coupling \eqref{examphi3} with $\Delta=3/2$ in AdS$_3$, for which we have:
\begin{equation}
    \gamma= -\frac{8\,S\,g^2}{\pi^2}\,\int_{\mathbb{R}^2}\underbrace{\frac{\nu \bar{\nu} \sinh (\pi  \nu) \sinh (\pi  \bar{\nu})}{\left(4 \nu^2+1\right) \left(4 \bar{\nu}^2+1\right) (\cosh (2 \pi  \nu)+\cosh (2 \pi  \bar{\nu}))}}_{I^{(\mu=0)}(\nu,\bar{\nu})/4}\,.
\end{equation}
Like in the previous example, also in this case using a $\zeta$-function regulator we can split the above integral into a convergent piece which we can directly evaluate at $\mu=0$ and a divergent piece which we can analytically continue. Considering the same change of variables $x=\nu+{\bar \nu}$ and $y=\nu-{\bar \nu}$, we have:
\begin{equation}
    {\cal F}^{\text{2pt bubble}}\left(\nu,{\bar \nu}\right)\qquad
    \rightarrow\qquad I^{(\mu)}(x,y)=I_1^{(\mu)}(x,y)+I_2^{(\mu)}(x,y)\,,
\end{equation}
with
\begin{align}
I_1^{(0)}(x,y)&=\frac{y^2 \text{sech}(\pi  x)}{4\left(y^2+1\right)^2}+\frac{x^2 \text{sech}(\pi  y)}{4 \left(x^2+1\right)^2}-\frac{\left(e^{\pi  y}-e^{\pi  x}\right) \left(e^{\pi  (y+x)}-1\right) (y-x) (y+x)}{2\left(e^{2 \pi  y}+1\right) \left(e^{2 \pi  x}+1\right) \left((y-x)^2+1\right) \left((y+x)^2+1\right)}\,,\\
I_2^{(\mu)}&=\frac14\left(x^2 \text{sech}(\pi  y) \left(x^2+1\right)^{-2 (\mu +1)}-y^2 \text{sech}(\pi  x)\left(y^2+1\right)^{-2 (\mu +1)}\right)\,.
\end{align}
The first integral can be evaluated numerically and gives:
\begin{equation}
    \int_0^\infty dx\,\int_0^{\infty} dy\, I_1^{(0)}(x,y)=0.0278017\,,
\end{equation}
while the second can be evaluated explicitly as
\begin{equation}
    \int_0^\infty dx\, \int_0^\infty dy\,I_2^{(\mu)}(x,y)=-\frac{\sqrt{\pi }\, \Gamma \left(2 \mu +\frac{1}{2}\right)}{16 \Gamma (2 \mu +2)}\sim -\frac{\pi }{16}+O\left(\mu\right)\,.
\end{equation}
The final numerical result for the anomalous dimension is:
\begin{equation}
    \gamma=-0.13662\; \times \; S\,g^2.
\end{equation}

\subsection{Summing over residues}
\label{subsec::sumoverresidues}

In this section we explain in detail the application of the standard analytic approach to Mellin Barnes integrals (as prescribed e.g. in \cite{Mellin}) to evaluate the bubble spectral integrals of the type \eqref{anomscalar}.\footnote{We thank Lorenzo Di Pietro for discussions which motivated us to give details on this approach.} This entails summing over residues. Setting for definiteness the dimension of the external legs to be equal $\Delta_1=\Delta_2=\Delta$ (for $\Delta_1\ne\Delta_2$ the result is vanishing) and re-labelling the dimension of the internal leg as $\Delta\to\Delta_1$ and $\bar{\Delta}\to\Delta_2$, we want to evaluate the following spectral integral:
\begin{subequations}\label{anomscalar2}
\begin{align}
   & \gamma= -g {\bar g}\,S \frac{\Gamma \left(\Delta-\frac{d}{2}\right)}{64 \pi ^{\frac{d+8}{2}} \Gamma \left(\frac{d}{2}\right) \Gamma \left(d-\Delta\right)}\frac{1}{ \Gamma (\Delta) \Gamma \left(-\frac{d}{2}+\Delta+1\right)} \int^\infty_{-\infty} d\nu d{\bar \nu}\,{\cal F}^{\text{2pt bubble}}\left(\nu,{\bar \nu}\right)\,\\ \label{anomscalar22}
    & {\cal F}^{\text{2pt bubble}}\left(\nu,{\bar \nu}\right)=
\frac{\nu {\bar \nu} \sinh (\pi  \nu) \sinh (\pi  {\bar \nu})}{\left[(\tfrac{d}{2}-\Delta_2 )^2+{\bar \nu}^2\right] \left[(\tfrac{d}{2}-\Delta_1)^2+\nu^2\right]}\\ \nonumber & \hspace*{3cm}
\times \Gamma \left(\tfrac{d-\Delta-i (\nu-{\bar \nu})}{2} \right)\Gamma \left(\tfrac{d-\Delta+i (\nu+{\bar \nu})}{2}\right)
\Gamma \left( \tfrac{\Delta-i (\nu-{\bar \nu})}{2}\right) \Gamma \left(\tfrac{\Delta+i (\nu+{\bar \nu})}{2}\right) 
 \\ \nonumber & \hspace*{3cm} 
\times  \Gamma \left(\tfrac{d-\Delta+i (\nu-{\bar \nu})}{2} \right)\Gamma \left(\tfrac{d-\Delta-i (\nu+{\bar \nu})}{2}\right) \Gamma \left(\tfrac{\Delta+i (\nu-{\bar \nu})}{2}\right)
\Gamma \left(\tfrac{\Delta-i (\nu+{\bar \nu})}{2}\right).
\end{align}
\end{subequations}
As before, it is convenient to change variables as
\begin{align}
    \nu&=\frac{x+y}2\,,& {\bar \nu}&=\frac{x-y}2\,.
\end{align}
In this way all $\Gamma$-functions arguments in the second and third lines of \eqref{anomscalar22} disentangle and the only place where $x$ and $y$ talk to each other is through the spectral functions of the propagators in the first line, which simplifies the extraction of residues. To wit,
\begin{multline}\label{gammanalxy}
    \gamma=-g\bar{g}S\frac{\pi ^{-\frac{d}{2}-4} \Gamma \left(\Delta -\frac{d}{2}\right)}{64 \Gamma \left(\frac{d}{2}\right) \Gamma (\Delta ) \Gamma (d-\Delta ) \Gamma \left(-\frac{d}{2}+\Delta +1\right)}\\\times \int_{-\infty}^\infty dx\, dy\,\frac{(x-y) (x+y) (\cosh (\pi  x)-\cosh (\pi  y))}{\left[(d-2 {\Delta_1})^2+(x+y)^2\right] \left[(d-2 \Delta_2)^2+(x-y)^2\right]}\\\times \Gamma \left(\tfrac{\Delta -i x}{2}\right) \Gamma \left(\tfrac{i x+\Delta}{2}\right) \Gamma \left(\tfrac{\Delta -i y}{2}\right) \Gamma \left(\tfrac{i y+\Delta}{2}\right) \Gamma \left(\tfrac{d-i x-\Delta }{2}\right) \Gamma \left(\tfrac{d+i x-\Delta}{2}\right) \Gamma \left(\tfrac{d-i y-\Delta}{2}\right) \Gamma \left(\tfrac{d+i y-\Delta}{2}\right)\,.
\end{multline}
It should be understood that the integration contours encircle all poles from a given $\Gamma$-function while separating the poles of pairs of $\Gamma$-functions whose arguments are of the type $A-ix$ and $A+ix$. In the following we shall assume that the parameters $\Delta$ and $\Delta_i$ are tuned so that the two series of poles from each such pair of $\Gamma$-functions are divided by the integration contour $x\in\mathbb{R}$.\footnote{Otherwise the contour of integration must be deformed in order to respect the separation of poles among different $\Gamma$-functions (this is standard with Mellin integrals of the type \eqref{GenHyp}, see e.g. \cite{Mellin}). This corresponds to an analytic continuation of the result obtained when no pole crosses the real axis.} The result for more general configurations of $\Delta$ and $\Delta_i$ can then be obtained by analytic continuation of the latter result. Studying the poles of the above integrand in the variable $x$, for those which sit below the integration contour we have for $n\geq 0$, $\Delta_i>\tfrac{d}{2}$ and $\Delta>\tfrac{d}{2}$:
\begin{subequations}
\begin{align}
    &\bf{A}_1&   x&=i (-d+\Delta -2 n)\\
    &\bf{A}_2&   x&=i (-\Delta -2 n)\\
    &\bf{B}&   x&=-y-i (2 \Delta_1-d)\\
    &\bf{C}&   x&=y-i (2 \Delta_2-d),
\end{align}
\end{subequations}
which have the following residues:
{\allowdisplaybreaks
\begin{subequations}
\begin{align}
    {\bf{A}}_1:&\quad -\tfrac{4 \pi  (-1)^n \left((d-\Delta +2 n)^2+ y^2\right)}{n! (d-\Delta -2 \Delta_1+2 n+i y) (d-\Delta +2 \Delta_1+2 n+i y) (d-\Delta -2 \Delta_2+2 n-i y) (d-\Delta +2 \Delta_2+2 n-i y)}\\\nonumber
    &\times\,\Gamma (d+n-\Delta ) \Gamma \left(-\tfrac{d}{2}-n+\Delta \right)\Gamma \left(\tfrac{d}{2}+n\right) \Gamma \left(\tfrac{\Delta -i y}{2}\right) \Gamma \left(\tfrac{i y+\Delta }{2}\right) \Gamma \left(\tfrac{d-i y-\Delta}{2}\right) \Gamma \left(\tfrac{d+i y-\Delta }{2}\right) \\\nonumber
    &\times[\cos (\pi  (d-\Delta ))-\cosh (\pi  y)]\\
    {\bf{A}}_2:&\quad -\tfrac{4 \pi  (-1)^n \left((\Delta +2 n)^2+y^2\right)}{n! (\Delta -2 \Delta_1+2 n+i y) (\Delta +2 \Delta_1+2 n+i y) (\Delta -2 \Delta_2+2 n-i y) (\Delta +2 \Delta_2+2 n-i y)}\\\nonumber
    &\times \,\Gamma (n+\Delta ) \Gamma \left(\tfrac{d}{2}-n-\Delta \right)\Gamma \left(\tfrac{d}{2}+n\right) \Gamma \left(\tfrac{\Delta -i y}{2}\right) \Gamma \left(\tfrac{i y+\Delta }{2}\right) \Gamma \left(\tfrac{d-i y-\Delta}{2}\right) \Gamma \left(\tfrac{d+i y-\Delta }{2}\right)\\\nonumber
    &\times [\cos (\pi  \Delta )-\cosh (\pi  y)]\\
    {\bf{B}}:&\quad -\frac{\pi  \sin (\pi  \Delta_1) (y+i \Delta_1) \sinh (\pi  (y+i \Delta_1))}{\Delta_2^2+(y+i \Delta_1)^2}\\\nonumber
    &\times \Gamma \left(\tfrac{\Delta -i y}{2}\right) \Gamma \left(\tfrac{i y+\Delta}{2}\right) \Gamma \left(\tfrac{d-i y-\Delta}{2}\right) \Gamma \left(\tfrac{d+i y-\Delta}{2}\right)\\\nonumber
    &\times \Gamma \left(\tfrac{i y+\Delta -2 \Delta_1}{2}\right) \Gamma \left(\tfrac{\Delta -i y+2\Delta_1}{2} \right) \Gamma \left(\tfrac{d+i y-\Delta -2 \Delta_1}{2}\right) \Gamma \left(\tfrac{d-i y-\Delta +2\Delta_1}{2}\right)\\\nonumber
    &\times\, [-\cosh (\pi  y)+\cos (\pi  (d-2 \Delta_1+i y))]\\
    {\bf C}:&\qquad -\frac{\pi  \sin (\pi  \Delta_2) (y-i \Delta_2) \sinh (\pi  (y-i \Delta_2))}{\Delta_1^2+(y-i \Delta_2)^2}\\\nonumber
    &\times
    \Gamma \left(\tfrac{\Delta -i y}{2}\right) \Gamma \left(\tfrac{i y+\Delta}{2}\right) \Gamma \left(\tfrac{d-i y-\Delta}{2}\right) \Gamma \left(\tfrac{d+i y-\Delta }{2}\right)\\\nonumber
    &\times\, \Gamma \left(\tfrac{-i y+\Delta -2 \Delta_2}{2}\right) \Gamma \left(\tfrac{i y+\Delta +2\Delta_2}{2}\right) \Gamma \left(\tfrac{d-i y-\Delta -2 \Delta_2}{2}\right) \Gamma \left(\tfrac{d+i y-\Delta +2\Delta_2}{2}\right)\\\nonumber
    &\times [-\cosh (\pi  y)+\cos (\pi  (d-2 \bar{\Delta}-i y))].
\end{align}
\end{subequations}}
The above reduces the double-integral in \eqref{gammanalxy} to a single integral in $y$, which can be evaluated using standard methods or again by extracting the $y$ residues. In  even-dimensions there are some simplifications due to the cancellation of an infinite number of poles in the residues of $\bf{A}_i$. In particular, in $d=2$ and $d=4$:
\begin{subequations}
\begin{align}
    d&=2\,:& \Gamma \left(\tfrac{\Delta -i y}{2}\right) \Gamma \left(\tfrac{i y+\Delta}{2}\right) \Gamma \left(\tfrac{d-i y-\Delta}{2}\right) \Gamma \left(\tfrac{d+i y-\Delta }{2}\right) (\cos (\pi  \Delta )-\cosh (\pi  y))&=-2\pi^2\,,\\
    d&=4\,:& \Gamma \left(\tfrac{\Delta -i y}{2}\right) \Gamma \left(\tfrac{i y+\Delta}{2}\right) \Gamma \left(\tfrac{d-i y-\Delta}{2}\right) \Gamma \left(\tfrac{d+i y-\Delta }{2}\right)(\cos (\pi  \Delta )-\cosh (\pi  y))&\\\nonumber
    &&=-\frac{\pi ^2}{2} &\left[(\Delta -2)^2+y^2\right]\,.
\end{align}
\end{subequations}

It is convenient to focus on dimensions in which UV divergences do not arise. Since the result does not depend on any regularisation, this also allows for straightforward comparison with other approaches. An example is given by AdS$_3$, which  in our conventions corresponds to $d=2$. We focus on this case in the following.

Defining $\delta_i=\Delta_i-\tfrac{d}2>0$, in this case the spectral integral simplifies to
\begin{align}
    \mathcal{F}(x,y)&=\frac{(x-y) (x+y) (\cosh (\pi  x)-\cosh (\pi  y))}{\left(4 \delta_1^2+(x-y)^2\right) \left(4 \delta_2^2+(x+y)^2\right)}\\\nonumber
    &\times\Gamma \left(\tfrac{-2 i x-2 \delta +2}{4}\right) \Gamma \left(\tfrac{2 i x-2 \delta +2}{4}\right) \Gamma \left(\tfrac{-2 i x+2 \delta +2}{4}\right) \Gamma \left(\tfrac{2 i x+2 \delta +2}{4}\right) \\\nonumber&\times\Gamma \left(\tfrac{-2 i y-2 \delta +2}{4}\right) \Gamma \left(\tfrac{2 i y-2 \delta +2}{4}\right) \Gamma \left(\tfrac{-2 i y+2 \delta +2}{4}\right) \Gamma \left(\tfrac{2 i y+2 \delta +2}{4}\right)\,.
\end{align}
The $x$ poles give:
\begin{subequations}
\begin{align}
    {\bf{A}}_1:&\quad \tfrac{(-\delta +2 n-i y+1) (-\delta +2 n+i y+1)}{8 \pi ^2 \delta ^2 (-\delta -2 \delta_1+2 n-i y+1) (-\delta +2 \delta_1+2 n-i y+1) (-\delta -2 \delta_2+2 n+i y+1) (-\delta +2 \delta_2+2 n+i y+1)}\\
    {\bf{A}}_2:&\quad -\tfrac{(\delta +2 n-i y+1) (\delta +2 n+i y+1)}{8 \pi ^2 \delta ^2 (\delta -2 \delta_1+2 n-i y+1) (\delta +2 \delta_1+2 n-i y+1) (\delta -2 \delta_2+2 n+i y+1) (\delta +2 \delta_2+2 n+i y+1)}\\
    {\bf{B}}:&\quad -\tfrac{\Gamma (\delta ) \sin (\pi  \delta_1) (y-i \delta_1)\sinh (\pi  (y-i \delta_1))}{64 \pi ^4 \Gamma (1-\delta ) \Gamma (\delta +1)^2 (-i \delta_1-i \delta_2+y) (-i \delta_1+i \delta_2+y)}\\\nonumber
    &\times \Gamma \left(\tfrac{-2 i y-2 \delta +2}{4}\right) \Gamma \left(\tfrac{2 i y-2 \delta +2}{4}\right) \Gamma \left(\tfrac{-i y+\delta +1}{2}\right) \Gamma \left(\tfrac{i y+\delta +1}{2}\right) \\\nonumber
    &\times\Gamma \left(\tfrac{-i y-\delta -2 \delta_1+1}{2}\right) \Gamma \left(\tfrac{-i y+\delta -2 \delta_1+1}{2}\right) \Gamma \left(\tfrac{i y+2\delta_1-\delta+1}{2}\right) \Gamma \left(\tfrac{i y+\delta +2 \delta_1+1}{2}\right)\\
    {\bf{C}}:&\quad -\tfrac{\Gamma (\delta ) \sin (\pi  \delta_2) (y+i \delta_2) \sinh (\pi  (y+i \delta_2))}{64 \pi ^4 \Gamma (1-\delta ) \Gamma (\delta +1)^2 (-i \delta_1+i \delta_2+y) (i \delta_1+i \delta_2+y)}\\\nonumber
    &\times \Gamma \left(\tfrac{-2 i y-2 \delta +2}{4}\right) \Gamma \left(\tfrac{2 i y-2 \delta +2}{4}\right) \Gamma \left(\tfrac{-i y+\delta +1}{2}\right) \Gamma \left(\tfrac{i y+\delta +1}{2}\right) \\\nonumber
    &\times\Gamma \left(\tfrac{i y-\delta -2 \delta_2+1}{2}\right) \Gamma \left(\tfrac{i y+\delta -2 \delta_2+1}{2}\right) \Gamma \left(\tfrac{-i y+2\delta_2-\delta+1}{2}\right) \Gamma \left(\tfrac{-i y+\delta +2 \delta_2+1}{2}\right)\,.
\end{align}
\end{subequations}
Taking the residue of the poles in $y$ for each of the above following the same prescription for separating the poles of each $\Gamma$-functions, we arrive to the following result for the anomalous dimension \eqref{anomscalar2} as an infinite sum:
\begin{align}
    \gamma&=-g\bar{g}S\sum_{n=0}^\infty\Bigg\{\frac1{16 \pi  \delta ^2}\Big[\tfrac{\delta -\delta_1+2 n+1}{(\delta -\delta_1+2 n+1)^2-\delta_2^2}-\tfrac{-\delta +\delta_1+2 n+1}{(-\delta +\delta_1+2 n+1)^2-\delta_2^2}+\tfrac{\delta +\delta_1+2 n+1}{(\delta +\delta_1+2 n+1)^2-\delta_2^2}+\tfrac{\delta +\delta_1-2 n-1}{(\delta +\delta_1-2 n-1)^2-\delta_2^2}\Big]\\\nonumber
    &\hspace{70pt}+\tfrac{(2 n+1) (\delta_1+\delta_2)}{2 \pi  \delta  (\delta -\delta_1-\delta_2+2 n+1) (-\delta +\delta_1+\delta_2+2 n+1) (\delta +\delta_1+\delta_2-2 n-1) (\delta +\delta_1+\delta_2+2 n+1)}\Bigg\}\\\nonumber
    &-\frac{g\bar{g}S}{{64 \delta ^2}}\sin (\pi  \delta ) \sin (\pi  \delta_1) \sin (\pi  \delta_2) \csc \left(\tfrac{-\delta +\delta_1+\delta_2+1}{2} \pi\right) \sec \left(\tfrac{\delta +\delta_1-\delta_2}{2} \pi \right) \sec \left(\tfrac{\delta -\delta_1+\delta_2}{2} \pi \right) \sec \left(\tfrac{\delta +\delta_1+\delta_2}{2} \pi \right)\,.
\end{align}
The above sums can be performed with Mathematica and give the following remarkably simple result:
\begin{multline}
    \gamma=-\frac{g\bar{g}\,S}{8\delta^2}\Big[\frac{\sin (\pi  \delta )}{\cos (\pi  \delta )+\cos (\pi  (\delta_1+\delta_2))}\\+\frac1{2\pi}\Big(\psi ^{(0)}\left(\tfrac{1-\delta -\delta_1-\delta_2}{2}\right)+\psi ^{(0)}\left(\tfrac{1+\delta +\delta_1+\delta_2}{2}\right)-\psi ^{(0)}\left(\tfrac{1+\delta -\delta_1-\delta_2}{2}\right)-\psi ^{(0)}\left(\tfrac{1-\delta +\delta_1+\delta_2}{2}\right)\Big)\Big],
\end{multline}
in terms of the polygamma function. After replacing $\delta=\Delta-\frac{d}{2}$, we then get\footnote{This formula agrees with the result independently obtained in the forthcoming \cite{pietro}, which instead employs a Hamiltonian approach for scalar fields in AdS. We thank D. Carmi, L. Di Pietro and S. Komatsu for exchanging with us examples of their independent result for a few specific values of $\Delta_1=\Delta_2=\Delta$.}
\begin{empheq}[box=\fbox]{multline}
\label{gammaSpec}
    \gamma=-\frac{g\bar{g}\, S}{8(\Delta-1)^2}\,\Big[\frac{\sin (\pi  \Delta )}{\cos (\pi  \Delta )-\cos (\pi  (\Delta_1+\Delta_2))}\\+\frac1{2\pi}\Big(H_{\frac{\Delta +\Delta_1+\Delta_2-4}{2}}+H_{\frac{2-\Delta -\Delta_1-\Delta_2}{2}}-H_{\frac{\Delta -\Delta_1-\Delta_2}{2}}-H_{\frac{-\Delta +\Delta_1+\Delta_2-2}{2}}\Big)\Big]\,,
\end{empheq}
which we also rewrote in terms of Harmonic numbers.
In particular, for $\Delta_1=\Delta_2=\Delta=3/2$ we obtain:
\begin{equation}
    \gamma=-g\bar{g}\,S\,\left(-\frac1{2}+\frac{2}{\pi}\right)\sim-\,0.13662\,\times\,g\bar{g}\,S\,,
\end{equation}
in perfect agreement with the numerical evaluation of the integral considered in \S\tcb{\ref{Delta32}}. We have checked many other (also complex) values and they precisely agree with the numerical evaluation. Note that for $\Delta>2$ one has to carefully take into account the poles that cross the real axis and that would not be included when performing the naive numerical integral just along the real axis. When such crossing of poles happens, the contour needs to be deformed to ensure that the analytic continuation is done properly. In this respect, it is also interesting to note that the above explicit result is not singular for integer values of $\Delta>2$ for which the prefactor $\frac1{\Gamma(d-\Delta)}$ would naively give zero. In this case the integral over the real line does indeed give a vanishing answer, however the correct analytic continuation must take into account also those poles which crossed the real line. Therefore the even $d$ result is simply given by a finite number of residues which crossed the real line in both directions for a given value of $\Delta$. We have explicitly checked that indeed defining the integral as an analytic continuation from the region where the poles are below the real line we recover the result \eqref{gammaSpec}.

\subsection{2pt tadpole}
\label{subsec::2pttad}
 \begin{figure}[h]
\centering
\captionsetup{width=0.8\textwidth}
\includegraphics[scale=0.3]{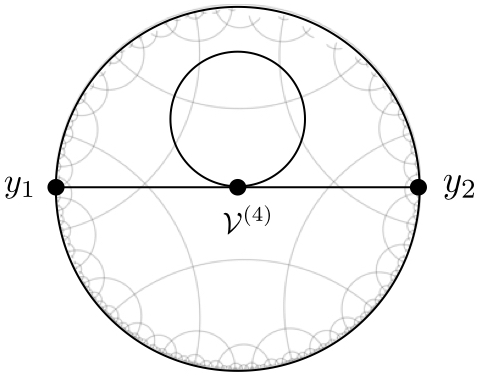}
\caption{Scalar two-point one-loop tadpole diagram generated by the quartic interaction ${\cal V}^{(4)}$.}
\label{fig::2pttadscalar}
\end{figure}

We now move onto two-point tadpole diagrams $\bep(14,10)\put(0,0){\line(1,0){14}}\put(7,4.25){\circle{8}}\eep$ illustrated in figure \ref{fig::2pttadscalar}. We begin in \S \tcb{\ref{subsubsec::phi4vertex}} with diagrams where the quartic coupling ${\cal V}^{\left(4\right)}$ is a non-derivative quartic interaction. In \S \tcb{\ref{subsubsec::derivativeint}} we generalise the latter for ${\cal V}^{\left(4\right)}$ involving any number of derivatives.

\subsubsection{$\phi^4$ tadpole}
\label{subsubsec::phi4vertex}

Consider the loop amplitude generated by the quartic coupling\footnote{In the following discussion we do not display explicitly the standard symmetry factors associated to the diagram $\bep(14,10)\put(0,0){\line(1,0){14}}\put(7,4.25){\circle{8}}\eep$ which depend on how many indistinguishable legs are present in a given coupling.
We recall that in the case of $\frac{g}{4!}\phi^4$ coupling all result obtained in this section should be multiplied by the symmetry factor $S=\frac12$. In the case of $O(N)$ model on AdS space with coupling $\frac14(\phi^a\phi^a)^2$ the corresponding multiplying factor is instead:
\begin{equation}
    S= g\,(N+2)\,.
\end{equation}}
\begin{equation}
{\cal V}^{\left(4\right)} = g \phi_1 \phi_2 \phi^2,
\end{equation}
given by
\begin{align}\label{onelphi4}
{\cal M}^{\text{1-loop tad.}}\left(P_1,P_2\right)  = -\,g \int_{\text{AdS}} dX\, K_{\Delta_1,0}\left(X,P_1\right) G_{\bar{\Delta},0}\left(X,X\right)K_{\Delta_2,0}\left(X,P_2\right). 
\end{align}
In this case the spectral representation \eqref{scalarbubu} of the bulk-to-bulk propagator allows to express the diagram \eqref{onelphi4} in terms of a tree-level three-point amplitude with a single the external leg integrated over the boundary, as illustrated in figure \ref{fig::genapp2}:

In particular, for the bulk-to-bulk propagator at coincident bulk points we have
{\small \begin{align} \nonumber
    G_{\Delta,0}\left(X;X\right) & = \int^\infty_{-\infty} \frac{\nu^2 d\nu}{\pi \left[\nu^2+\left(\bar{\Delta}-\frac{d}{2}\right)^2\right]} \int_{\partial \text{AdS}}dP\,K_{\frac{d}{2}+i\nu,0}\left(X;P\right) K_{\frac{d}{2}-i\nu,0}\left(X;P\right)\\ \label{coinibulkposc}
    & =  \frac{\Gamma\left(\frac{d}{2}+1\right)}{2\pi^{\frac{d}{2}+1}\Gamma\left(d\right)} \int^{\infty}_{-\infty} \frac{d\nu}{\left[\nu^2+\left(\bar{\Delta}-\frac{d}{2}\right)^2\right]} \frac{\Gamma\left(\frac{d}{2}+i\nu\right)\Gamma\left(\frac{d}{2}-i\nu\right)}{\Gamma\left(i\nu\right)\Gamma\left(-i\nu\right)}  \int_{\partial \text{AdS}}dP\, K_{d,0}\left(X;P\right),
\end{align}}
\!\!where the gamma function factor in the $\nu$ integrand comes from the normalisation of the bulk-to-boundary propagators on the first line. For the tadpole diagram, upon interchanging AdS and boundary integration, this yields: 
\begin{multline}\label{phi33pt}
{\cal M}^{\text{1-loop tad.}}\left(P_1,P_2\right) = -\,g\frac{\Gamma\left(\frac{d}{2}+1\right)}{2\pi^{\frac{d}{2}+1}\Gamma\left(d\right)} \int^{\infty}_{-\infty}  \frac{d\nu}{\left[\nu^2+\left(\bar{\Delta}-\frac{d}{2}\right)^2\right]} \frac{\Gamma\left(\frac{d}{2}+i\nu\right)\Gamma\left(\frac{d}{2}-i\nu\right)}{\Gamma\left(i\nu\right)\Gamma\left(-i\nu\right)} \\ \times \int_{\partial \text{AdS}}dP \,{\cal M}_{\Delta_1,\Delta_2,d}^{\text{3pt tree}}\left(P_1,P_2,P\right), 
\end{multline}
in terms of the three-point amplitude \eqref{123scal} with an external leg integrated over the boundary. Inserting the explicit expression result for the amplitude ${\cal M}_{\Delta_1,\Delta_2,d}^{\text{3pt tree}}$, one obtains
\begin{multline}\label{phi4specfun}
{\cal M}^{\text{1-loop tad.}}\left(P_1,P_2\right) = -\,g
\frac{\Gamma\left(\frac{d}{2}+1\right)}{2\pi^{\frac{d}{2}+1}\Gamma\left(d\right)}{\sf B}\left(\Delta_1,\Delta_2,d;0\right)
\int^{\infty}_{-\infty} d\nu\, {\cal F}^{\text{1-loop tad.}}\left(\nu\right)\\ \times
\underbrace{P^{-\frac{1}{2}\left(\Delta_1+\Delta_2-d\right)}_{12} \int_{\partial\text{AdS}} \frac{dP}{\left(-2 P_1 \cdot P\right)^{\frac{1}{2}\left(d+\Delta_1-\Delta_2\right)} \left(-2 P_2 \cdot P\right)^{\frac{1}{2}\left(d+\Delta_2-\Delta_1\right)}}}_{=I^{\text{1-loop}}_{\text{2pt}}\left(y_1,y_2\right)\ \eqref{M12}},
\end{multline}
in terms of the two-point conformal integral \eqref{int2ptconf} whose divergences regulated in dimensional regularisation generates the $\log$ contribution. The spectral function is given by:
\begin{equation}\label{fphi4nod}
    {\cal F}^{\text{1-loop tad.}}\left(\nu\right) =  \frac{1}{\left[\nu^2+\left(\bar{\Delta}-\frac{d}{2}\right)^2\right]}\frac{\Gamma\left(\frac{d}{2}+i\nu\right)\Gamma\left(\frac{d}{2}-i\nu\right)}{\Gamma\left(i\nu\right)\Gamma\left(-i\nu\right)}.
\end{equation}
Combining the above with the dimensionally regularised form of the boundary integral \eqref{M12} and keeping track of the normalisation of 2-pt functions, we obtain the following spectral representation for the anomalous dimension:
\begin{equation}\label{anomphi4spec}
    \gamma=g\, \delta_{\Delta_1,\Delta_2} \frac{\pi ^{\frac{d}{2}-1}\,d\, \Gamma \left(\Delta_1 +1-\frac{d}{2}\right)}{\Gamma (d) \Gamma (\Delta_1 )}\, {\sf B}\left(\Delta_1,\Delta_2,d;0\right)\int_{-\infty}^\infty d\nu\,{\cal F}^{\text{1-loop tad.}}\left(\nu\right)\,.
\end{equation}

In the following we explain how to evaluate the spectral integral in \eqref{anomphi4spec}. In even dimensions $d$ we have
\begin{equation}\label{evendphi4}
    {\cal F}^{\text{1-loop tad.}}\left(\nu\right)= \frac{1}{\left[\nu^2+\left(\bar{\Delta}-\frac{d}{2}\right)^2\right]} \prod^{\frac{d-2}{2}}_{j=0}\left(\nu^2+j^2\right),
\end{equation}
while in odd $d$
\begin{equation}\label{odddphi4}
     {\cal F}^{\text{1-loop tad.}}\left(\nu\right)=\frac{\nu \tanh{\pi\nu}}{\left[\nu^2+\left(\bar{\Delta}-\frac{d}{2}\right)^2\right]} \prod^{\frac{d-2}{2}}_{j=\frac{1}{2}}\left(\nu^2+j^2\right).
\end{equation}
Let us note that, as expected, the above gives the same spectral integral as the $\zeta$-function $\xi_{(\Delta,0)}(1)$. This can be made manifest performing first the integration over the boundary than the integral over AdS (see Appendix \ref{app:ccp}). Commute the AdS integral with boundary and spectral integrals, however, makes manifest the analogy with momentum space Feynman rules where the integral over space time is commuted with the momentum space integrals and performed once and for all. Divergences are then encoded into momentum space integrals. This remarkable analogy become more apparent considering that the analogue of flat space harmonic function can be defined in terms of plane waves as $\Omega_\nu(x)=\nu\int d^dk\, e^{ik\cdot x}\delta(k^2-\nu^2)$. We thus see that the split representation provides a close analogue to momentum space for AdS Feynman diagrams.

\subsubsection*{Tadpole in even dimensions}

The UV divergence in \eqref{evendphi4} can be taken care of by introducing a regulator $\mu$:
\begin{equation}
    \zeta^{\phi^4}_{\Delta}\left(\mu\right)= \int^{\infty}_{-\infty}\,  \frac{d\nu}{\left[\nu^2+\left(\Delta-p\right)^2\right]^{\mu+1}}\prod^{\frac{d-2}{2}}_{j=0}\left(\nu^2+j^2\right).
\end{equation}
Evaluating the above for $\mu$ complex and $\Delta>\tfrac{d}{2}$, one then obtains
\begin{equation}
\zeta^{\phi^4}_{\Delta}\left(\mu \rightarrow 0\right)=    (-1)^{d/2}\,\pi^2 \frac{\Gamma\left(\Delta\right)}{\Gamma\left(\Delta-d+1\right)}.
\end{equation}

Combining the above $\zeta$-function with the formula for anomalous dimensions, we arrive to the following expression for the anomalous dimension in even dimensions:
\begin{equation}
    \gamma=g\frac{(-1)^{d/2+1}}{2^{d+2} \pi ^{\frac{d-1}{2}}}\,\frac{ \Gamma (\Delta )}{(\Delta-\frac{d}{2}) \Gamma \left(\frac{1+d}{2}\right) \Gamma (\Delta-d +1)}\,.
\end{equation}
It is interesting to consider the case of a conformally coupled scalar field for which (assuming $\Delta>\tfrac{d}{2}$) $\Delta=\frac{d+1}2$:
\begin{align}
\gamma^{\text{conf.}}=g(-1)^{d/2+1}\,\frac{\pi ^{\frac{1-d}{2}}}{2^{d+1}\Gamma \left(\frac{3-d}{2}\right)}\,.
\end{align}
This is non vanishing in any even dimension $d$. Note that this effect is, however, an IR effect which does not enter in the flat space result where the first non-trivial contribution arises at 2 loops for massless scalar. The counterpart in AdS of the absence of UV divergences in flat space is the absence of single poles in the $\zeta$-function regulator $\mu$.

\subsubsection*{Tadpole in odd dimensions}
The $\zeta$-function tadpole computation is a bit more involved in odd CFT dimension $d$, in particular since the integrand does not reduce to a rational function. The result can still be given implicitly upon splitting the hyperbolic tangent in the spectral function \eqref{odddphi4} for the anomalous dimension \eqref{odddphi4} into a piece which is formally divergent and should be regularised, and a convergent piece:
\begin{equation}
    \gamma=\gamma^{\text{reg.}}+\gamma^{\text{fin.}}\,,
\end{equation}
with 
\begin{subequations}
\begin{align}
\gamma^{\text{reg.}}&=-\,g\frac{2^{-d} \pi ^{-\frac{d}{2}-\frac{1}{2}}}{(d-2 \Delta ) \Gamma \left(\frac{d+1}{2}\right)}\int_0^\infty d\nu\frac{\nu\, p^{(d)}(\nu^2)}{\left[\left(\Delta-\tfrac{d}{2}\right)^2+\nu^2\right]^{1+\mu}}\,,\\ \label{fing123}
    \gamma^{\text{fin.}}&=g\frac{2^{-d+1} \pi ^{-\frac{d}{2}-\frac{1}{2}}}{(d-2 \Delta ) \Gamma \left(\frac{d+1}{2}\right)}\int_0^\infty d\nu\frac{\nu\, p^{(d)}(\nu^2)}{(1+e^{2\pi\nu})\left[\left(\Delta-\tfrac{d}{2}\right)^2+\nu^2\right]}\,,
\end{align}
\end{subequations}
where the polynomial $p^{(d)}(\nu^2)$ is given by the product:
\begin{equation}
    p^{(d)}(\nu^2)=\prod_{i=0}^{\tfrac{d-3}2}\left[\left(i+\tfrac12\right)^2+\nu^2\right]=\sum_{n=0}^{\tfrac{d-3}2}\lambda_n^{(d)}\nu^{2n}\,.
\end{equation}
The integral giving $\gamma^{\text{reg.}}$ can thus be performed using the standard identity:
\begin{multline}\label{NoExpInt}
    \int_0^\infty d\nu\frac{\nu^{2n+1}\, }{\left[\left(\Delta-\tfrac{d}{2}\right)^2+\nu^2\right]^{1+\mu}}=\left(\Delta-\tfrac{d}{2}\right)^{2(n-\mu)}\frac{\Gamma (n+1) \Gamma (\mu -n)}{2 \Gamma (\mu +1)}\\\sim\frac{(-1)^n \left(\Delta -\frac{d}{2}\right)^{2 n}}{2}\left[\frac{1}{\mu}+H_n-2 \log \left(\Delta -\frac{d}{2}\right)\right]\,,
\end{multline}
in terms of the harmonic numbers $H_n$. This yields:
\begin{equation}
    \gamma^{\text{reg.}}=-\,g\frac{2^{-d} \pi ^{-\frac{d}{2}-\frac{1}{2}}}{(d-2 \Delta ) \Gamma \left(\frac{d+1}{2}\right)}\sum_{n=0}^{\tfrac{d-3}2}\lambda_n^{(d)}\,\frac{(-1)^n \left(\Delta -\frac{d}{2}\right)^{2 n}}{2}\left[H_n-2 \log \left(\Delta -\frac{d}{2}\right)\right]
\end{equation}
To tackle the integral \eqref{fing123} for the finite part $\gamma^{\text{fin.}}$, we rewrite part of the integrand as
\begin{multline}
    \frac{p^{(d)}(\nu^2)}{\left(\Delta-\tfrac{d}{2}\right)^2+\nu^2}=\frac{\Gamma (\Delta )}{\Gamma (-d+\Delta +1)}\frac{1}{\left(\Delta-\tfrac{d}{2}\right)^2+\nu^2}+\tilde{p}^{(d)}(\nu^2)\\\equiv \frac{\Gamma (\Delta )}{\Gamma (\Delta-d +1)}\frac{1}{\left(\Delta-\tfrac{d}{2}\right)^2+\nu^2}+\sum_{n=0}^{\tfrac{d-3}2}\bar{\lambda}^{(d)}_n\,\nu^{2n}\,,
\end{multline}
where the final equality defines the coefficients $\bar{\lambda}^{(d)}_k$. One can then evaluate the $\nu$ integrals analytically using the following identities valid for $\Delta>\tfrac{d}2$:
\begin{subequations}\label{ExpInt}
\begin{align}
\int_0^\infty d\nu\frac{\nu}{(1+e^{2\pi\nu})\left[\left(\Delta-\tfrac{d}{2}\right)^2+\nu^2\right]}&=\frac{1}{2} \left(\psi\left(\Delta-\frac{d}{2} +\frac{1}{2}\right)-\log \left(\Delta -\frac{d}{2}\right)\right)\,,\\
\int_0^\infty d\nu\frac{\nu^{n}}{(1+e^{2\pi\nu})}&=\left(1-2^{-n}\right) (2 \pi )^{-n-1} \zeta (n+1) \Gamma (n+1)\,,
\end{align}
\end{subequations}
where $\psi(z)$ is the digamma function and $\zeta(z)$ is the $\zeta$-function. Combining all the above ingredients we arrive to the following expression for the finite part of the anomalous dimension, valid in any odd CFT dimension $d$:
\begin{multline}
    \gamma^{\text{fin}}=g\frac{2^{-d+1} \pi ^{-\frac{d}{2}-\frac{1}{2}}}{(d-2 \Delta ) \Gamma \left(\frac{d+1}{2}\right)}\left[\frac{1}{2}\frac{\Gamma (\Delta )}{\Gamma (\Delta-d +1)} \left(\psi\left(\Delta-\frac{d}{2} +\frac{1}{2}\right)-\log \left(\Delta -\frac{d}{2}\right)\right)\right.\\\left.+\sum_{n=0}^{\tfrac{d-3}2}\bar{\lambda}_n^{(d)}\,\frac{\left(1-2^{-2 n-1}\right) \left| B_{2 (n+1)}\right| }{4 (n+1)}\right]\,.
\end{multline}
Below, for convenience, we give the coefficients $\bar{\lambda}_n^{(d)}$ in some examples $\{\bar{\lambda}_0^{(d)},\bar{\lambda}_1^{(d)},\bar{\lambda}_2^{(d)},\ldots\}$:
\begin{align}
\bar{\lambda}_n^{(3)}&=\left\{1\right\}\,,\\
\bar{\lambda}_n^{(5)}&=\left\{-(\Delta -5) \Delta -\frac{15}{4},1\right\}\,,\\
\bar{\lambda}_n^{(7)}&=\left\{\frac{1}{4} (\Delta -7) \Delta  (4 (\Delta -7) \Delta +63)+\frac{945}{16},\frac{1}{2} (-2 (\Delta -7) \Delta -7),1\right\}\,,\\
\bar{\lambda}_n^{(9)}&=\left\{\frac{1}{64} (-4 (\Delta -9) \Delta  (4 (\Delta -9) \Delta  (4 (\Delta -9) \Delta +159)+8049)-127295),\right.\\&\hspace{20pt}\left.\frac{1}{16} (8 (\Delta -9) \Delta  (2 (\Delta -9) \Delta +39)+1731),\frac{1}{4} (3-4 (\Delta -9) \Delta ),1\right\}\,,\nonumber
\end{align}
which give the following expression for anomalous dimensions:
\begin{subequations}
\begin{align}
\gamma^{(1)}&=-\,g\frac{1}{2\pi}\,\frac{\log \left(\Delta -\frac{1}{2}\right)}{2\Delta-1}\,,\\
\gamma^{(3)}&=-\,g\frac{3 (\Delta -3) \Delta +7}{48 \pi ^2 (2 \Delta -3)}+g\frac{6 (\Delta -2) (\Delta -1) \psi(\Delta -1)}{48 \pi ^2 (2 \Delta -3)}\,,\\
\gamma^{(5)}&=+\,g\frac{5 (\Delta -5) \Delta  (9 (\Delta -5) \Delta +98)+1298}{3840 \pi ^3 (2 \Delta -5)}-g\frac{(\Delta -4) (\Delta -3) (\Delta -2) (\Delta -1) \psi (\Delta -2)}{64 \pi ^3 (2 \Delta -5)}\,,\\
\gamma^{(7)}&=-\,g\frac{21 (\Delta -7) \Delta  (5 (\Delta -7) \Delta  (11 (\Delta -7) \Delta +326)+15638)+1010368}{967680 \pi ^4 (2 \Delta -7)}\nonumber\\&\hspace{50pt}+g\frac{(\Delta -6) (\Delta -5) (\Delta -4) (\Delta -3) (\Delta -2) (\Delta -1) \psi(\Delta -3)}{768 \pi ^4 (2 \Delta -7)}\,,\\
\gamma^{(9)}&=-\,g\frac{(\Delta -8) (\Delta -7) (\Delta -6) (\Delta -5) (\Delta -4) (\Delta -3) (\Delta -2) (\Delta -1) \psi(\Delta -4)}{12288 \pi ^5 (2 \Delta -9)}\\&+\,g\tfrac{(\Delta -9) \Delta  (21 (\Delta -9) \Delta  (5 (\Delta -9) \Delta  (25 (\Delta -9) \Delta +1564)+178516)+36755072)+129256824}{30965760 \pi ^5 (2 \Delta -9)}\,,\nonumber
\end{align}
\end{subequations}
together with similar results in higher dimensions. For the case of the conformally coupled scalar ($\Delta=\tfrac{d+1}2$) the above gives:
\begin{align}
\gamma^{(1)}&=g\frac{\log (2)}{2 \pi }\,,&\gamma^{(3)}&=-\,g\frac{1}{48 \pi ^2}\,,& \gamma^{(5)}&=-\,g\frac{11}{1920 \pi ^3}\,,\\ \gamma^{(7)}&=-\,g\frac{359}{120960 \pi ^4}\,,& \gamma^{(9)}&=-\,g\frac{8777}{3870720 \pi ^5}\,.
\end{align}
It is also interesting to notice that in the conformally coupled case the $\tfrac1{\mu}$ pole in the $\zeta$-function regulator is cancelled, in agreement with the expected absence of UV divergences in the flat space result. In general, in odd dimensions the regulator pole is proportional to:
\begin{equation}
   \sim \frac1{\mu}\,\prod_{i=0}^{d-2}(\Delta-1-i)\,,
\end{equation}
and vanishes for integer conformal dimensions $\Delta<d$. Still, there is a IR contribution to the anomalous dimension.

\subsubsection{Wilson-Fisher fixed point in AdS$_4$}

A possible application of the results obtained in this section is to consider the Wilson-Fisher fixed point \cite{Wilson:1971dc,Wilson:1973jj} for the $O(N)$ model in hyperbolic space with $N$ real conformally coupled scalar fields:
\begin{equation}
    S=\int d^{d+1}x\sqrt{-g}\left(\frac12(\pl\phi^a)^2+\frac{M_d}2(\phi^a)^2+\frac{g}{4}(\phi^a\phi^a)^2\right)\,,
\end{equation}
and conformal mass:
\begin{equation}
    M_d=\frac{\Lambda}4(d+1)(d-1)\,.
\end{equation}
In this case the one loop $\beta$-function in $d=4-\epsilon$ dimensions obtained from standard epsilon expansion reads:
\begin{equation}
    \beta=\frac{N+8}{8 \pi^2}\,g^2-\epsilon\, g\,,
\end{equation}
and the fixed point sits at
\begin{equation}
    g^\star=\frac{8\pi^2}{N+8}\,\epsilon\,.
\end{equation}
One can then plug the above value of the fixed point coupling into the anomalous dimension for the conformally coupled scalar on hyperbolic space obtaining the following prediction (with $\zeta$-function regularisation) for the anomalous dimension of the dual operator of dimension $\Delta=\frac{5-\epsilon}2$:\footnote{If we use $\tfrac{g}{4!}\,\phi^4$ the result below should be redefined with $N=1$ and $\epsilon\to 6\epsilon$}
\begin{equation}
    \gamma=-\,\frac{\epsilon}{6(N+8)}\,.
\end{equation}
 It is natural to interpret this result as the anomalous dimension of an operator in a ``defect CFT" on the boundary of AdS$_4$.

\subsubsection{General 2pt tadpole with derivatives}
\label{subsubsec::derivativeint}

\begin{figure}[h]
\captionsetup{width=0.8\textwidth}
\centering
\includegraphics[scale=0.4]{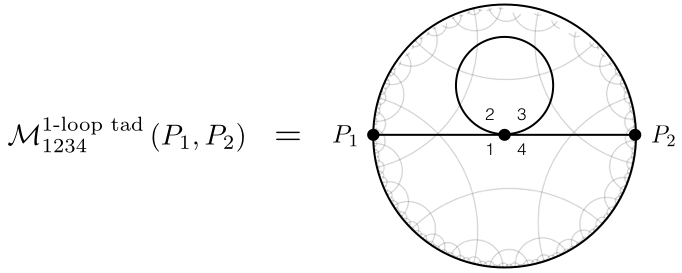}
\caption{One-loop tadpole diagram ${\cal M}^{\text{1-loop tad}}_{1234}$ generated by the quartic vertex \eqref{basisquart}. The point split fields $\phi_1$ and $\phi_4$ are external, while $\phi_2$ and $\phi_3$ propagate in the loop. The other diagrams \eqref{gencondiagrs} permute the positions of the point-split fields $\phi_i$.}
\label{fig::gentad}
\end{figure}

Here we generalise the results in \S \tcb{\ref{subsubsec::phi4vertex}} to tadpole diagrams for an arbitrary quartic scalar self-interaction dressed with derivatives. Using the ambient space framework (\S \tcb{\ref{subsec::ambnotconv}}), a complete basis for the latter is given by 
\begin{multline}\label{basisquart}
{\cal V}^{\left(4\right)}_{k,m}\left(X\right) = \frac{g}{\left(k+m\right)!}\left[ \phi\left(X\right) \left(\partial_U \cdot  \partial_X\right)^k \phi\left(X\right)\right]\\ \times \left(\partial_U \cdot  \partial_X\right)^m \phi\left(X\right)\left(U \cdot  \partial_X\right)^{k+m} \phi\left(X\right), \quad k \ge 2 m \ge 0.
\end{multline}

In this case there are four distinct contributing diagrams. To label the possibilities, we employ the point-splitting notation:

{\small \begin{align}\label{pointsplit}
{\cal V}^{\left(4\right)}_{k,m}\left(X\right) & =  \frac{g}{\left(k+m\right)!}\left[ \phi_1\left(X\right) \left(\partial_U \cdot  \partial_X\right)^k \phi_2\left(X\right)\right]\left(\partial_U \cdot  \partial_X\right)^m \phi_3\left(X\right)\left(U \cdot  \partial_X\right)^{k+m} \phi_4\left(X\right)\Big|_{\phi_i=\phi},
\end{align}}

and denote the contributing diagrams by:
\begin{equation}\label{gencondiagrs}
{\cal M}^{\text{1-loop tad}}_{1234}, \qquad {\cal M}^{\text{1-loop tad}}_{1342}, \qquad {\cal M}^{\text{1-loop tad}}_{3142}, \qquad {\cal M}^{\text{1-loop tad}}_{4132}. 
\end{equation}
The subscript labels the positions of the scalar fields in the point-split vertex \eqref{pointsplit}, and is illustrated in figure \ref{fig::gentad}. 

In this more general case, the scalar propagators are acted on by ambient partial derivatives -- which are straightforward to manage. For bulk-to-boundary propagators for instance, we have
\begin{equation}\label{derivscbubou}
    \left(U \cdot \partial_X\right)^n K_{\Delta,0}\left(X;P\right)=2^n\left(\Delta+1-\tfrac{d}{2}\right)_n \left(U \cdot P\right)^n K_{\Delta+n,0}\left(X;P\right).
\end{equation}
This in particular leads to a shift in the argument of the gamma functions in the spectral function compared to the $\phi^4$ case \eqref{fphi4nod}, and can be seen simply from:

{\small \begin{align} \nonumber
  &  \left(U_1 \cdot \partial_{X_1}\right)^p
\left(U_2 \cdot \partial_{X_2} \right)^{q} G_{\Delta,0}\left(X_1,X_2\right)\Big|_{X_i=X}  \\ \nonumber
& \hspace*{1cm} =  \int^\infty_{-\infty} \frac{\nu^2 d\nu}{\pi \left[\nu^2+\left(\Delta-\frac{d}{2}\right)^2\right]}  \int_{\partial \text{AdS}}dP\,  \left(U_1 \cdot \partial_{X_1}\right)^p K_{\frac{d}{2}+i\nu,0}\left(X_1;P\right) \left(U_2 \cdot \partial_{X_2} \right)^{q} K_{\frac{d}{2}-i\nu,0}\left(X_2;P\right)\Big|_{X_i=X}\\
    & \hspace*{1cm} =  \frac{2^{p+q}\Gamma\left(\frac{d}{2}+1+p+q\right)}{2\pi^{\frac{d}{2}+1}\Gamma\left(d\right)}\int^\infty_{-\infty} \frac{d\nu}{\left[\nu^2+\left(\Delta-\frac{d}{2}\right)^2\right]} \frac{\Gamma\left(\tfrac{d}{2}+i\nu+p\right)\Gamma\left(\tfrac{d}{2}-i\nu+q\right)}{\Gamma\left(i\nu\right)\Gamma\left(-i\nu\right)}  \nonumber \\ \label{derivpropsplit}
    & \hspace*{7cm} \times \int_{\partial \text{AdS}}dP\, \left(P \cdot U_1\right)^p\left(P \cdot U_2\right)^q K_{d+p+q,0}\left(X;P\right),
\end{align}}

where we used point splitting to restrict the action of each derivative to only one of either of the two ends of the propagator and the identity \eqref{derivscbubou}. Generalising \eqref{fphi4nod}, the spectral function in the case of derivative interactions \eqref{basisquart} is thus of the form:
\begin{equation}
    {\cal F}^{\text{1-loop tad.}}_{p,q}\left(\nu\right)= \frac{1}{\left[\nu^2+\left(\Delta-\frac{d}{2}\right)^2\right]}\frac{\Gamma\left(\tfrac{d}{2}+i\nu+p\right)\Gamma\left(\tfrac{d}{2}-i\nu+q\right)}{\Gamma\left(i\nu\right)\Gamma\left(-i\nu\right)}.
\end{equation}
We discuss the evaluation of the corresponding spectral integral at the end of this section.

The expression \eqref{derivpropsplit} allows one to immediately conclude that the diagram ${\cal M}^{\text{1-loop tad}}_{1342}$ is vanishing for $m>0$: In this case we have $U_1 = \partial_U$ and $U_1 = \partial_U$, and  \eqref{derivpropsplit} vanishes since $P$ is a null vector: $P^2=0$. For $m=0$, ${\cal M}^{\text{1-loop tad}}_{1342}$ is the same as ${\cal M}^{\text{1-loop tad}}_{3142}$. We give the remaining diagrams below.

Using \eqref{derivpropsplit} and together with the identity \eqref{derivscbubou} for ambient derivatives of bulk-to-boundary propagators, we have

{\small \begin{align}
&{\cal M}^{\text{1-loop tad}}_{1234}\left(P_1,P_2\right) \\ \nonumber
& \hspace*{1cm}= -\,\frac{g}{\left(k+m\right)!}\int_{\text{AdS}} dX\, K_{\Delta,0}\left(X,P_1\right) \left(\partial_U \cdot \partial_{X_1}\right)^k  \left(\partial_U \cdot \partial_{X_2}\right)^m G_{\Delta,0}\left(X_1,X_2\right)\Big|_{X_i=X} \left(U \cdot \partial_X \right)^{k+m} K_{\Delta,0}\left(X,P_2\right) \\ \nonumber
& \hspace*{1cm}= -\,\frac{g(-2)^{k+m}}{\left(k+m\right)!}\frac{\Gamma\left(\frac{d}{2}+1+k+m\right)}{2\pi^{\frac{d}{2}+1}\Gamma\left(d\right)} \left(\Delta+1-\tfrac{d}{2}\right)_{k+m}
\int^{\infty}_{-\infty}d\nu {\cal F}^{\text{1-loop tad.}}_{k,m}\left(\nu\right) \\ \nonumber
& \hspace*{6cm} \times  \int_{\partial \text{AdS}} dP\, \left(-2 P \cdot P_2\right)^{k+m} {\cal M}^{\text{3pt tree}}_{\Delta,\Delta+k+m,d+k+m}\left(P_1,P_2,P\right) 
\end{align}}

 Inserting the expression \eqref{123scal} for the three-point amplitude yields:
 
{\small \begin{multline}
{\cal M}^{\text{1-loop tad}}_{1234}\left(P_1,P_2\right) = -\,\frac{g(-2)^{k+m}}{\left(k+m\right)!}\frac{\Gamma\left(\frac{d}{2}+1+k+m\right)}{2\pi^{\frac{d}{2}+1}\Gamma\left(d\right)} \left(\Delta+1-\tfrac{d}{2}\right)_{k+m}{\sf B}\left(\Delta,\Delta+k+m,d+k+m;0\right) 
 \\ 
 \times 
{\cal M}^{\text{1-loop}}\left(P_1,P_2\right)\int^{\infty}_{-\infty}d\nu {\cal F}^{\text{1-loop tad.}}_{k,m}\left(\nu\right),
\end{multline}}
with spectral representation for the anomalous dimension:
\begin{multline}
\gamma_{1234}= -\,\frac{g(-2)^{k+m+1}\pi^{\frac{d}{2}-1}}{\left(k+m\right)!}\frac{\Gamma\left(\frac{d}{2}+1+k+m\right)}{ \Gamma\left(d\right)\Gamma\left(\frac{d}{2}\right)\Gamma\left(\Delta\right)}  \Gamma \left(\Delta+1-\tfrac{d}{2}+k+m\right)\\
\times {\sf B}\left(\Delta,\Delta+k+m,d+k+m;0\right) \int^{\infty}_{-\infty}d\nu {\cal F}^{\text{1-loop tad.}}_{k,m}\left(\nu\right)
\end{multline}
Similarly, for the other diagrams we have
{\small \begin{align}
&{\cal M}^{\text{1-loop tad}}_{3142}\left(P_1,P_2\right)\\ \nonumber
& \hspace*{1cm} = -\,\frac{g}{\left(k+m\right)!}\int_{\text{AdS}} dX\, \left(\partial_U \cdot \partial_{X}\right)^m K_{\Delta,0}\left(X,P_1\right) \left(\partial_U \cdot \partial_{X}\right)^k K_{\Delta,0}\left(X,P_2\right)    
\left(U \cdot \partial_{X_2} \right)^{k+m} G_{\Delta,0}\left(X,X_2\right) \\ \nonumber
& \hspace*{1cm}= -\,\frac{g(-2)^{k+m}}{\left(k+m\right)!}\frac{\Gamma\left(\frac{d}{2}+1+k+m\right)}{2\pi^{\frac{d}{2}+1}\Gamma\left(d\right)} \left(\Delta+1-\tfrac{d}{2}\right)_{k}\left(\Delta+1-\tfrac{d}{2}\right)_{m} {\sf B}\left(\Delta+m,\Delta+k,d+m+k;0\right) 
 \\ \nonumber
& \hspace*{2cm} \times 
{\cal M}^{\text{1-loop}}\left(P_1,P_2\right)\int^{\infty}_{-\infty}d\nu {\cal F}^{\text{1-loop tad.}}_{0,k+m}\left(\nu\right),
 \end{align}}
with anomalous dimension:
{\small \begin{multline}
\gamma_{3142}= -\,\frac{g(-2)^{k+m+1}\pi^{\frac{d}{2}-1}}{\left(k+m\right)!}\frac{\Gamma\left(\frac{d}{2}+1+k+m\right)}{\pi \Gamma\left(\frac{d}{2}\right)\Gamma\left(d\right)\Gamma\left(\Delta\right)} \Gamma \left(\Delta+1-\tfrac{d}{2}+k\right)\left(\Delta+1-\tfrac{d}{2}\right)_{m}  \\ 
 \times {\sf B}\left(\Delta+m,\Delta+k,d+m+k;0\right)
\int^{\infty}_{-\infty}d\nu {\cal F}^{\text{1-loop tad.}}_{0,k+m}\left(\nu\right).
 \end{multline}}
And finally
{\small \begin{align}
&{\cal M}^{\text{1-loop tad}}_{4132}\left(P_1,P_2\right) \\ \nonumber & \hspace*{1cm} = -\,\frac{g}{\left(k+m\right)!}\int_{\text{AdS}} dX\,  \left(\partial_U \cdot \partial_{X}\right)^k K_{\Delta,0}\left(X,P_2\right) 
\left(\partial_U \cdot \partial_{X_2}\right)^m G_{\Delta,0}\left(X;X_2\right)
\left(U \cdot \partial_{X} \right)^{k+m} K_{\Delta,0}\left(X,P_1\right)  \\ \nonumber
& \hspace*{1cm}= -\,\frac{g(-2)^{k+m}}{\left(k+m\right)!}\frac{\Gamma\left(\frac{d}{2}+1+m\right)}{2\pi^{\frac{d}{2}+1}\Gamma\left(d\right)} \left(\Delta+1-\tfrac{d}{2}\right)_{k}\left(\Delta+1-\tfrac{d}{2}\right)_{k+m} {\sf B}\left(\Delta+m+k,\Delta+k,d+m;0\right)
 \\ \nonumber
& \hspace*{2cm} \times 
 {\cal M}^{\text{1-loop}}\left(P_1,P_2\right)\int^{\infty}_{-\infty}d\nu {\cal F}^{\text{1-loop tad.}}_{0,m}\left(\nu\right),
 \end{align}}
with anomalous dimension:
{\small \begin{multline}
\gamma_{4132}= -\,\frac{g(-2)^{k+m+1}\pi^{\frac{d}{2}-1}}{\left(k+m\right)!}\frac{\Gamma\left(\frac{d}{2}+1+m\right)}{\Gamma\left(\frac{d}{2}\right)\Gamma\left(d\right)\Gamma\left(\Delta\right)} \Gamma \left(\Delta+1-\tfrac{d}{2}+k\right)\left(\Delta+1-\tfrac{d}{2}\right)_{k+m} 
 \\ 
 \times 
 {\sf B}\left(\Delta+m+k,\Delta+k,d+m;0\right) \int^{\infty}_{-\infty}d\nu {\cal F}^{\text{1-loop tad.}}_{0,m}\left(\nu\right).
 \end{multline}}

To conclude this section let us discuss the evaluation of the spectral integrals. The integrals are of a similar type to those \eqref{fphi4nod} arising in $\phi^4$ theory, and can be divided into two parts:
\begin{multline}\label{genspint0}
    \int_{-\infty}^\infty d\nu {\cal F}^{\text{1-loop tad.}}_{m,n}(\nu)\\=\int_0^\infty d\nu\frac{\nu\, p(\nu^2)+r(\nu^2)}{\left[\left(\Delta-\tfrac{d}{2}\right)^2+\nu^2\right]^{1+\mu}}\,-2\int_0^\infty d\nu\frac{\nu\, \left[a+q(\nu^2)\left[\left(\Delta-\tfrac{d}{2}\right)^2+\nu^2\right]\right]}{(1+e^{2\pi\nu})\left[\left(\Delta-\tfrac{d}{2}\right)^2+\nu^2\right]}\,,
\end{multline}
in terms of polynomials $p(\nu^2)\equiv\sum_i\xi_i \,\nu^{2i}$, $r(\nu^2)\equiv\sum_i r_i \,\nu^{2i}$ and $q(\nu^2)\equiv\sum_i\zeta_i\,\nu^{2i}$ which are defined by the above equality for integer dimensions. The polynomial $r(\nu^2)$ appears in even dimensions, while $p(\nu^2)$ and $q(\nu^2)$ are non-vanishing in odd dimensions and satisfy the relation 
\begin{equation}
  p(\nu^2)=\eta+q(\nu^2)\left[\left(\Delta-\tfrac{d}{2}\right)^2+\nu^2\right]\,, 
\end{equation}
with $\eta$ a constant.
One can thus in full generality evaluate the corresponding spectral integrals in $\zeta$-function regularisation using \eqref{ExpInt} and \eqref{NoExpInt}, obtaining the result as a linear combination of the constants $\xi_n$ and $\zeta_n$:
\begin{align}\label{genspint}
   \int_{-\infty}^\infty &d\nu {\cal F}^{\text{1-loop tad.}}_{m,n}(\nu)\\\nonumber &=\left[\sum_{i=0}\xi_i\,\frac{(-1)^i \left(\Delta -\frac{d}{2}\right)^{2 i}}{2}\left[H_i-2 \log \left(\Delta -\frac{d}{2}\right)\right]\right]\\
   &-\left[\sum_i\zeta_i\,\frac{\left(1-2^{-2 i-1}\right) \left| B_{2 (i+1)}\right| }{2 (i+1)}\right]-\eta \left[\psi\left(\Delta-\frac{d}{2} +\frac{1}{2}\right)-\log \left(\Delta -\frac{d}{2}\right)\right]\nonumber\\
   &-\left[\sum_{i}\pi\,r_i \left(-\frac{1}{4}\right)^i (d-2 \Delta )^{2 i-1}\right]\,,\nonumber
\end{align}
which is expressed in terms of Bernoulli numbers $B_i$, harmonic numbers $H_i$ and digamma function $\psi(z)$. Similar results can also be obtained using Mellin-Barnes regularisation.

\subsection{One-point bulk tadpole}
\label{subsec::onepttad}

\begin{figure}[h]
\captionsetup{width=0.8\textwidth}
\centering
\includegraphics[scale=0.5]{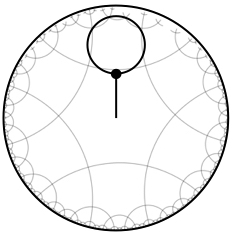}
\caption{Scalar one-point tadpole diagram with off-shell external leg, generated by the cubic vertex \eqref{lollycusc}.}
\label{fig::scalar1pttadbulk01}
\end{figure}

In this section we consider the one-point tadpole diagram with a single off-shell external leg in the bulk, generated by the cubic coupling:
\begin{equation}\label{lollycusc}
    {\cal V}^{(3)} = g {\bar \phi} \phi^2. 
\end{equation}
It is given by the bulk integral:
\begin{equation}\label{1ptsctad}
\mathfrak{T}^{\text{1pt tadpole}}\left(X_1\right)=-\,g \int_{\text{AdS}} dX\, G_{{\bar \Delta},0}\left(X_1;X\right) G_{\Delta,0}\left(X;X\right),
\end{equation}
and depicted in figure \ref{fig::scalar1pttadbulk01}. In the following we argue that this is vanishing. Using the spectral representation \eqref{scalarbubu} of the scalar bulk-to-bulk propagator, the diagram factorises as:
 \begin{multline}
\mathfrak{T}^{\text{1pt tadpole}}\left(X_1\right)=-\,\frac{g}{4\pi^{d+1}} \int^\infty_{-\infty}  \frac{d{\bar \nu} }{\left[{\bar \nu}^2+\left({\bar \Delta}-\frac{d}{2}\right)^2\right]} \frac{\Gamma\left(\frac{d}{2}+i{\bar \nu}\right)\Gamma\left(\frac{d}{2}-i{\bar \nu}\right)}{\Gamma\left(i{\bar \nu}\right)\Gamma\left(-i{\bar \nu}\right)} \int_{\partial \text{AdS}}d{\bar P}\,\frac{1}{\left(-2 X_1 \cdot {\bar P}\right)^{\frac{d}{2}+i{\bar \nu}}} \\ \times 
\int_{\text{AdS}} dX\, \frac{1}{\left(-2 X \cdot {\bar P}\right)^{\frac{d}{2}-i{\bar \nu}}} G_{\Delta,0}\left(X;X\right),\label{facttadpolesc}
\end{multline}
\begin{figure}[h]
\captionsetup{width=0.8\textwidth}
\centering
\includegraphics[scale=0.45]{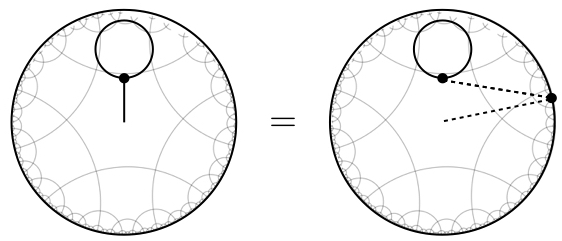}
\caption{The factorisation \eqref{facttadpolesc} of the tadpole diagram \eqref{1ptsctad} into a tadpole connected to the boundary and a bulk-to-boundary propagator, integrated over their common boundary point.}
\label{fig::scalar1pttadbulk002}
\end{figure}
which is shown in figure \ref{fig::scalar1pttadbulk002}. Concentrating on the tadpole factor on the second line which is connected to the boundary point ${\bar P}$: Using the identity \eqref{coinibulkposc} for the bulk-to-bulk propagator at coincident points, we have
\begin{multline}\label{tapofactsc}
\int_{\text{AdS}} dX\, \frac{1}{\left(-2 X \cdot {\bar P}\right)^{\frac{d}{2}-i{\bar \nu}}} G_{\Delta,0}\left(X;X\right)=\frac{1}{4\pi^{d+1}}\int^\infty_{-\infty} \frac{d\nu }{\left[\nu^2+\left(\Delta-\frac{d}{2}\right)^2\right]} \frac{\Gamma\left(\frac{d}{2}+i\nu\right)\Gamma\left(\frac{d}{2}-i\nu\right)}{\Gamma\left(i\nu\right)\Gamma\left(-i\nu\right)}\\ \times \int_{\partial \text{AdS}}dP\int_{\text{AdS}} dX\,\frac{1}{\left(-2 X \cdot P\right)^{d}}\frac{1}{\left(-2 X \cdot {\bar P}\right)^{\frac{d}{2}-i{\bar \nu}}}.
\end{multline}
The two-point bulk integrals of the type on the second line are given by:\footnote{This equation is the AdS analogue of the orthogonality relation $\int d^dx\,e^{ix\cdot(p_1-p_2)}=\delta^{(d)}(p_1-p_2)$.}
\begin{multline}\label{2ptbulkint}
\int_{\text{AdS}} dX\,\frac{1}{\left(-2 X \cdot P_1\right)^{\Delta_1}}\frac{1}{\left(-2 X \cdot P_2\right)^{\Delta_2}}=2\pi^{d/2+1}\frac{\Gamma(\Delta_1-\tfrac{d}2)}{\Gamma(\Delta_1)}\,\frac{1}{P_{12}^{\Delta_1}}\,\delta(\Delta_1-\Delta_2)\\ +2\pi^{d+1}\frac{\Gamma(\tfrac{d}{2}-\Delta_1)\Gamma(\tfrac{d}{2}-\Delta_2)}{\Gamma(\Delta_1)\Gamma(\Delta_2)}\delta^{(d)}(P_1,P_2)\,\delta(\Delta_1+\Delta_2-d)\,,
\end{multline}
which implies
\begin{multline}\label{tadsc1ptaux}
\int_{\partial \text{AdS}}dP \int_{\text{AdS}} dX\,\frac{1}{\left(-2 X \cdot P\right)^{d}}\frac{1}{\left(-2 X \cdot {\bar P}\right)^{\frac{d}{2}-i{\bar \nu}}} \\ = 2 \pi^{\frac{d}{2}+1} \frac{\Gamma\left(\frac{d}{2}\right)}{\Gamma\left(d\right)}A \delta\left(\tfrac{d}{2}+i{\bar \nu}\right) + 2 \pi^{d+1} \frac{\Gamma\left(-\frac{d}{2}\right)\Gamma\left(i {\bar \nu}\right)}{\Gamma\left(d\right)\Gamma\left(\frac{d}{2}-i {\bar \nu}\right)} \delta\left(\tfrac{d}{2}-i {\bar \nu}\right).
\end{multline}
The constant $A$ is given by the divergent integral 
\begin{equation}
    A = \int_{\partial \text{AdS}}dP \frac{1}{\left(- 2 P \cdot {\bar P}\right)^d},
\end{equation}
which vanishes in dimensional regularisation.
Since the integration over the parameter ${\bar \nu}$ in \eqref{facttadpolesc} is also restricted to real values, the tadpole factor \eqref{tapofactsc} connected to the boundary is zero. It thus appears that, as expected, the tadpole is vanishing when regularising the bulk IR divergences (which maps to a UV boundary divergence):
\begin{equation}
    \mathfrak{T}^{\text{1pt tadpole}}\left(X_1\right)\equiv 0.
\end{equation}
We may thus argue that such diagrams do not contribute to bulk amplitudes.

\section{Spinning diagrams}
\label{sec::sbd}
Having illustrated the evalution of two-point one-loop diagrams for the simplest case of scalar field theories, we now turn to theories of spinning fields. We mostly focus on two-point bubble diagrams, but in \S \tcb{\ref{subsec::oneptspintad}} at the end of this section we also discuss tadpole diagrams with a single off-shell bulk external leg.

The bulk-to-boundary propagator for a totally symmetric field of spin $s$ and mass $m^2 R^2=\Delta\left(\Delta-d\right)-s$ is most simply expressed in the ambient space formalism, where it is given by \cite{Mikhailov:2002bp,Costa:2014kfa}:\footnote{For ease of notation our definition of mass is based on the wave operator $(\nabla_\mu\nabla^\mu+m^2)\varphi_{\mu(s)}=0$ acting on symmetric traceless and transverse filed where $\nabla$ is the AdS covariant derivative. This definition allows to simplify various formulas in the radial reduction. Note that this mass is not zero for gauge fields.}
\begin{equation}
    K_{\Delta,s}\left(X,U;P,Z\right)=
    \left(U \cdot Z-\frac{U \cdot P Z \cdot X}{P \cdot X}\right)^s\frac{C_{\Delta,s}}{\left(-2P \cdot X\right)^{\Delta}},
\end{equation}
with normalisation
\begin{equation}
    C_{\Delta,s}=\frac{\left(\Delta+s-1\right)\Gamma\left(\Delta\right)}{2\pi^{d/2}\left(\Delta-1\right)\Gamma\left(\Delta+1-\frac{d}{2}\right)}.
\end{equation}
It is often convenient to express the bulk-to-boundary propagator in the form \cite{Sleight:2016hyl}
\begin{equation} \label{bubos}
     K_{\Delta,s}\left(X, U;P, Z \right) = \frac{1}{\left(\Delta-1\right)_s}\left({\cal D}_P\left(Z;U\right)\right)^s K_{\Delta,0}\left(X;P\right),
\end{equation}
with differential operator
\begin{equation}
 {\cal D}_P\left(Z;U\right) = \left(Z \cdot U\right)\left(Z \cdot \frac{\partial}{\partial Z} - P \cdot \frac{\partial}{\partial P}\right) + \left(P \cdot U \right)\left(Z \cdot \frac{\partial}{\partial P}\right),
\end{equation}
acting on a scalar bulk-to-boundary propagator \eqref{scbubou} of the same dimension. This in particular leads to identities that generalise \eqref{derivscbubou}:
\begin{equation}
  \left(U_i \cdot \partial_X\right)^n  K_{\Delta,s}\left(X, U;P, Z \right) = \frac{2^n\left(\Delta+1-\frac{d}{2}\right)_n}{\left(\Delta-1\right)_s}\left({\cal D}_P\left(Z;U\right)\right)^s\left(U_i \cdot P\right)^n K_{\Delta+n,0}\left(X;P\right),
\end{equation}
which are useful to evaluate Witten diagrams with derivative interactions.

The spectral representation of the bulk-to-bulk propagator takes the form\footnote{For concision we define:\begin{equation}
    \sum_{{\bf p}}=\sum^{\left[s/2\right]}_{p_1=0}\sum^{s-2p_1}_{p_3=0}\sum^{\left[p_3/2\right]+p_1}_{p_2=0}.
\end{equation}}
\begin{multline}\label{genbubuprop}
G_{\Delta,s}\left(x_1,x_2\right)=\sum_{\bf{p}}\int^\infty_{-\infty}d\nu\,g^{(s)}_{p_1,p_2,p_3}\left(\nu\right) \left(u^2_1\right)^{p_1} \left(u^2_2\right)^{p_2} \\ \times  \left(u_1 \cdot \nabla_1\right)^{p_3} \left(u_2 \cdot \nabla_2\right)^{p_3+2\left(p_1-p_2\right)}\Omega_{\nu,s-2p_2-p_3}\left(x_1,x_2\right),
\end{multline}
for some functions {\small $g^{(s)}_{p_1,p_2,p_3}\left(\nu\right)$} whose properties we discuss later on. Symmetry in $\left(x_1,u_1\right) \leftrightarrow \left(x_2,u_2\right)$ imposes: {\small $g^{(s)}_{p_2,p_1,p_3+2\left(p_1-p_2\right)}\left(\nu\right)=g^{(s)}_{p_1,p_2,p_3}\left(\nu\right)$}. This way of representing bulk-to-bulk propagators has so far been applied in the literature for totally symmetric massive spin-$s$ fields \cite{Costa:2014kfa} and spin-$s$ gauge fields \cite{Bekaert:2014cea}.\footnote{For other works on spinning bulk-to-bulk propagators, see \cite{Leonhardt:2003qu,Francia:2007qt,Francia:2008hd,Manvelyan:2008ks}.} The totally symmetric spin-$J$ harmonic function $\Omega_{\nu,J}$ is traceless and divergenceless regular bi-tensor, with equation of motion
\begin{equation}
    \left(\Box_1+\left(\tfrac{d}{2}\right)^2+\nu^2+J\right)\Omega_{\nu,J}\left(x_1;x_2\right)=0.
\end{equation}
Like for the scalar harmonic functions \eqref{scafacharm}, they factorise into a product of bulk-to-boundary propagators:
\begin{equation}\label{spinharmfact}
      \Omega_{\nu,J}\left(x_1;x_2\right)= \frac{\nu^2}{\pi }  \int_{\partial \text{AdS}}dP K_{\frac{d}{2}+i\nu,J}\left(X_1;P\right) \cdot K_{\frac{d}{2}-i\nu,J}\left(X_2;P\right).
\end{equation}
Combining \eqref{spinharmfact} with the representation \eqref{genbubuprop} of the bulk-to-bulk propagators, a one-loop bubble diagram {\small ${\cal M}^{\text{2pt bubble}}_{s;s_1,s_2}$} with spin-$s$ external fields of mass {\small$m^2 R^2=\Delta\left(\Delta-d\right)-s$} and fields of spins $s_1$ and $s_2$ propagating in the loop has a decomposition of the form 
\begin{multline}\label{decomspin}
{\cal M}^{\text{2pt bubble}}_{s;s_1,s_2}\left(y_1,y_2\right) = \sum_{{\bf p},{\bf q}}\frac{1}{\pi^2} \int^\infty_{-\infty} \nu^2 d\nu {\bar \nu}^2 d{\bar \nu}\, g^{(s_1)}_{p_1,p_2,p_3}\left(\nu\right)g^{(s_2)}_{q_1,q_2,q_3}\left({\bar \nu}\right) \\  \times \int d^dyd^d{\bar y}\,{\cal M}^{\text{3pt tree-level}}_{s,s^\prime_1,s^\prime_2;\Delta,\frac{d}{2}+i\nu,\frac{d}{2}+i{\bar \nu}}\left(y_1,y,{\bar y}\right) \cdot {\cal M}^{\text{3pt tree-level}}_{s,s^\prime_1,s^\prime_2;\Delta,\frac{d}{2}-i\nu,\frac{d}{2}-i{\bar \nu}}\left(y_2,y,{\bar y}\right),
\end{multline}
in terms tree-level spinning three-point amplitudes {\small ${\cal M}^{\text{3pt tree-level}}_{s,s^\prime_1,s^\prime_2;\Delta,\frac{d}{2}\pm i\nu,\frac{d}{2}\pm i{\bar \nu}}$}, which generalises the scalar case \eqref{2ptbubampscafact} and is illustrated in figure \ref{fig::genapp1}. For concision we introduced: {\small $s^\prime_i=s_i-2p_{i+1}-p_{i-1}$} where $i \cong i+3$. 

For totally symmetric fields, all tree level three-point amplitudes are known for arbitrary cubic coupling constants \cite{Sleight:2016dba,Sleight:2016hyl,Sleight:2017fpc}. The task is then to evaluate the three- and two-point spinning conformal integrals in each term of the decomposition \eqref{decomspin}. We explain how to do this in \S \tcb{\ref{subsec::confint}}. We first review the evaluation of tree-level three-point Witten diagrams for spinning fields in the following section.

\subsection{Review: Cubic couplings and 3pt Witten diagrams}
\label{subsec::revcubic} 

For a generic triplet of spinning fields on AdS$_{d+1}$, the possible couplings respecting the AdS isometry are in general not unique. In the ambient space formalism, a basis of on-shell cubic vertices for totally symmetric fields $\varphi_{s_i}$ of spins $s_i$ and mass $m^2_i R^2 = \Delta_i\left(\Delta_i-d\right)-s_i$, is given by \cite{Sleight:2017fpc}\footnote{For concision we define: $\sum\limits_{m_i} = \sum\limits^{\text{min}\left\{s_1,s_2,n_3\right\}}_{m_3=0}  \sum\limits^{\text{min}\left\{s_1-n_3,s_3,n_2\right\}}_{m_2=0}\sum\limits^{\text{min}\left\{s_2-n_3,s_3-n_2,n_1\right\}}_{m_1=0}$.}
\begin{multline}\label{Ical}
{\cal I}_{s_1,s_2,s_3}^{n_1,n_2,n_3}=\sum_{m_i} C_{s_1,s_2,s_3;m_1,m_2,m_3}^{n_1,n_2,n_3}{\cal Y}^{s_1-m_2-m_3}_1{\cal Y}^{s_2-m_3-m_1}_2{\cal Y}^{s_3-m_1-m_2}_3  {\cal H}^{m_1}_1{\cal H}^{m_2}_2{\cal H}^{m_3}_3\, \\
\times \varphi_{s_1}\left(X_1,U_1\right)\varphi_{s_2}\left(X_2,U_2\right)\varphi_{s_s}\left(X_3,U_3\right)\Big|_{X_i=X},
\end{multline}
with coefficients 
\begin{multline}
   C_{s_1,s_2,s_3;m_1,m_2,m_3}^{n_1,n_2,n_3}= \left(\tfrac{d-2 (s_1+s_2+s_3-1)-\left(\tau_1+\tau_2+\tau_3\right)}{2} \right)_{m_1+m_2+m_3}\\\times\, \prod_{i=1}^3 \left[2^{m_i}\,\binom{n_i}{m_i} (n_i+\delta_{(i+1)(i-1)}-1)_{m_i}\right]\,,
\end{multline}
and $\delta_{(i-1)(i+1)}=\frac12(\tau_{i-1}+\tau_{i+1}-\tau_i)$, $i \cong i+3$. This is built from six basic $SO\left(d+1,1\right)$-covariant contractions (see e.g. \cite{Joung:2011ww,Joung:2012rv,Taronna:2012gb,Joung:2012hz}):
\begin{subequations}\label{6cont}
\begin{align}
    \mathcal{Y}_1&=\pl_{U_1}\cdot\pl_{X_2}\,,&\mathcal{Y}_2&=\pl_{U_2}\cdot\pl_{X_3}\,,&\mathcal{Y}_3&=\pl_{U_3}\cdot\pl_{X_1}\,,\\
    \mathcal{H}_1&=\pl_{U_2}\cdot\pl_{U_3}\,,&\mathcal{H}_2&=\pl_{U_3}\cdot\pl_{U_1}\,,&\mathcal{H}_3&=\pl_{U_1}\cdot\pl_{U_2}\,.
\end{align}
\end{subequations}

The basis \eqref{Ical} is convenient for Witten diagram computations, in particular because the three-point amplitude generated by each basis element is given by simple three-point conformal structure on the boundary \cite{Sleight:2017fpc}:
\begin{equation}\label{IntegralBasisAdS}
{\cal M}^{n_1,n_2,n_3}_{s_1,s_2,s_3;\tau_1,\tau_2,\tau_3}\left(y_1,y_2,y_3\right)={\sf B}(s_i;n_i;\tau_i)\,[[{\cal O}_{\Delta_1,s_1}(y_1){\cal O}_{\Delta_2,s_2}(y_2) {\cal O}_{\Delta_3,s_3}(y_3)]]^{(\text{{\bf n}})}\,,
\end{equation}
with\footnote{Recall the six three-point conformally covariant building blocks are given by ($i \cong i+3$) 
\begin{subequations}
\begin{align}
 {\sf Y}_{i,\left(i-1\right)\left(i+1\right)} & =\frac{z_i\cdot y_{(i-1)i}}{y_{(i-1)i}^2}-\frac{z_i\cdot y_{(i+1)i}}{y_{(i+1)i}^2}, \\
 {\sf H}_{\left(i-1\right)\left(i+1\right)} &=\frac{1}{y_{(i-1)(i+1)}^2}\left(z_{i-1}\cdot z_{i+1}+\frac{2 z_{i-1}\cdot y_{(i-1)(i+1)}\, z_{i+1}\cdot y_{(i+1)(i-1)}}{y_{(i+1)(i-1)}^2}\right).
\end{align}
\end{subequations}
Note that we adopt a different notation to \cite{Sleight:2017fpc}, which can be obtained through the replacements: ${\sf Y}_{i,\left(i-1\right)\left(i+1\right)}\rightarrow{\sf Y}_i$, ${\sf H}_{\left(i-1\right)\left(i+1\right)} \rightarrow {\sf H}_{i}$, ${\sf q}_{i,\left(i-1\right)\left(i+1\right)}\rightarrow {\sf q}_i$.
} 
 \begin{multline}
 \left[\left[{\cal O}_{\Delta_1,s_1}(y_1){\cal O}_{\Delta_2,s_2}(y_2) {\cal O}_{\Delta_3,s_3}(y_3)\right]\right]^{(\text{{\bf n}})} \\ \equiv\frac{{\sf H}_{32}^{n_1}{\sf H}_{13}^{n_2}{\sf H}_{21}^{n_3}}{(y_{12})^{\delta_{12}}(y_{23})^{\delta_{23}}(y_{31})^{\delta_{31}}} 
\left[\prod_{i=1}^32^{\tfrac{\delta_{(i+1)(i-1)} }{2}+n_i-1} \Gamma \left(\tfrac{\delta_{(i+1)(i-1)} }{2}+n_i\right)\right]
\\ \times\left[\prod_{i=1}^3 {\sf q}_{i,\left(i-1\right)\left(i+1\right)}^{\frac{1-n_i}{2}-\frac{\delta_{(i+1)(i-1)} }{4}} J_{(\delta_{(i+1)(i-1)}+2n_i -2)/2}\left(\sqrt{{\sf q}_{\left(i-1\right)\left(i+1\right)}}\right)\right]
\,{\sf Y}_{1,32}^{s_1-n_2-n_3}{\sf Y}_{2,13}^{s_2-n_3-n_1}{\sf Y}_{3,21}^{s_3-n_1-n_2}, \label{nicebasis}
 \end{multline}
and we define
\begin{equation}
    {\sf q}_{i,\left(i-1\right)\left(i+1\right)} = 2 {\sf H}_{\left(i-1\right)\left(i+1\right)}\,\partial_{{\sf Y}_{i+1,i\left(i-1\right)}} \cdot \partial_{{\sf Y}_{i-1,(i+1)i}}.
\end{equation}
 The coefficients ${\sf B}(s_i;n_i;\tau_i)$ are given by
\begin{multline}\label{beeeeeee}
{\sf B}(s_i;n_i;\tau_i)=\pi ^{-d}  (-2)^{(s_1+s_2+s_3)-(n_1+n_2+n_3)-4}\,\Gamma \left(\tfrac{\tau_1+\tau_2+\tau_3-d+2 (s_1+s_2+s_3)}{2} \right)\\  \times\,\prod_{i=1}^3 \frac{\Gamma \left(s_i- n_{i+1}+ n_{i-1}+\frac{\tau_i+\tau_{i+1}-\tau_{i-1}}{2}\right)\Gamma \left(s_i+n_{i+1}- n_{i-1}+\frac{\tau_i+\tau_{i-1}-\tau_{i+1}}{2}\right)}{\Gamma \left(2 n_i+\frac{\tau_{i+1}+\tau_{i-1}-\tau_i}{2}\right)}\\ 
\times \prod_{i=1}^3 \frac{\Gamma (s_i+n_{i+1}+n_{i-1}+\tau_i-1)}{\Gamma \left(s_i+\tau_i-\tfrac{d}{2}+1\right)\Gamma (2 s_i+\tau_i-1)}. 
\end{multline}
The expression \eqref{IntegralBasisAdS} for the amplitude is to be compared with the comparably more involved amplitude \cite{Sleight:2016dba} generated by the canonical basis of cubic couplings given by monomials in ${\cal Y}_{i,\left(i-1\right)\left(i+1\right)}$ and ${\cal H}_{\left(i-1\right)\left(i+1\right)}$.

Employing the basis \eqref{Ical} of cubic couplings and bulk-to-bulk propagators \eqref{genbubuprop}, the spectral decomposition of spinning bubble diagrams \eqref{decomspin} will contain terms of the generic form
\begin{equation}\label{gnericterm}
\int^\infty_{-\infty}  d\nu d{\bar \nu}\,\nu^2{\bar \nu}^2 g^{(s_1)}_{p_1,p_2,p_3}\left(\nu\right)g^{(s_2)}_{q_1,q_2,q_3}\left({\bar \nu}\right) F^{{\bf n},{\bf m}}_{s,s^\prime_1,s^\prime_2;\tau_s}\left(\nu,{\bar \nu};y_1,y_2\right),
\end{equation}
where, 
{\footnotesize \begin{multline}\label{gnericterm2}
 F^{{\bf n},{\bf m}}_{s,s^\prime_1,s^\prime_2;\tau_s}\left(\nu,{\bar \nu};y_1,y_2\right) \\ \propto
\int_{\partial \text{AdS}} d^dy d^d{\bar y} \, {\cal M}^{n_1,n_2,n_3}_{s,s_1,s_2;\tau_s,\tfrac{d}{2}+i\nu-s_1,\tfrac{d}{2}+i{\bar \nu}-s_2}\left(y_1,y,{\bar y}\right) \cdot {\cal M}^{m_1,m_2,m_3}_{s,s_1,s_2;\tau_s,\tfrac{d}{2}+i\nu-s_1,\tfrac{d}{2}+i{\bar \nu}-s_2}\left(y_2,y,{\bar y}\right).
\end{multline}}

Inserting in \eqref{gnericterm2} the explicit expressions \eqref{IntegralBasisAdS} for the three-point amplitudes, we see that a key step is then to evaluate conformal integrals of the type:
\begin{multline}\label{genconfint0}
    \mathfrak{K}^{({\bf n,m})}(\nu,{\bar \nu}\,;y_1,y_2)=\int d^dyd^d{\bar y}\,[[{\cal O}_{\Delta,s}(y_1,z_1){\cal O}_{\frac{d}{2}+i\nu,s_1}(y,\hat{\pl}_{z}) {\cal O}_{\frac{d}{2}+i{\bar \nu},s_2}({\bar y},\hat{\pl}_{{\bar z}})]]^{(\text{{\bf n}})}\\\times[[{\cal O}_{\frac{d}{2}-i{\bar \nu},s_2}({\bar y},{\bar z}) {\cal O}_{\frac{d}{2}-i \nu,s_1}(y,z){\cal O}_{\Delta,s}(y_2,z_2)]]^{(\text{{\bf m}})}\,,
\end{multline}
which we discuss in the following.

\subsection{Conformal Integrals}
\label{subsec::confint}

As explained in the previous section, by employing the basis \eqref{Ical} of on-shell cubic vertices, the task of computing one-loop bubble diagrams is reduced to evaluating conformal integrals of the form 
\begin{multline}\label{genconfint}
     \mathfrak{K}^{({\bf n,m})}_{s;s_1,s_2}(\nu,{\bar \nu}\,;y_1,y_2)=\int d^dyd^d{\bar y}\,[[{\cal O}_{\Delta,s}(y_1,z_1){\cal O}_{\frac{d}{2}+i\nu,s_1}(y,\hat{\pl}_{z}) {\cal O}_{\frac{d}{2}+i{\bar \nu},s_2}({\bar y},\hat{\pl}_{{\bar z}})]]^{(\text{{\bf n}})}\\\times[[{\cal O}_{\frac{d}{2}-i{\bar \nu},s_2}({\bar y},{\bar z}) {\cal O}_{\frac{d}{2}-i \nu,s_1}(y,z){\cal O}_{\Delta,s}(y_2,z_2)]]^{(\text{{\bf m}})}\,,
\end{multline}
for external fields of spin $s$ and mass $m^2R^2 = \Delta\left(\Delta-d\right)-s$, and internal spins $s_1$ and $s_2$. 

The integral \eqref{genconfint} can be expanded in terms of the basic conformal integrals: 
\begin{equation}\label{basicconfi}
    \mathfrak{I}^{a_1,a_2,b_1,b_2}_{\alpha_1,\alpha_2,\gamma,\beta_1,\beta_2}\equiv\int d^dy d^d{\bar y}\frac{(z_1\cdot \left(y_{1}-y\right))^{a_1}(z_2\cdot \left(y_{2}-y\right))^{a_2}(z_1\cdot\left(y_{1}-{\bar y}\right))^{b_1}(z_2\cdot \left(y_{2}-{\bar y}\right))^{b_2}}{\left[\left(y_{1}-y\right)^2\right]^{\alpha_1}\left[\left(y_{2}-y\right)^2\right]^{\alpha_2}\left[\left(y-{\bar y}\right)^2\right]^{\gamma}\left[\left(y_{1}-{\bar y}\right)^2\right]^{\beta_1}\left[\left(y_{2}-{\bar y}\right)^2\right]^{\beta_2}}\,,
\end{equation}
where conformal invariance requires:
\begin{align}
\alpha_1-a_1+\alpha_2-a_2+\gamma&=d\,,& \beta_1-b_1+\beta_2-b_2+\gamma&=d\,.
\end{align}
This decomposition of \eqref{genconfint} is shown in \S \tcb{\ref{subsec::expansion}}. Direct evaluation of \eqref{basicconfi} gives:\footnote{Without loss of generality we set $z_1\cdot z_2=0$, since terms proportional to $z_1 \cdot z_2$ can be recovered by conformal symmetry.}
\begin{multline}
     \mathfrak{I}^{a_1,a_2,b_1,b_2}_{\alpha_1,\alpha_2,\gamma,\beta_1,\beta_2}=\frac{\pi^{d/2}}{(y_{12}^2)^{d/2-\gamma}}\sum_{n=0}^{a_1}\sum_{m=0}^{a_2}\binom{a_1}{n}\binom{a_2}{m}\left(\frac{z_1\cdot y_{12}}{y_{12}^2}\right)^{a_1-n}\left(\frac{z_2\cdot y_{21}}{y_{12}^2}\right)^{a_2-m}\\\nonumber
    \times\frac{\Gamma(\alpha_1+\gamma-a_1+n-\tfrac{d}{2})\Gamma(\alpha_2+\gamma-a_2+m-\tfrac{d}{2})\Gamma(\tfrac{d}{2}-\gamma+a_1+a_2-n-m)}{\Gamma(\alpha_1)\Gamma(\alpha_2)\Gamma(\gamma)}\\\nonumber
    \times\frac{\Gamma(\beta_1+\alpha_1+\gamma-a_1-b_1-\tfrac{d}{2})\Gamma(\beta_2+\alpha_2+\gamma-a_2-b_2-\tfrac{d}{2})}{\Gamma(\beta_1+\alpha_1+\gamma-a_1+n-\tfrac{d}{2})\Gamma(\beta_2+\alpha_2+\gamma-a_2+m-\tfrac{d}{2})}\\\nonumber
    \times(-\tfrac12 z_1\cdot\pl_{y_1})^{n+b_1}(-\tfrac12 z_2\cdot\pl_{y_2})^{m+b_2}{\cal M}^{\text{1-loop}}\left(y_1,y_2\right)\,,
\end{multline}

Using conformal symmetry to recover the full CFT structure and evaluating the derivatives in $y_1$ and $y_2$, we arrive to the following expression for the $\log$ term:
\begin{multline}
      \mathfrak{I}^{a_1,a_2,b_1,b_2}_{\alpha_1,\alpha_2,\gamma,\beta_1,\beta_2}\Big|_{\log}=\frac{2\pi^d}{(y_{12})^{d-\gamma}}\left(\frac{z_1\cdot y_{12}}{y_{12}^2}\right)^{a_1+b_1}\left(\frac{z_2\cdot y_{12}}{y_{12}^2}\right)^{a_2+b_2}\,\log(y_{12}^2)\ \sum_{n=0}^{a_1}\sum_{m=0}^{a_2}\binom{a_1}{n}\binom{a_2}{m}\\\times\tfrac{\Gamma \left(-a_1+n+\alpha_1+\gamma -\frac{d}{2}\right) \Gamma \left(-a_2+m+\alpha_2+\gamma -\frac{d}{2}\right) \Gamma \left(b_1+b_2+\frac{d}{2}+m+n\right) \Gamma \left(a_1+a_2+\frac{d}{2}-m-n-\gamma \right)}{\Gamma (\alpha_1) \Gamma (\alpha_2) \Gamma (\gamma ) \Gamma \left(b_1+\frac{d}{2}+n\right) \Gamma \left(b_2+\frac{d}{2}+m\right)}\,.
\end{multline}
One can then combine this result with the expansion of \eqref{genconfint} in terms of the basic conformal integrals \eqref{basicconfi} derived in \S \tcb{\ref{subsec::expansion}} to obtain the log contribution to $\mathfrak{K}^{({\bf n,m})}_{s;s_1,s_2}$.

\subsection{$s-\left(s^\prime\,0\right)-s$ bubble}
\label{subsec::ssprime0}

\begin{figure}[h]
\captionsetup{width=0.8\textwidth}
\centering
\includegraphics[scale=0.325]{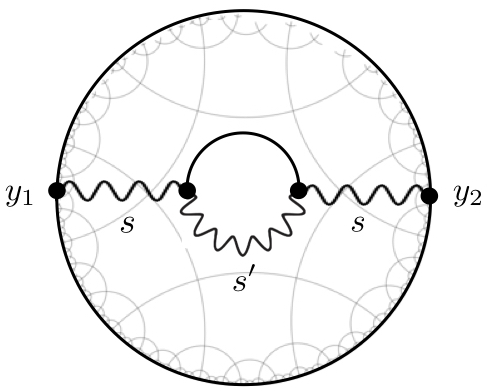}
\caption{One-loop bubble diagram with a gauge spin-$s$ field and a scalar propagating internally between two external gauge fields of spin $s$. Throughout we represent gauge fields with wavy lines.}
\label{fig::ssprimescalar}
\end{figure}

Let us now use this approach to extract the log contribution to bubble diagrams with a spin $s^\prime$ gauge field and a scalar field propagating internally between two external spin-$s$ gauge fields, illustrated in figure \ref{fig::ssprimescalar}. Owing to the scalar propagating in the loop, in this case there is no contribution from ghosts. Ghosts will be required only when gauge fields are propagating in the loop, as we do in \S \tcb{\ref{subsec::oneptspintad}} where tadpole diagrams with spin-$s$ gauge fields in the loop are considered.

In this subsection, we restrict ourselves to the contributions generated by the traceless and transverse part of the bulk-to-bulk propagators, which in the spectral representation \eqref{genbubuprop} corresponds to the term with $p_1=p_2=p_3=0$. This is the universal part of the propagator, which encodes the exchanged single-particle state. The spectral representation of the traceless and transverse part of a spin-$s$ bulk-to-bulk propagator for a field of mass $m^2 R^2 = \Delta\left(\Delta-d\right)-s$ is given by:
\begin{subequations}\label{ttbubu}
\begin{align}
    G^{TT}_{\Delta,s}\left(x_1;x_2\right) & = \int^\infty_{-\infty} d\nu\, g^{(s)}_{0,0,0}\left(\nu\right)  \Omega_{\nu,s}\left(x_1;x_2\right),\\ 
   g^{(s)}_{0,0,0}\left(\nu\right) &=\frac{1}{\left[\nu^2+\left(\Delta-\frac{d}{2}\right)^2\right]}.
\end{align}
\end{subequations}
The notation $TT$ signifies the restriction to the traceless and transverse part. The other terms in the propagators (i.e. terms in \eqref{genbubuprop} with at least one $p_i>0$) generate purely contact contributions to Witten diagrams, which in contrast are not universal and are dependent on the choice of field frame. In particular, contact contributions collapse in the bubble to  $\bep(14,10)\put(0,0){\line(1,0){14}}\put(7,4.25){\circle{8}}\eep$-type tadpole diagrams. This can be understood by noting that these contact contributions are related to $\bep(14,10)\put(0,0){\line(1,0){14}}\put(7,4.25){\circle{8}}\eep$ one-loop diagrams generated by quartic couplings under field re-definitions. In \S \tcb{\ref{subsec::gravloops}}, and also \S \tcb{\ref{secLLfullsinglecut}}, in some examples we shall compute bubble diagrams using the full bulk-to-bulk propagators which includes such contact terms.

The cubic vertex for spin-$s$, $s^\prime$ gauge fields with a scalar is given in de Donder gauge by \eqref{dedondver}, whose TT part reads: 
\begin{equation}\label{cubicssprime}
    {\cal V}^{\left(3\right)}_{s,s^\prime,0} = g {\cal Y}^{s}_1 {\cal Y}^{s^\prime}_2 \varphi_{s}\left(X_1,U_1\right)\varphi_{s^\prime}\left(X_2,U_2\right) \phi\left(X_3\right) \Big|_{X_i=X},
\end{equation}
for some coupling constant $g$. Recall that there are no contributions from Ghost vertices in this case owing to the scalar propagating in the loop. 
Via the factorisation \eqref{spinharmfact}, the bubble diagram generated by \eqref{cubicssprime} decomposes as
\begin{equation}
    {\cal M}^{\text{2pt bubble}}\left(P_1,P_2\right) = g^2 \int^\infty_{-\infty} \frac{\nu^2 {\bar \nu}^2 d\nu d{\bar \nu}}{\pi^2[\nu^2+(\Delta_{s^\prime}-\frac{d}{2})^2][{\bar \nu}^2+( \Delta-\frac{d}{2})^2]} F^{{\bf 0},{\bf 0}}_{s,s^\prime,0;\tau_s}\left(\nu,{\bar \nu};P_1,P_2\right),
\end{equation}
where $F^{{\bf 0},{\bf 0}}_{s,s^\prime,0;\tau_s}$ is the product of tree-level three-point amplitudes \eqref{gnericterm2}. Plugging in the explicit expressions \eqref{IntegralBasisAdS} for the latter, one obtains 
 \begin{multline}\label{ttanomspec}
    {\cal M}^{\text{2pt bubble}}\left(P_1,P_2\right) = g^2 \int^\infty_{-\infty} \frac{\nu^2 {\bar \nu}^2 d\nu d{\bar \nu}}{\pi^2[\nu^2+(\Delta_{s^\prime}-\frac{d}{2})^2][{\bar \nu}^2+( \Delta-\frac{d}{2})^2]}   \\ \times 
{\sf B}\left(s,s^\prime,0;0;\Delta_s-s,\tfrac{d}{2}+i\nu-s^\prime,\tfrac{d}{2}+i{\bar \nu}\right){\sf B}\left(s,s^\prime,0;0;\Delta_s-s,\tfrac{d}{2}-i\nu-s^\prime,\tfrac{d}{2}-i{\bar \nu}\right) \\ \times \mathfrak{K}^{\left({\sf 0},{\sf 0}\right)}_{s;s^\prime,0}(\nu,{\bar \nu};y_1,y_2),
\end{multline}
where $\mathfrak{K}_{s;s^\prime,0}^{({\sf 0,0})}$ is the conformal integral \eqref{subsec::confint}, with log contribution (see \S \tcb{\ref{subsec::confint}}) whose explicit evaluation yields the remarkably simple result:
\begin{multline}\label{ksso}
   \mathfrak{K}_{s;s^\prime,0}^{({\sf 0,0})}(\nu,{\bar \nu};y_1,y_2)\Big|_{\log(y_{12}^2)}=\frac{\pi ^{d+\frac{1}{2}} 2^{-d-s^\prime+6} s! \Gamma (d+s^\prime-2) \Gamma (d+2 s-4)}{(d+2 s-2) \Gamma \left(\frac{d-1}{2}\right) \Gamma \left(\frac{d}{2}+s^\prime-1\right) \Gamma (d+s-3)}\\\times \tfrac{\Gamma \left(\tfrac{s^\prime+2+i (\nu- \bar{\nu})}{2} \right) \Gamma \left(\tfrac{s^\prime+2-i (\nu- \bar{\nu})}{2} \right) \Gamma \left(\tfrac{-d+s^\prime-2 s+4+i (\nu- \bar{\nu})}{2}\right) \Gamma \left(\tfrac{-d+s^\prime-2 s+4-i (\nu+ \bar{\nu})}{2} \right)}{\Gamma \left(\tfrac{d+s^\prime-2+i (\nu-\bar{\nu})}{2}\right) \Gamma \left(\tfrac{-d+s^\prime+4+i (\nu-\bar{\nu})}{2}\right) \Gamma \left(\tfrac{-d+s^\prime+4-i (\nu- \bar{\nu})}{2} \right) \Gamma \left(\tfrac{d+s^\prime-2-i (\nu-\bar{\nu})}{2}\right)
   }\,\frac{\log(y_{12}^2)}{(y_{12}^2)^{d-2}}\,\left(\frac{{\sf H}_{21}}{2}\right)^s\,.
\end{multline}
Recall that in this section we take $\Delta_s = s+d-2$ for a spin-$s$ gauge field, which is substituted in \eqref{ksso} above.

Putting everything together gives the following spectral representation of the contribution to the anomalous dimension of a spin-$s$ higher-spin current on the boundary: 
\begin{multline}\label{gensosprimett}
   \gamma_{TT} = -\, g^2_{s,0,s^\prime}
\frac{\pi ^{-\frac{7+d}{2}} s! 2^{-d+s^\prime+s-2} \Gamma (d+s^\prime-2)}{(d+2 s-4) \Gamma \left(\frac{d-1}{2}\right) \Gamma \left(\frac{d}{2}+s^\prime-1\right) \Gamma \left(\frac{d}{2}+s\right) \Gamma (d+2 s-2)}\\\times \int^\infty_{-\infty} d\nu d{\bar \nu}\, {\cal F}^{\text{2pt bubble}}_{TT}\left(\nu,{\bar \nu}\right), 
\end{multline}
and
{\small \begin{align}\label{specfunspinbubtypa0}
& {\cal F}^{\text{2pt bubble}}_{TT}\left(\nu,{\bar \nu}\right) =
\frac{\nu\bar{\nu}\,\sinh (\pi  \nu) \sinh (\pi  \bar{\nu})}{\left[\nu^2+\left(\Delta_s-\tfrac{d}{2}\right)^2\right]\left[\bar{\nu}^2+\left(\Delta-\tfrac{d}{2}\right)^2\right]}\frac{\Gamma \left(\frac{d}{2}-i \nu-1\right) \Gamma \left(\frac{d}{2}+i \nu-1\right) }{\Gamma \left(\frac{d}{2}+s^\prime-i \nu-1\right) \Gamma \left(\frac{d}{2}+s^\prime+i \nu-1\right)}\nonumber\\& \hspace*{1cm} \times \Gamma \left(\tfrac{d+s^\prime+2 s-2+i (\nu- \bar{\nu})}{2}\right) \Gamma \left(\tfrac{d+s^\prime+2 s-2-i (\nu- \bar{\nu})}{2}\right) \Gamma \left(\tfrac{d+s^\prime+2 s-2-i (\nu+\bar{\nu})}{2}\right) \Gamma \left(\tfrac{d+s^\prime+2 s-2+i (\nu+\bar{\nu})}{2}\right)\nonumber\\& \hspace*{1cm}\times \Gamma \left(\tfrac{s^\prime+2+i(\nu-\bar{\nu})}{2} \right) \Gamma \left(\tfrac{s^\prime+2-i (\nu- \bar{\nu})}{2}\right) \Gamma \left(\tfrac{s^\prime+2+i (\nu+\bar{\nu})}{2}\right) \Gamma \left(\tfrac{s^\prime+2-i (\nu+\bar{\nu})}{2}\right) \,.
\end{align}}

\noindent
A consistency check is the recovery of the spectral function \eqref{phi3specfun} from \eqref{specfunspinbubtypa0} for the bubble in $\phi^3$ theory when one sets $s=s^\prime=0$, and $\Delta_1=\Delta_2=d-2$ in \eqref{phi3specfun}.

\subsubsection*{Pole structure}

It is also interesting to study the pole structure of the spectral function \eqref{specfunspinbubtypa0}. At fixed $\bar{\nu}$, apart from the single poles at $\nu=\pm i(\Delta_s-\tfrac{d}{2})$, which is usually uplifted to a branch cut in $\zeta$-function regularisation, the above displays $8$ series of poles -- one for each gamma functions factor in the numerator -- labelled by non-negative integers:
\begin{align}\label{nupoles}
\pm i\nu&= \pm i \bar{\nu}+d+s^\prime+2s-2+2n\,,& \pm i \nu&=\pm i \bar{\nu}+s^\prime+2+2n\,,
\end{align}
for all possible uncorrelated permutations of the $\pm$. On top of the above poles \eqref{nupoles}, we also have a finite number of additional (spurious) poles at:
\begin{align}
    \pm i\nu&=1-\tfrac{d}{2}-n\,,& \pm i\nu&-1+\tfrac{d}{2}+s^\prime>0\,,
\end{align}
coming from the $\Gamma$-function factor on the first line of \eqref{specfunspinbubtypa0}, which arise for $s^\prime>n$ and are absent for $s^\prime=0$. Their effect is compensated by the contact contributions in the bulk-to-bulk propagator, see e.g. \cite{Cornalba:2007fs,Costa:2012cb}.
Upon introducing regulators $\mu$ and ${\bar \mu}$ one can perform the above integral with Mellin-Barnes techniques defining:
\begin{equation}
    H(\mu,{\bar \mu})=\int_{-\infty}^\infty d\nu\,d\bar{\nu}\,{\cal F}^{\text{2pt bubble}}_{TT}\left(\nu,{\bar \nu}\right)\,\mu^{i\nu}{\bar \mu}^{i\bar{\nu}}\,,
\end{equation}
which is analytic in $\mu$ and ${\bar \mu}$ for an appropriate domain in the complex $\mu$ and ${\bar \mu}$ plane. As mentioned in the introduction, the above function defines a generalised hypergeometric function whose analyticity properties regulate the spectral integral. After closing the contour in the appropriate domain and performing the $\nu$ integration, one is left with a function of $\bar{\nu}$ with a pole at $\bar{\nu}=\pm i(\Delta-\tfrac{d}{2})$ and some leftover single poles which can be obtained from \eqref{nupoles} upon substituting the location of the $\nu$ pole. For instance, when sitting on the pole $\nu=\pm i(\Delta_s-\tfrac{d}{2})$ the corresponding $\bar{\nu}$ poles are located at:
\begin{align}
\pm i\bar{\nu}&= \pm(\Delta_s-\tfrac{d}{2})+d+s^\prime+2s-2+2n\,,& \pm i \bar{\nu}&=\pm(\Delta_s-\tfrac{d}{2})+s^\prime+2+2n\,.
\end{align}

It should also be noted that for integer values of $\nu$ and $\bar{\nu}$ the $\sinh$ has zeros which cancel possible poles at these location.

A relatively simple and interesting case is $d=3$, which is relevant for higher-spin gauge theories on AdS$_4$. In this case the structure of the spectral function drastically simplifies:
\begin{multline}\label{3dspct}
{\cal F}^{\text{2pt bubble}}_{TT}\left(\nu,{\bar \nu}\right) =
\frac{\pi\nu\,\pi\bar{\nu}\,\sinh (\pi  \nu) \sinh (\pi  \bar{\nu})}{\left[\nu^2+\left(\Delta_s-\tfrac{3}{2}\right)^2\right]\left[\bar{\nu}^2+\left(\Delta-\tfrac{3}{2}\right)^2\right]}\,\frac{\Gamma \left(\frac{3}{2}-i \nu-1\right) \Gamma \left(\frac{3}{2}+i \nu-1\right) }{\Gamma \left(\frac{3}{2}+s^\prime-i \nu-1\right) \Gamma \left(\frac{3}{2}+s^\prime+i \nu-1\right)}\\\times P(\nu-\bar{\nu})P(\nu+\bar{\nu})\,\frac{\pi(\nu+\bar{\nu})\,\pi(\nu-\bar{\nu})}{\sinh [\pi (\nu+\bar{\nu})] \sinh [\pi  (\nu-\bar{\nu})]}\,,
\end{multline}
in terms of a polynomial function $P$ which depends only on the internal and external spins $s$ and $s^\prime$:
\begin{equation}
    P(\alpha)=\left[\prod_{i=0}^{s-1}\left[\left(\tfrac{s^\prime+1}2+i\right)^2+\left(\tfrac{\alpha}{2}\right)^2\right]\right]\left[\prod_{j=1}^{s^\prime}\left[\left(\tfrac{j}2\right)^2+\left(\tfrac{\alpha}{2}\right)^2\right]\right]\,.
\end{equation}
Apart from the spurious poles coming from the $\Gamma$-function factors on the first line of \eqref{3dspct}, one can see that all physical poles are resummed into the simple factor:
\begin{equation}
    \frac{\pi(\nu+\bar{\nu})\,\pi(\nu-\bar{\nu})}{\sinh [\pi (\nu+\bar{\nu})] \sinh [\pi  (\nu-\bar{\nu})]}\,,
\end{equation}
dressed by a polynomial factor at fixed $s$ and $s^\prime$.

\subsection{One-point bulk tadpoles}
\label{subsec::oneptspintad}

\begin{figure}[h]
\centering
\begin{subfigure}{.5\textwidth}
  \centering
  \captionsetup{width=0.8\textwidth}
  \includegraphics[scale=0.45]{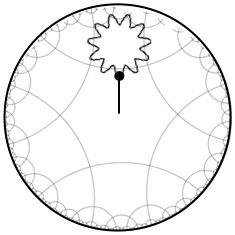}
  \caption{One-point tadpole with off-shell scalar external leg and spin-$s$ gauge field propagating in the loop.}
  \label{fig:sub1}
\end{subfigure}%
\begin{subfigure}{.5\textwidth}
  \centering
  \includegraphics[scale=0.45]{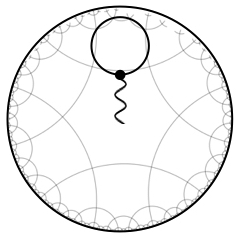}
  \caption{One-point tadpole with off-shell external spin-$s$ gauge field and a scalar propagating in the loop.}
  \label{fig:sub2}
\end{subfigure}
\caption{}
\label{fig:test}
\end{figure}

Let us also discuss the contribution from tadpole diagrams generated by the coupling \eqref{cubicssprime}, with a single bulk external leg. There are two cases, which are depicted in figure \ref{fig:test}. As in the preceding section, we focus on the contributions generated by the traceless and transverse part of the bulk-to-bulk propagators. Like for the scalar one-point tadpole diagrams considered in \S \tcb{\ref{subsec::onepttad}}, we can argue that they give vanishing contributions.

We first consider the case of a scalar external leg and a spin-$s$ field propagating in the loop, displayed in figure \ref{fig:test} (a). In this case, there is in principle a contribution from ghost fields whose cubic vertex is given by the second term in \eqref{dedonphgho} below, in de Donder gauge. 

The corresponding generalisation of the tadpole factor \eqref{tapofactsc} connected to the boundary associated to a $0$-$s$-$s$ vertex in type A theory is, for both physical and ghost fields:

{\small \begin{multline}
   \mathcal{T}_s(\bar{P})=-\,g_{s,s,0}\,\frac{C_{\bar{\Delta},0}}{\pi}\,q_s({\bar{\Delta}})\left[\int_{-\infty}^\infty d\nu\underbrace{\frac{\nu^2}{\nu^2+(\Delta-\tfrac{d}2)^2}\,C_{\tfrac{d}2-i \nu,s}C_{\tfrac{d}2+i \nu,s}}_{f_s(\nu)}\right]\,\\\times\underbrace{\int dP\,dX\,\frac{(-2P\cdot {\bar P})^s}{(-2P\cdot X)^{d+s}(-2{\bar P}\cdot X)^{{\bar \Delta}+s}}}_{\mathcal{I}_{s}}\,,
\end{multline}}
with
\begin{equation}
    q_s({\bar{\Delta}})=(-2)^s \frac{(d+2 s-2)\,(d+s-3)!}{(d-2)!} \frac{\Gamma (s+\bar{\Delta} )}{ \Gamma (\bar{\Delta} )}\,.
\end{equation}
The latter result holds for both ghost and physical vertex \cite{Sleight:2017cax} (see also \S \ref{secLLfullsinglecut}) which read in this case:
\begin{equation}\label{dedonphgho}
    \mathcal{V}=g_{0,s,s}\left[\mathcal{Y}_1^s\mathcal{Y}_2^s\varphi_{1}\varphi_2\phi_3+s(d-4+2s)\,\mathcal{Y}_1^{s-1}\mathcal{Y}_2^{s-1}\bar{c}_{1}c_2\phi_3\right]\,,
\end{equation}
and which are both polynomials in the $\mathcal{Y}_i$ structures.
The coupling constant $g_{0,s,s}$ for the type A theory reads:
\begin{equation}
    g_{0,s,s}=\frac{\pi ^{\frac{d-3}{4}} 2^{\frac{3 d}{2}+s-\frac{1}{2}} \sqrt{\Gamma \left(\frac{d-1}{2}\right)} \Gamma \left(\frac{d-1}{2}+s\right)}{s!\, \Gamma (d+2 s-3)}\,,
\end{equation}
The UV divergent spectral integral in $\nu$ coming from the spin-$s$ bulk-to-bulk propagator is completely factorised from the bulk and boundary integral, and the integrand reads more explicitly:
\begin{equation}
     f_s(\nu)=\frac{\nu^2}{4\pi^{d}}\frac{\nu^2+(s+\tfrac{d}{2}-1)^2}{\nu^2+(\Delta-\tfrac{d}2)^2}\Gamma(\tfrac{d}{2}-1-i\nu)\Gamma(\tfrac{d}{2}-1+i\nu)\frac{\sinh{\pi\nu}}{\pi\nu}\,,
\end{equation}
where for a spin-$s$ gauge field one chooses $\Delta^{\text{ph.}}=d-2+s$ and for spin $s-1$ ghosts one chooses $\Delta^{\text{gh.}}=d-1+s$. We have also introduced the function $q_{s}({\bar \Delta)}$ which encodes the result of vertex contractions in terms of the dimension ${\bar \Delta}=\frac{d}{2}-i{\bar \nu}$ of the external leg to the tadpole. In $d=3$ the latter simplifies to
\begin{equation}
    f_s(\nu)=\frac1{4\pi^3}\frac{4\nu^2+(2s+1)^2}{4\nu^2+(2\Delta-3)^2}\,\nu\tanh{\pi\nu}\,,
\end{equation}
which can be regularised via $\zeta$-function regularisation after splitting it into two pieces as:
\begin{equation}
     \frac1{4\pi^3}\int\,\nu\,\frac{[4\nu^2+(2s+1)^2]}{\left[4\nu^2+(2\Delta-3)^2\right]^{1+\mu}}\,- \frac1{4\pi^3}\int\frac{4\nu^2+(2s+1)^2}{4\nu^2+(2\Delta-3)^2}\frac{2\nu}{1+e^{2\pi \nu}}\,,
\end{equation}
with the second integral convergent. The above integrals, being of the general type \eqref{genspint0}, can also be explicitly evaluated via \eqref{genspint}.

Using the expression \eqref{2ptbulkint} for a generic two-point bulk integral, in this case we have (for $s>0$\footnote{In the $s>0$ case the second term in eq.~\eqref{2ptbulkint} is proportional to $\int dx^d (x^2)^s\delta(x)=0$ and therefore vanishes identically.}):
\begin{equation}
    \mathcal{I}_s=2\pi^{d/2+1}\frac{\Gamma(\tfrac{d}{2}+s)}{\Gamma(d+s)}A\,\delta(d-\bar{\Delta}).
\end{equation}
and combining all the ingredients we can then write down the following expression for the tadpole:
\begin{multline}
    \mathcal{T}^{\text{ph.}}_s=-\,\frac{2^{\frac{d+5}{2}} \pi ^{\frac{3 (d+1)}{4}} (-1)^s (d+2 s-3) (d+2 s-2)^2 \sqrt{\Gamma \left(\frac{d-1}{2}\right)} }{(d+s-2) (d+s-1) \Gamma (d-1) s!}\frac{\Gamma (s+\bar{\Delta})}{\Gamma (\bar{\Delta}) }\\\times\left(\int_{-\infty}^{\infty}d\nu f^{\text{ph.}}_s(\nu)\right)\,C_{\bar{\Delta},0}\,A\,\delta(\bar{\Delta}-d)\,,
\end{multline}
for physical fields together with
\begin{multline}
    \mathcal{T}^{\text{gh.}}_s=-\,\frac{2^{\frac{d+5}{2}} \pi ^{\frac{3 (d+1)}{4}} (-1)^s (d+2 s-4)^2 (d+2 s-3) \sqrt{\Gamma \left(\frac{d-1}{2}\right)} }{(d+s-3) (d+s-2) \Gamma (d-1) (s-1)!}\frac{\Gamma (s+\bar{\Delta}-1)}{\Gamma (\bar{\Delta}) }\\\times\left(\int_{-\infty}^{\infty}d\nu f_{s-1}^{\text{gh.}}(\nu)\right)\,C_{\bar{\Delta},0}\,A\,\delta(\bar{\Delta}-d),
\end{multline}
for the ghost contribution. We recall that the constant $A$ is given by $A=\int d^d x\frac{1}{(x^2)^{d}}$ and vanishes in our modified dimensional regularisation scheme (see \S \tcb{\ref{app::two2ptreg}}). Still, the above UV divergent coefficient can be straightforwardly evaluated using the methods of section \eqref{genspint}. Like for the scalar case presented in \S \tcb{\ref{subsec::onepttad}}, noticing also that $\bar{\Delta}=\tfrac{d}{2}-i\bar{\nu}$ with $\bar{\nu}$ restricted to real values, this contribution is vanishing.\footnote{Also the scalar cut vanishes for analogous reasons, since the corresponding real dimension for the conformally coupled scalar is also outside the domain in which the $\delta$-function is concentrated.}

To summarise, regulating the AdS IR divergences automatically recover the vanishing of the tadpole. The UV divergence is instead controlled by a factorised spectral integral which depends explicitly on $\bar{\Delta}$.

Let us now consider the diagram in figure \ref{fig:test} (b), with a spin-$s$ external leg and scalar propagating in the loop. In this case there is no contribution from ghosts. The diagram is given by: 
\begin{align}
\mathfrak{T}^{\text{1pt tadpole}}\left(X_1;U_1\right)& =-\,g_{s,0,0} \int_{\text{AdS}}dX\, \left(\partial_{U_2} \cdot \partial_{X_2}\right)^s G_{d-2,0}\left(X,X_2\right)\Big|_{X_2=X} G_{d-2,s}\left(X_1,U_1;X,U_2\right).
\end{align}
Focusing on the traceless and transverse part of the spin-$s$ bulk-to-bulk propagator, this factorises as
\begin{align}
\mathfrak{T}^{\text{1pt tadpole}}\left(X_1;U_1\right)\Big|_{TT}&=-\,g_{s,0,0} \int^\infty_{-\infty} \frac{{\bar \nu}^2d{\bar \nu}}{\pi\left[{\bar \nu}^2+\left(s+\frac{d}{2}-2\right)^2\right]}\int_{\partial \text{AdS}}d{\bar P}\,K_{\frac{d}{2}+i{\bar \nu},s}\left(X_1,U_1;{\bar P},{\hat \partial}_Z\right)\nonumber \\  & \hspace*{0.25cm} \times \int_{\text{AdS}}dX\, \left(\partial_{U_2} \cdot \partial_{X_2}\right)^s G_{d-2,0}\left(X,X_2\right)\Big|_{X_2=X} K_{\frac{d}{2}-i{\bar \nu},s}\left(X,U_2;{\bar P},Z\right).
\end{align}
Using the identity \eqref{derivpropsplit} for derivatives of bulk-to-bulk propagators at coincident points and \eqref{bubos} for spinning bulk-to-boundary propagators, the tadpole factor in the second line gives:
\begin{multline}
\int_{\text{AdS}}dX\, \left(\partial_{U_2} \cdot \partial_{X_2}\right)^s G_{d-2,0}\left(X,X_2\right)\Big|_{X_2=X} K_{\frac{d}{2}-i{\bar \nu},s}\left(X,U_2;{\bar P},Z\right)\\= \frac{2^sC_{\frac{d}{2}-i{\bar \nu},0}}{\left(\frac{d}{2}-i{\bar \nu}-1\right)_s}\int^\infty_{-\infty} \frac{C_{\frac{d}{2}+i\nu,0}C_{\frac{d}{2}-i\nu+s,0}\nu^2d\nu}{\pi \left[\nu^2+\left(\frac{d}{2}-2\right)^2\right]}\left(-i\nu+\tfrac{d}{2}\right)_s\\\times \int_{\partial\text{AdS}}dP\,\left({\cal D}_{{\bar P}}\left(Z;P\right)\right)^s\int_{\text{AdS}}dX\, \frac{1}{\left(-2 X \cdot P\right)^{d+s}}\frac{1}{\left(-2 X \cdot {\bar P}\right)^{\frac{d}{2}-i{\bar \nu}}}.
\end{multline}
In the same way as for the diagram (a), we can argue that in dimensional regularisation
\begin{equation}
    \mathfrak{T}^{\text{1pt tadpole}}\left(X_1;U_1\right)\Big|_{TT} \equiv 0\,.
\end{equation}
Considering other regularisations one can still argue that the latter vanishes using \eqref{2ptbulkint}:
\begin{multline}
    \int_{\text{AdS}}dX\, \frac{1}{\left(-2 X \cdot P\right)^{d+s}}\frac{1}{\left(-2 X \cdot {\bar P}\right)^{\frac{d}{2}-i{\bar \nu}}}= 2\pi^{d/2+1}\frac{\Gamma(\tfrac{d}2+s)}{\Gamma(d+s)}\,\frac{1}{(-2P\cdot\bar{P})^{d+s}}\delta(\tfrac{d}{2}+i{\bar \nu}+s)\\+2\pi^{d+1}\frac{\Gamma(-\tfrac{d}{2}-s)\Gamma(i {\bar \nu})}{\Gamma(d+s)\Gamma(\frac{d}{2}-i{\bar \nu})}\delta^{(d)}(P,\bar{P})\,\delta(s+\tfrac{d}{2}-i{\bar \nu}),
\end{multline}
and the fact that ${\bar \nu}$ is restricted to real values when considering a bulk to bulk propagator attached to a point in AdS.

\section{Applications}\label{sec::applications}

\subsection{Graviton bubble}
\label{subsec::gravloops}

In this section we consider the bubble diagram generated by the minimal coupling of scalar fields to gravity. In this case we shall use the full graviton propagator, which in de-Donder gauge reads \cite{Sleight:2017cax}:\footnote{In terms of the decomposition \eqref{genbubuprop}, we have
\begin{align}
\quad & g^{(2)}_{1,1,0}\left(\nu\right)= - \tfrac{1}{d\left(d-1\right)\left[\nu^2+\left(\frac{d}{2}+1\right)^2\right]},  \quad  && 
g^{(2)}_{1,0,0}\left(\nu\right)= \tfrac{1}{d\left[\nu^2+\left(\frac{d}{2}+1\right)^2\right]\left[\nu^2+\frac{d}{2}\left(\frac{d}{2}+4\right)\right]},
\\
 & g^{(2)}_{0,0,2}\left(\nu\right)=- \tfrac{\left(d-1\right)}{d\left[\nu^2+\left(\frac{d}{2}+1\right)^2\right]\left[\nu^2+\frac{d}{2}\left(\frac{d}{2}+4\right)\right]^2}, \quad && g^{(2)}_{0,0,1}\left(\nu\right)=0,
\end{align}
 and the traceless and transverse part, which is the same in any gauge, is: $g^{(2)}_{0,0,0}\left(\nu\right)= \frac{1}{\left[\nu^2+\left(\frac{d}{2}\right)^2\right]}$.
} 

{\small \begin{multline}\label{gravdedon}
G_{d,2}\left(x_1,x_2\right) = \int^\infty_{-\infty} \frac{d\nu}{\nu^2+\left(\tfrac{d}{2}\right)^2}\Omega_{\nu,2}\left(x_1,x_2\right)- \int^\infty_{-\infty} d\nu\, u^2_1u^2_2 \frac{1}{d\left(d-1\right)\left[\nu^2+\left(\tfrac{d}{2}+1\right)^2\right]}\Omega_{\nu,0}\left(x_1,x_2\right)\\
+ \int^\infty_{-\infty} d\nu  \frac{1}{d\left[\nu^2+\left(\tfrac{d}{2}+1\right)^2\right]\left[\nu^2+\tfrac{d}{2}\left(\tfrac{d}{2}+4\right)\right]}\left[  u^2_1 \left(u_2\cdot \nabla_2\right)^2+  u^2_2 \left(u_1\cdot \nabla_1\right)^2 \right]\Omega_{\nu,0}\left(x_1,x_2\right)\\
 - \int^\infty_{-\infty} d\nu  \frac{\left(d-1\right)}{d\left[\nu^2+\left(\tfrac{d}{2}+1\right)^2\right]\left[\nu^2+\tfrac{d}{2}\left(\tfrac{d}{2}+4\right)\right]^2}   \left(u_1\cdot \nabla_1\right)^2\left(u_2\cdot \nabla_2\right)^2\Omega_{\nu,0}\left(x_1,x_2\right)
\end{multline}}

The cubic coupling of scalars $\phi_1$ and $\phi_2$ to gravity is given in de Donder gauge by \cite{Sleight:2017fpc}
\begin{equation}
    \mathcal{V}^{(3)}_{2,0,0}\left(X\right)=g\,{\cal Y}_3^2\,\phi_1(X_1)\phi_2(X_2)\varphi_3(X_3,U_3)+g\,\frac12(d-2)\phi_1(X_1)\phi_2(X_2)\varphi^\prime_3(X_3)\Big|_{X_i=X}\,.
\end{equation}

In the following we compute the bubble diagram with $\phi_1$ on the external legs. This is given by the four terms,
\begin{equation}
    {\cal M}^{\text{2pt-bubble}}={\cal M}^{\text{2pt-bubble}}_{1,0;1,0}+\tfrac{1}{2}\left(d-2\right){\cal M}^{\text{2pt-bubble}}_{1,0;0,1}+\tfrac{1}{2}\left(d-2\right){\cal M}^{\text{2pt-bubble}}_{0,1;1,0}+\tfrac{1}{4}\left(d-2\right)^2{\cal M}^{\text{2pt-bubble}}_{0,1;0,1},
\end{equation}
where we defined:\footnote{Note that: $\left(U \cdot {\cal P} \cdot U\right)=u^2$.}
\begin{multline}\label{4gravexch}
{\cal M}^{\text{2pt-bubble}}_{a,c;b,d}\left(P_1,P_2\right) =g\, \int_{\text{AdS}}dX_1dX_2\,K_{\Delta_1,0}\left(X_1;P_1\right)K_{\Delta_2,0}\left(X_2;P_2\right)G_{d,2}\left(X_1,\partial_{U_1};X_2,\partial_{U_2}\right)\\
\times  \left(U_1 \cdot {\cal P}_1 \cdot U_1\right)^{c}\left(U_2 \cdot {\cal P}_2 \cdot U_2\right)^{d} \left(U_1 \cdot \partial_{X_1}\right)^{2a} \left(U_2 \cdot \partial_{X_2}\right)^{2b} G_{\Delta,0}\left(X_1,X_2\right).
\end{multline}
The spectral representation of the graviton \eqref{gravdedon} and scalar \eqref{scalarbubu} bulk-to-bulk propagators, via the factorisation \eqref{spinharmfact} of harmonic functions, leads to the following decomposition of the bubble diagram:
 \begin{multline}\label{factgravbub}
{\cal M}^{\text{2pt-bubble}}_{a,c;b,d}\left(y_1,y_2\right) = \frac{g^2}{\pi^2}\sum_{{\bf p}} \int^\infty_{-\infty}\nu^2 d\nu\, {\bar \nu}^2 d{\bar \nu}\, g^{(2)}_{p_1,p_2,p_3}\left(\nu\right)g^{(0)}_{0,0,0}\left({\bar \nu}\right) \\ \times \int_{\partial \text{AdS}}dPd{\bar P}\,{\cal A}^{a,c;p_1,p_3}_{\Delta_1,\frac{d}{2}+i{\bar \nu},\frac{d}{2}+i\nu}\left(P_1,{\bar P},P\right) \cdot {\cal A}^{b,d;p_2,p_3+2(p_1-p_2)}_{\Delta_2,\frac{d}{2}-i{\bar \nu},\frac{d}{2}-i\nu}\left(P_2,{\bar P},P\right),
\end{multline}
in terms of the tree-level three-point diagrams:
\begin{multline}\label{a3ptint}
   \mathcal{A}^{a,c;p_1,p_3}_{\Delta_1,\Delta_2,\Delta_3}\left(P_1,P_2,P_3;Z\right) =\int_{\text{AdS}}dX\, K_{\Delta_1,0}(X,P_1) \left(\partial_U \cdot {\cal P} \cdot \partial_U\right)^c \left(\partial_U \cdot \partial_X\right)^{2a} K_{\Delta_2,0}\left(X,P_2\right) \\
  \times \left(U \cdot {\cal P} \cdot U\right)^{p_1} \left(U \cdot \nabla \right)^{p_3}K_{\Delta_3,s-2p_1-p_3}\left(X,U;P_3,Z\right).
\end{multline}
In \S \tcb{\ref{sec::gravbub}} we show how to bring \eqref{factgravbub} into the form \eqref{gnericterm}. This gives the spectral representation:
\begin{multline}
{\cal M}^{\text{2pt-bubble}}\left(y_1,y_2\right) = g^2 \int^\infty_{-\infty} \frac{\nu^2 {\bar \nu}^2 d\nu d{\bar \nu}}{\pi^2[\nu^2+(\frac{d}{2})^2][{\bar \nu}^2+( \Delta-\frac{d}{2})^2]}   \\ \times 
{\sf B}\left(0,2,0;0;\Delta_1,\tfrac{d}{2}+i\nu-2,\tfrac{d}{2}+i{\bar \nu}\right){\sf B}\left(0,2,0;0;\Delta_2,\tfrac{d}{2}-i\nu-2,\tfrac{d}{2}-i{\bar \nu}\right) \mathfrak{K}^{\left({\sf 0},{\sf 0}\right)}_{0;2,0}(\nu,{\bar \nu};y_1,y_2)
\\ + \int^\infty_{-\infty} d\nu d{\bar \nu}\, {\cal G}^{\text{2pt-bubble}}_{\text{contact}}\left(\nu,{\bar \nu }\right)\mathfrak{K}^{({\bf 0},{\bf 0})}_{0;0,0}\left(\nu,{\bar \nu};y_1,y_2\right).
\end{multline}
The first line is the traceless and transverse contribution, which coincides with the previous result \eqref{ttanomspec} for $s=0$, $s^\prime=2$ and $\Delta_1=\Delta_2=d-2$. The second line is the contribution from the contact terms in the propagator \eqref{gravdedon}, which involve traces and gradients. The function ${\cal G}^{\text{2pt-bubble}}_{\text{contact}}\left(\nu,{\bar \nu }\right)$ is rather involved, and is given in \S \tcb{\ref{sec::gravbub}} together with its derivation.

The corresponding form for the contribution to the anomalous dimension is given by:
\begin{equation}
    \gamma=\gamma_{TT}+\gamma_{\text{contact}},
\end{equation}
where the tracless and transverse contribution $\gamma_{TT}$ is given by \eqref{gensosprimett} with $s=0$ and $s^\prime=2$, while:
\begin{multline}
\gamma_{\text{contact}}= -g^2\delta_{\Delta_1 \Delta_2}\frac{\pi ^{d+\frac{1}{2}} 2^{-d+4} \Gamma \left(\Delta_1 -\frac{d}{2}\right) \Gamma (d-2) }{\Gamma \left(\frac{d}{2}\right) \Gamma \left(\frac{d-1}{2}\right) \Gamma (d-\Delta_1 ) \Gamma \left(\frac{d}{2}-1\right)} \frac{1}{\sqrt{C_{\Delta_1,0}C_{\Delta_2,0}}} \\
\times \int^\infty_{-\infty} d\nu d{\bar \nu}\,  \frac{\Gamma \left(\frac{d-\Delta_1 +i (\nu-{\bar \nu})}{2}\right) \Gamma \left(\frac{d-\Delta_1 -i( \nu-{\bar \nu})}{2} \right)}{ \Gamma \left(\frac{\Delta_1 -i (\nu-{\bar \nu})}{2}\right) \Gamma \left(\frac{\Delta_1 +i (\nu-{\bar \nu})}{2}\right)}{\cal G}^{\text{2pt-bubble}}_{\text{contact}}\left(\nu,{\bar \nu }\right).
\end{multline}

\subsection{Type A higher-spin gauge theory}
\label{subcsec::criton}

The spectrum of the minimal type A higher-spin gauge theory on AdS$_{d+1}$ consists of an infinite tower of gauge fields $\varphi_s$ of spins $s=2, 4, 6, ...$ and a parity even scalar $\phi$ of fixed mass $m^2_0=-2\left(d-2\right)/R^2$. The results of \S \tcb{\ref{sec::sbd}} can be employed to compute the $s-\left(s^\prime 0\right)-s$ bubble diagrams in the theory, focusing on the contribution from the traceless and transverse part of the bulk-to-bulk propagators. 

The traceless and transverse cubic couplings of the interacting theory are given in ambient space by \cite{Sleight:2016dba,Sleight:2016xqq}:\footnote{See \cite{Boulanger:2008tg,Bekaert:2010hk,Joung:2011ww,Joung:2013doa,Joung:2013nma} for previous studies and classifications of metric-like cubic vertices of totally symmetric higher-spin gauge fields in AdS, as relevant for this work.} 
\begin{equation}
{\cal V}_{s_1,s_2,s_3} = g_{s_1,s_2,s_3} {\cal I}^{0,0,0}_{s_1,s_2,s_3},
\end{equation}
where ${\cal I}^{0,0,0}_{s_1,s_2,s_3}$ was defined in equation \eqref{Ical} and the coupling constants are:
\begin{equation}\label{typeacc}
   g_{s_1,s_2,s_3}=\frac{1}{\sqrt{N}}\frac{\pi ^{\frac{d-3}{4}}2^{\tfrac{3 d-1+s_1+s_2+s_3}{2}}}{ \Gamma (d+s_1+s_2+s_3-3)}  \prod_{i=1}^3\sqrt{\frac{\Gamma(s_i+\tfrac{d-1}{2})}{\Gamma\left(s_i+1\right)}}
\end{equation}
for canonically normalised kinetic terms. 

In generic space-time dimensions, the spectral form of the contribution from the traceless and transverse part of the propagators to the anomalous dimension is simply given by \eqref{gensosprimett} with couplings $g=g_{s,0,s^\prime}$:
\begin{multline}\label{typabub}
   \gamma_{TT} = -\, g^2_{s,0,s^\prime}
\frac{\pi ^{-\frac{7+d}{2}} s! 2^{-d+s^\prime+s-2} \Gamma (d+s^\prime-2)}{(d+2 s-4) \Gamma \left(\frac{d-1}{2}\right) \Gamma \left(\frac{d}{2}+s^\prime-1\right) \Gamma \left(\frac{d}{2}+s\right) \Gamma (d+2 s-2)}\\\times \int^\infty_{-\infty} d\nu d{\bar \nu}\, {\cal F}^{\text{2pt bubble}}_{TT}\left(\nu,{\bar \nu}\right), 
\end{multline}
and
{\small \begin{align}\label{specfunspinbubtypa}
&{\cal F}^{\text{2pt bubble}}_{TT}\left(\nu,{\bar \nu}\right) =
\frac{\nu\bar{\nu}\,\sinh (\pi  \nu) \sinh (\pi  \bar{\nu})}{[\nu^2+\left(\Delta_s-\tfrac{d}{2}\right)^2][\bar{\nu}^2+\left(\Delta-\tfrac{d}{2}\right)^2]}\frac{\Gamma \left(\frac{d}{2}-i \nu-1\right) \Gamma \left(\frac{d}{2}+i \nu-1\right) }{\Gamma \left(\frac{d}{2}+s^\prime-i \nu-1\right) \Gamma \left(\frac{d}{2}+s^\prime+i \nu-1\right)}\nonumber \\  & \hspace*{2cm} \times  \Gamma \left(\tfrac{d+s^\prime+2 s-2+i (\nu- \bar{\nu})}{2}\right) \Gamma \left(\tfrac{d+s^\prime+2 s-2-i (\nu- \bar{\nu})}{2}\right) \Gamma \left(\tfrac{d+s^\prime+2 s-2-i (\nu+\bar{\nu})}{2}\right) \Gamma \left(\tfrac{d+s^\prime+2 s-2+i (\nu+\bar{\nu})}{2}\right)\nonumber\\& \hspace*{2cm}  \times \Gamma \left(\tfrac{s^\prime+2+i(\nu-\bar{\nu})}{2}\right) \Gamma \left(\tfrac{s^\prime+2-i (\nu- \bar{\nu})}{2}\right) \Gamma \left(\tfrac{s^\prime+2+i (\nu+\bar{\nu})}{2}\right)\Gamma \left(\tfrac{s^\prime+2-i (\nu+\bar{\nu})}{2}\right)\,,
\end{align}}
whose properties were discussed in \S \tcb{\ref{subsec::ssprime0}}.

Let us note that this result holds for the standard boundary condition on the scalar field near $z=0$:\footnote{Here we work in Poincar\'e co-ordinates $x^\mu=\left(z,y^i\right)$
\begin{align}
   ds^2 = \frac{R^2}{z^2}\left(dz^2+dy_idy^i\right),
\end{align}
where $z$ here should not be confused with the boundary auxiliary vector $z^i$. The boundary of AdS is located at $z=0$, with boundary directions $y^i$, $i = 1, ..., d$.
}
\begin{equation}\label{scalarbc}
    \phi\left(z,y\right) \:\sim \: z^{\Delta_+},
\end{equation}
where $\Delta_+$ is the largest root of the equation:\footnote{Which has solutions: \begin{equation}
   \Delta = \Delta_{\pm} = \frac{d}{2} \pm \sqrt{\frac{d^2}{4}+m^2R^2}.
\end{equation}}
\begin{equation}
    \Delta\left(\Delta-d\right)=m^2_0 R^2.\label{scdimeq}
\end{equation}
By definition, $\Delta_+ \geq \frac{d}{2}$. For $m^2_0 R^2 > -\frac{d^2}{4}+1$, \eqref{scalarbc} is the unique admissible boundary condition invariant under the symmetries of AdS space \cite{1982AnPhy.144..249B}. That the result \eqref{typabub} holds for this particular boundary condition can be seen by noting that the spectral representation \eqref{scalarbubu} only holds for square integrable functions, which requires $\Delta > \frac{d}{2}$.

On the other hand, if the scalar mass lies within the window
\begin{equation}\label{window}
    -\frac{d^2}{4} < m^2_0R^2 < -\frac{d^2}{4}+1,
\end{equation}
there is a second admissible boundary condition \cite{1982AnPhy.144..249B}: 
\begin{equation}\label{scalarbc2}
    \phi\left(z,y\right) \:\sim \: z^{\Delta_-},
\end{equation}
 where $\Delta_-$ is the smallest root of equation \eqref{scdimeq}. This choice of scalar boundary condition is possible for the type A higher-spin gauge theory on AdS$_4$, where the scalar mass $m^2_0R^2=-2\left(d-2\right)=-2$ falls within the range \eqref{window}. 
 While the result \eqref{typabub} holds in the type A theory for the boundary behaviour \eqref{scalarbc} with $\Delta_+=2$, in the following section we show how the bubble diagram can be evaluated for the alternative boundary condition \eqref{scalarbc2} with $\Delta_-=1$.

\subsubsection{Alternative quantization on AdS$_4$}
\label{subsubsec::altquantads4}

In this section we show how to evaluate the bubble diagrams with the alternative boundary condition \eqref{scalarbc2} on the bulk scalar. See e.g. \cite{Gubser:2002zh,Hartman:2006dy,Giombi:2011ya} for previous works on Witten diagrams for the alternative boundary conditions.

The bulk-to-bulk propagator of a spin-$J$ field of mass $m^2 R^2=\Delta\left(\Delta-d\right)-J$ with the alternative boundary condition is given by:\footnote{To obtain this expression one uses that harmonic functions can be expressed as a linear combination of the propagators with two different boundary conditions \cite{Leonhardt:2003qu}:
    \begin{equation}
    \Omega_{\frac{i}{2}\left(\Delta_- - \Delta_+\right),J}\left(x_1,x_2\right)=\frac{\left(\Delta_+-\Delta_-\right)}{4\pi}\left[G_{\Delta_+,J}\left(x_1,x_2\right)-G_{\Delta_-,J}\left(x_1,x_2\right)\right].
\end{equation}
}
\begin{align}\nonumber
   G_{\Delta_-,J}\left(x_1,x_2\right) & =G_{\Delta_+,J}\left(x_1,x_2\right)-\frac{4\pi}{\left(\Delta_+-\Delta_-\right)}\Omega_{\frac{i}{2}\left(\Delta_- - \Delta_+\right),J}\left(x_1,x_2\right)\\ \label{altbubu}
   & =G_{\Delta_+,J}\left(x_1,x_2\right)+\left(\Delta_+-\Delta_-\right)\int_{\partial \text{AdS}} d^dy\, K_{\Delta_+,J}\left(x_1;y\right) \cdot K_{\Delta_-,J}\left(y;x_2\right),
\end{align}
where in the second equality we inserted the factorised form \eqref{spinharmfact} of the harmonic function. From this expression for $J=0$, we see that the $s-\left(s^\prime 0\right)-s$ bubble diagrams with the alternative boundary condition on the scalar running in the loop can be obtained from those with the standard boundary condition \eqref{scalarbc}, supplemented by the additional diagrams generated by the rightmost term in the modified propagator \eqref{altbubu} -- to account for the difference in boundary condition. This is illustrated in figures \ref{minuses}, and we show how to evaluate the additional diagrams in the following. 

\begin{subfigures}\label{minuses}
\begin{figure}[h]
\captionsetup{width=0.8\textwidth}
\centering
\includegraphics[scale=0.45]{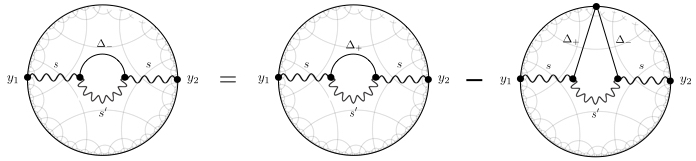}
\caption{Diagrammatic relation between bubble diagrams with different conformal boundary conditions on the scalar propagating inside the loop. For $s^\prime > 0$, they differ by a single-cut of the scalar internal line.}
\label{minuss}
\end{figure}
\begin{figure}[h]
\captionsetup{width=0.8\textwidth}
\centering
\includegraphics[scale=0.45]{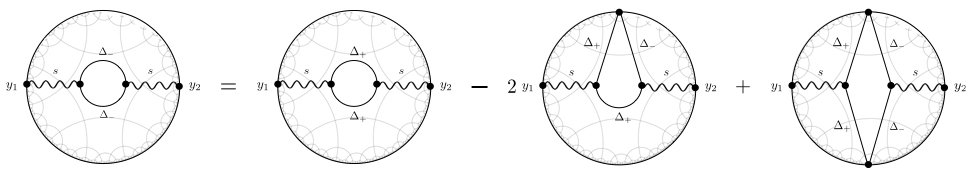}
\caption{For bubble diagrams with two scalars propagating in the loop, diagrams with different conformal boundary conditions on the scalar fields differ by both a single and double cut of the internal lines.}
\label{minusminus}
\end{figure}
\end{subfigures}

\subsection*{Single cut}

Let us first evaluate the additional diagram in figure \ref{minuss}, which for $s^\prime=0$ is equal to the left-most additional diagram in figure \ref{minusminus}.
This corresponds to ``cutting'' the scalar bulk-to-bulk propagator in the $s-\left(s^\prime 0\right)-s$ bubble diagram \eqref{typabub} -- i.e. going on-shell with respect to the internal scalar leg. Given the result \eqref{typabub}, the spectral form for the contribution to anomalous dimension from this diagram is easy to write down by fixing $\frac{d}{2}+i {\bar \nu} = \Delta_+ $: 

 \begin{multline}\label{typabubminus}
   \gamma^{\Delta_+ \Delta_-}_{s,s^\prime} = -\, g^2_{s,0,s^\prime}
\frac{\pi ^{-\frac{7+d}{2}} s!\, 2^{-d+s^\prime+s-2} \Gamma (d+s^\prime-2)}{(d+2 s-4) \Gamma \left(\frac{d-1}{2}\right) \Gamma \left(\frac{d}{2}+s^\prime-1\right) \Gamma \left(\frac{d}{2}+s\right) \Gamma (d+2 s-2)} \\ \times \int^\infty_{-\infty} d\nu\, {\cal F}^{\text{2pt bubble}}_{\Delta_+ \Delta_-}\left(\nu \right), 
\end{multline}
where
\begin{align}
{\cal F}^{\text{2pt bubble}}_{\Delta_+ \Delta_-}\left(\nu \right) & = 
\frac{2\pi}{i{\bar \nu}} [{\bar \nu}^2+( \Delta-\tfrac{d}{2})^2] \times 
{\cal F}^{\text{2pt bubble}}_{TT}\left(\nu,{\bar \nu}\right)\Big|_{{\bar \nu}=-i\left(\Delta_+-\frac{d}{2}\right)}.
\end{align}
The notation $\gamma^{\Delta_+ \Delta_-}_{s,s^\prime}$ is defined as
\begin{equation}
    \gamma^{\Delta_+ \Delta_-}_{s,s^\prime} = \gamma^{\Delta_+}_{s,s^\prime} - \gamma^{\Delta_-}_{s,s^\prime},
\end{equation}
where $\gamma^{\Delta_+}_{s,s^\prime}$ is the contribution to the anomalous dimension generated by the $s-\left(s^\prime 0\right)-s$ bubble diagram with the $\Delta_+$ boundary condition on the scalar (which was considered in the previous section), and $\gamma^{\Delta_-}_{s,s^\prime}$ is the same but with the $\Delta_-$ boundary condition.

In the present case of AdS$_4$ with $\Delta_+=2$, we have in particular 
\begin{multline}\label{diffsprimett}
{\cal F}^{s^\prime}_{TT}\left(\nu\right)= - \pi^4\,  2^{1-4 (s^\prime+s)}
 \left[\nu^2+( s^\prime+\tfrac{1}{2})^2\right] \frac{\nu\tanh (\pi  \nu) \text{sech}(\pi  \nu) }{\left[\nu^2+\left(\Delta_{s^\prime}-\frac{d}{2}\right)^2\right]} \\ \times \frac{\Gamma \left(s^\prime+2 s-i \nu+\frac{1}{2}\right) \Gamma \left(s^\prime+2 s+i \nu+\frac{1}{2}\right)}{\Gamma \left(\frac{1}{2}-i \nu\right) \Gamma \left(i \nu+\frac{1}{2}\right)}. 
\end{multline}
The $\nu$ integral in this case can be evaluated by expanding \eqref{diffsprimett} as a series in $\nu^2$:
\begin{equation}\label{diffsprimettseries}
{\cal F}^{s^\prime}_{TT}\left(\nu\right)= - \pi^4\, 2^{1-4(s+s^\prime)} \left(\sum_{n} c_{s,s^\prime}^{(n)}\nu^{2n+1}\right)\tanh (\pi  \nu) \text{sech}(\pi  \nu),
\end{equation}
which truncates to a polynomial in $\nu^2$ since the denominator of the first line cancels with one of the factors within the $\Gamma$-functions in the numerator of the second line. The coefficients are defined as:
\begin{equation}
c^{(n)}_{s,s^\prime} = \text{coeff.}\left[\frac{\nu^2+( s^\prime+\tfrac{1}{2})^2}{\left[\nu^2+\left(\Delta_{s^\prime}-\frac{d}{2}\right)^2\right]} \frac{\Gamma \left(i \nu+\frac{1}{2}+2 s+s^\prime\right)\Gamma \left(-i \nu+\frac{1}{2}+2 s+s^\prime\right)}{\Gamma \left(i \nu+\frac{1}{2}\right)\Gamma \left(-i \nu+\frac{1}{2}\right)},\nu^{2n}\right].
\end{equation}
Using the identity:
\begin{equation}
    \int_{-\infty}^{\infty}d\nu \,\nu^{2n+1}\tanh (\pi  \nu) \text{sech}(\pi  \nu)=\frac1{\pi}\left(-\frac{1}{4}\right)^{n}\,(2n+1) E_{2n}\,,
\end{equation}
where $E_{n}$ are the Euler numbers the integral can be analytically evaluated for any spins.\footnote{Notice that the single cut gives a convergent integral in $\nu$. This confirms the expectation that the UV divergences for $\Delta_+$ and $\Delta_-$ boundary conditions precisely cancel. The anomalous dimension then only receives finite IR contributions coming from the boundary conformal integrals.}

\begin{figure}[h]
\captionsetup{width=0.8\textwidth}
\centering
\includegraphics[scale=0.3]{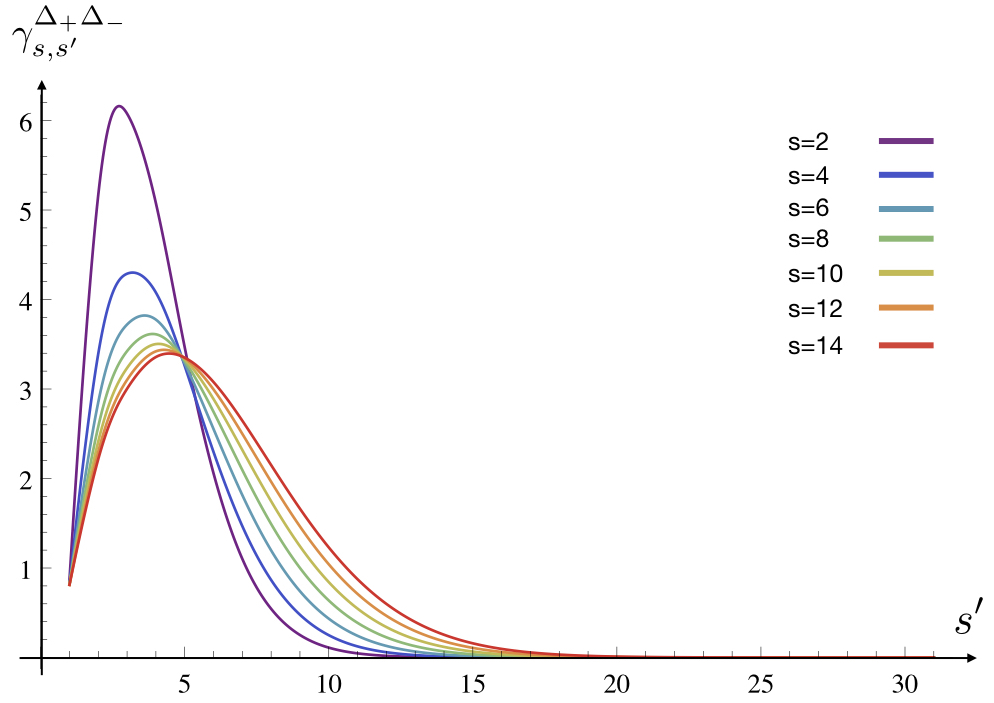}
\caption{Plot of the contributions to the anomalous dimension from a single cut of the $s-\left(s^\prime 0\right)-s$ bubble diagram on the internal scalar leg. On the horizontal axis we vary the internal spin $s^\prime$, while the colour gradient represents varying external spin $s$. The contributions are exponentially suppressed for large $s^\prime$}
\label{fig::plotsl}
\end{figure}

The final form for the contribution \eqref{typabubminus} to the anomalous dimension from the single cut of a $s-\left(s^\prime 0\right)-s$ bubble is thus:
\begin{multline}\label{typabubminus2}
\gamma^{\Delta_+ \Delta_-}_{s,s^\prime} = g^2_{s,0,s^\prime}
\frac{\pi ^{-\frac{1}{2}-\frac{d}{2}} s!\, 2^{-d-1-3 (s^\prime+ s)} \Gamma (d+s^\prime-2)}{(d+2 s-4) \Gamma \left(\frac{d-1}{2}\right) \Gamma \left(\frac{d}{2}+s^\prime-1\right) \Gamma \left(\frac{d}{2}+s\right) \Gamma (d+2 s-2)} \\ 
\times  \sum_{n}c^{(n)}_{s,s^\prime}\left(-\frac{1}{4}\right)^n\left(2n+1\right)E_{2n}.
\end{multline}
where for generality we have kept $d$ arbitrary in the overall prefactor. For the $s^\prime=0$ contribution we can evaluate the sum over $n$ exactly:
\begin{equation}\label{typabubminus20}
  \gamma^{\Delta_+ \Delta_-}_{s,0}=  
\frac{32 s^2}{N \pi ^2 (2 s-1) (2 s+1)}.
\end{equation}
We give a plot of the $s^\prime>0$ contributions in figure \ref{fig::plotsl}. It is interesting to notice that contributions from higher $s^\prime$ are exponentially suppressed in $s^\prime-s$, so that dropping terms with $s^\prime>2s$ gives only a small error when evaluating the sum over spins. One may verify for large $s^\prime$ that contributions for $s^\prime>>s$ are of order {\small $10^{-\tfrac{s^\prime}2+s}$}. This allows to obtain approximated analytic results with arbitrarily small errors.

\subsection*{Double cut}

For the bubble diagram $s-\left(00\right)-s$, with only scalars propagating in the loop, for the $\Delta_-$ boundary condition there is a further additional diagram given by the ``double cut'' of the scalar bulk-to-bulk propagators, which is the rightmost diagram shown in figure \ref{minusminus}. It is given by:
\begin{multline}\nonumber
 {\cal M}^{\Delta_+,\Delta_-}_{\Delta_+,\Delta_-}\left(y_1,y_2\right) = \frac{1}{2}g^2_{s,0,0} \left(\Delta_+-\Delta_-\right)^2 \\ \times \int_{\partial \text{AdS}}d^dyd^d{\bar y}\, {\cal M}^{0,0,0}_{s,0,0;d-2,\Delta_+,\Delta_+}\left(y_1,y,{\bar y}\right) \cdot  {\cal M}^{0,0,0}_{s,0,0;d-2,\Delta_-,\Delta_-}\left(y_2,y,{\bar y}\right). 
\end{multline}
The corresponding contribution $(\gamma_{s,0})^{\Delta_+\Delta_-}_{\Delta_+\Delta_-}$ to the anomalous dimension is very easy to extract, and can be done by simply setting $\frac{d}{2}+i\nu=\Delta_+$ and $s^\prime=0$ in the spectral representation \eqref{typabubminus} of the contribution for the anomalous dimension from the single cut diagram. The result reads: 
\begin{align}\label{dcutanom}
(\gamma_{s,0})^{\Delta_+\Delta_-}_{\Delta_+\Delta_-}&=\left[\left(\Delta_+-\Delta_-\right)^2C_{\Delta_+,0}C_{\Delta_-,0}\right]^2\\
&\hspace{50pt} \times \frac{2^6\, \pi ^{2 d}\,(d-4) \Gamma \left(2-\frac{d}{2}\right) \Gamma (s+1)}{N (d+2 s-4) (d+2 s-2) \Gamma \left(\frac{d}{2}-1\right) \Gamma (d+s-3)}\nonumber\\
&=-\frac{2^{2 d-2} (d-4) \Gamma \left(\frac{d-1}{2}\right)^2 s! \csc \left(\tfrac{\pi d}{2}\right)\sin ^2\left(\frac{\pi  d}{2}\right)}{\pi ^2 (d+2 s-4) (d+2 s-2) \Gamma (d+s-3)}\,.\nonumber
\end{align}
One can check that this agrees on the CFT side with the contribution to the anomalous dimension of the ``two-triangle" diagram (also known as ``Aslamazov-Larkin" diagram), see e.g. \cite{Hikida:2016wqj, Hikida:2016cla}, in agreement with the general arguments in \cite{Hartman:2006dy, Giombi:2011ya}.

Combining with the contribution \eqref{typabubminus20} from the single-cut diagram, the total additional contribution from $s-\left(00\right)-s$ one-loop diagrams for the $\Delta_+$ boundary condition with respect to the  $\Delta_-$ boundary condition is given by:
\begin{align}\label{gamma0cuts}
  \gamma_{s,0}  & \equiv \gamma^{\Delta_+\Delta_-}_{s,0}-(\gamma_{s,0})^{\Delta_+\Delta_-}_{\Delta_+\Delta_-}\\\nonumber &=\left(\frac{32 s^2}{\pi ^2 (2 s-1) (2 s+1)N}+\frac{16 s}{\pi ^2 \left(2 s+1\right)\left(2 s-1\right)N}\right)=\frac{16 s}{N \pi ^2 (2 s-1)}\,.
\end{align}

\subsubsection*{Total contribution}
\begin{figure}[h]
\captionsetup{width=0.8\textwidth}
\centering
\includegraphics[scale=0.35]{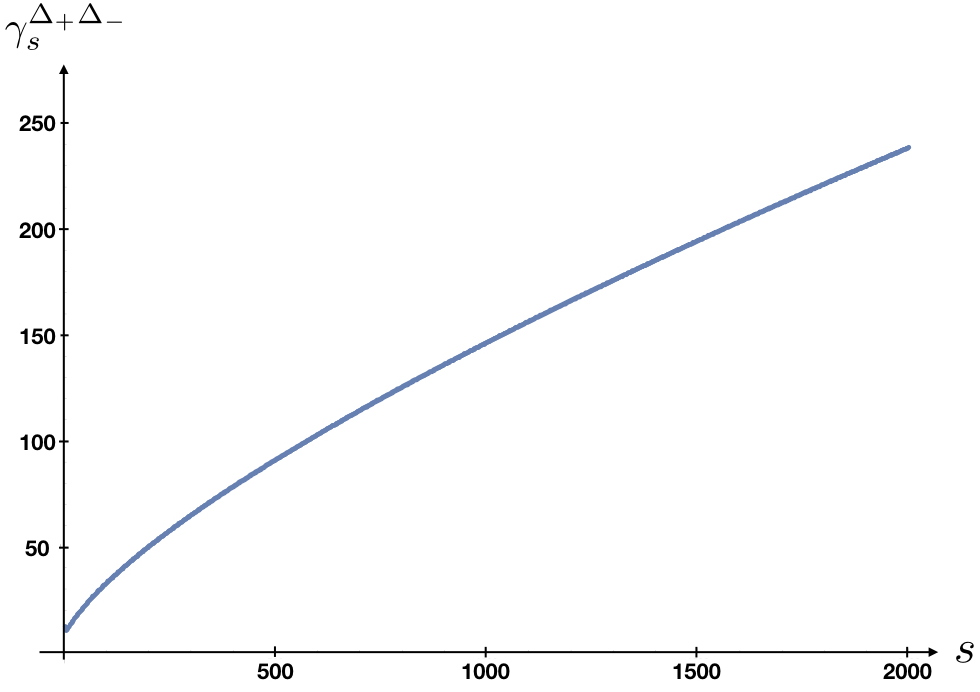}
\caption{Plot of the re-summation of the contributions to the anomalous dimension from the difference of $s-\left(s^\prime 0\right)-s$ bubble diagrams for the $\Delta_-$ and $\Delta_+$ boundary condition on the scalar field. The internal spin $s^\prime$ is summed over while the external spin $s$, which is displayed on the horizontal axis, is fixed.}
\label{fig::Resum}
\end{figure}

To obtain the total contribution from the additional diagrams for $s-\left(s^\prime 0\right)-s$ bubbles in the alternative quantisation of the type A higher-spin gauge theory, we need to sum over the exchanged spin $s^\prime$ in the spectrum. In particular, this is given by:
 \begin{equation}\label{totdiffbccont}
     \gamma^{\Delta_+\,\Delta_-}_s = \sum_{s^\prime \in 2 \mathbb{N}} \gamma^{\Delta_+\,\Delta_-}_{s,s^\prime}.
 \end{equation}
As anticipated, evaluating this sum analytically is rather complicated due to the involved form of expansion coefficients $c_{s,s^\prime}^{(n)}$. However, it is possible to obtain an analytic estimate of the result by truncating the summation over spin. This is possible owing to the exponential damping of the contributions for higher and higher exchanged spins, illustrated in figure \ref{fig::plotsl}. We plot the result in figure \ref{fig::Resum} for fixed external spin $s$, up to $s=2000$.

\subsubsection{Comparison with dual CFT}
\label{subsubsec::compwdcft}

\begin{figure}[h]
\captionsetup{width=0.8\textwidth}
\centering
\includegraphics[scale=0.4]{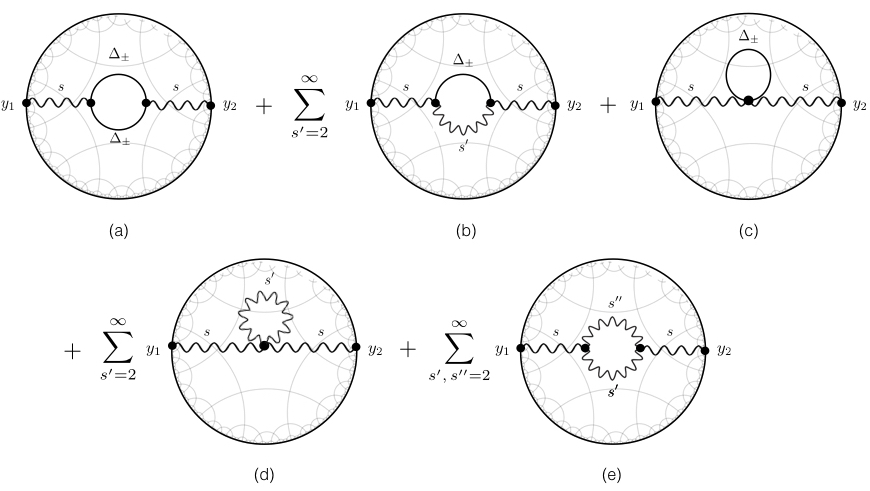}
\caption{Diagrams contributing to the one-loop two-point amplitude ${\cal M}^{\Delta_\pm,\text{total 1-loop}}_{s}\left(y_1,y_2\right)$ with external spin-$s$ gauge fields in the type A higher-spin gauge theory on AdS$_4$, for both the $\Delta_+$ and $\Delta_-$ boundary conditions on the bulk scalar. Diagrams (a) and (b) were considered in \S \tcb{\ref{subcsec::criton}} of this work.}
\label{tot1loop}
\end{figure}

In addition to the $s-\left(s^\prime 0\right)-s$ bubble diagrams considered so far in this section, there are other types of processes that contribute at one-loop to the total two-point amplitude in the type A minimal higher-spin gauge theory. For external spin-$s$ fields, all diagrams that contribute are shown in figure \ref{tot1loop}, for both boundary conditions on the bulk scalar field. Notice that we have not included $\bep(14,10)\put(0,0){\line(1,0){14}}\put(6,0){\line(0,1){4}}\put(6,7){\circle{6}}\eep$-type tadpole diagrams, since it was argued in \S \tcb{\ref{subsec::oneptspintad}} that, at least taken individually, such diagrams do not contribute.\footnote{It should however be noted that, in order to consider diagrams individually (i.e. for fixed spins propagating internally before summing over the spectrum), it needs to be investigated whether the infinite sum over spin commutes with the integration over AdS. This is a subtle issue, in particular since the sum over spin in higher-spin gauge theories has a finite radius of convergence \cite{Sleight:2017pcz} and the integration over boundary \eqref{m1loop} is divergent. We discuss this point further in \S \tcb{\ref{subsubsec::discussion}}.}

In the context of AdS/CFT, the diagrams displayed in figure \ref{tot1loop} give the holographic computation of the $1/N$ correction to the two-point CFT correlation function of the single-trace operator dual to a spin-$s$ gauge field on AdS. On AdS$_4$, the type A minimal higher-spin theory with $\Delta_-=1$ boundary condition \eqref{scalarbc2} is conjectured to be dual to the free scalar $O\left(N\right)$ model in three dimensions, restricted to the $O\left(N\right)$ singlet sector \cite{Sezgin:2002rt}. The spectrum of primary operators consists of a tower of even spin conserved currents
\begin{equation}
    \partial \cdot {\cal J}_s\;\approx\;0,
\end{equation}
 dual to a spin-$s$ gauge field $\varphi_s$ in the bulk, and a scalar ${\cal O}$ of scaling dimension $\Delta_-$ which is dual to the bulk parity even scalar $\phi$. Owing to the absence of $1/N$ corrections in free theory, the total of the diagrams in figure \ref{tot1loop} for the $\Delta_-$ boundary condition is then expected to vanish. 

Adding a double-trace deformation $\lambda {\cal O}^2$ to the free theory above induces a flow an IR fixed point where ${\cal O}$ has instead dimension $\Delta_+=2$, known as the critical $O\left(N\right)$ model. In the holographic picture, the double-trace deformation modifies the boundary condition on the dual bulk scalar field \cite{Witten:2001ua,Berkooz:2002ug}, requiring instead to impose the $\Delta_+$ boundary condition \eqref{scalarbc}. This bulk interpretation of multi-trace deformations inspired the conjectured duality between the type A minimal higher-spin gauge theory with $\Delta_+=2$ boundary condition and the critical $O\left(N\right)$ model in three dimensions \cite{Klebanov:2002ja}. At this interacting fixed point, the operators ${\cal J}_s$ are no-longer conserved and acquire an anomalous dimension:
\begin{equation}
    \Delta_s=s+d-2+\gamma_{s}.
\end{equation}
At the operator level, this statement reads as the non-conservation equation of the schematic form
\begin{equation}
     \partial \cdot {\cal J}_s = \frac{1}{\sqrt{N}} \sum {\cal J}{\cal J},
\end{equation}
which implies that the anomalous dimensions are $\gamma_s \sim {\cal O}\left(1/N\right)$. At leading order in $1/N$, they are given by \cite{Ruhl:2004cf,Lang:1991kp}
\begin{equation}\label{cftgammas}
    \gamma_s = \frac{16\left(s-2\right)}{3\pi^2 N\left(2s-1\right)},
\end{equation}
and to date have been determined using various approaches in CFT \cite{Skvortsov:2015pea,Giombi:2016hkj,Hikida:2016wqj,Gopakumar:2016wkt}. 

To date the anomalous dimensions \eqref{cftgammas} have not yet been extracted via a direct one-loop calculation in AdS. From the large $N$ expansion of the two-point function
\begin{equation}\label{2ptcrit}
\langle {\cal J}_s\left(y_1\right){\cal J}_s\left(y_2\right) \rangle = C_{{\cal J}_s} \frac{{\sf H}^s_{21}}{\left(y^2_{12}\right)^{d-2}}\left(1-\gamma_s \log \left(y^2_{12}\right) + ...\right), 
\end{equation}
where $C_{{\cal J}_s}$ is the ${\cal O}\left(1\right)$ normalisation and the $...$ contain ${\cal O}\left(1/N^2\right)$ terms and corrections to the normalisation, we see that the anomalous dimensions of the higher-spin operators may be computed holographically at ${\cal O}\left(1/N\right)$ by extracting the log contribution from the bulk two-point amplitude at one-loop for the $\Delta_+$ boundary condition, shown in figure \ref{tot1loop}. While in this work we have not evaluated all diagrams in the total one-loop amplitude (in particular, we have not evaluated diagrams (c)-(e)), with the results of \S \tcb{\ref{sec::sbd}} we can still however study how the different one-loop processes in figure \ref{tot1loop} contribute to the anomalous dimensions \eqref{cftgammas}:

\begin{figure}[h]
\captionsetup{width=0.8\textwidth}
\centering
\includegraphics[scale=0.4]{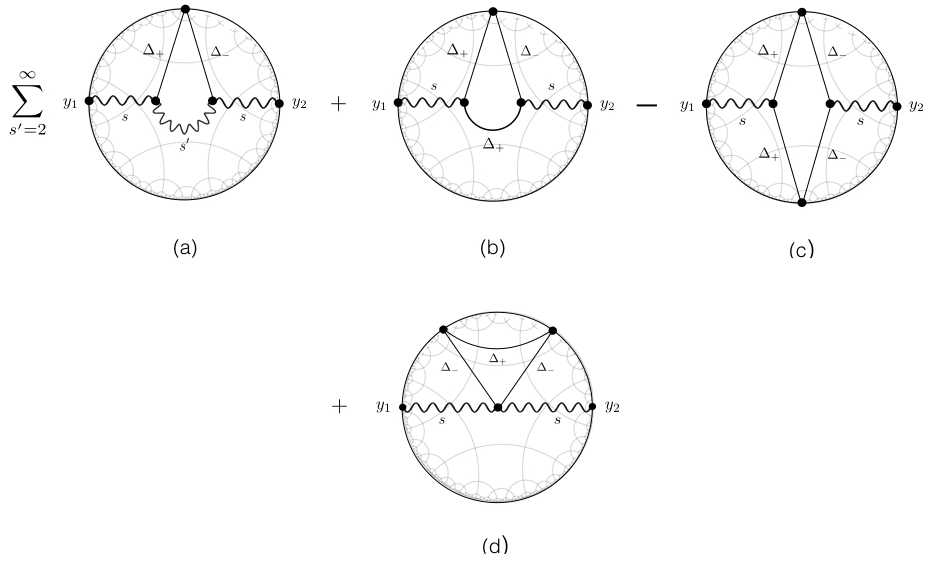}
\caption{Diagrams which contribute to the difference ${\cal M}^{\Delta_+,\text{total 1-loop}}_{s}\left(y_1,y_2\right)-{\cal M}^{\Delta_-,\text{total 1-loop}}_{s}\left(y_1,y_2\right)$ of two-point one-loop amplitudes for the $\Delta_+$ and $\Delta_-$ boundary conditions on the bulk scalar. Diagrams (a)-(c) on the first line were computed in \S \tcb{\ref{subsubsec::altquantads4}}.}
\label{diffbc}
\end{figure}

In order for the duality with the free scalar theory to hold, the two-point amplitude with $\Delta_-$ boundary condition should not generate anomalous dimensions. Under this assumption, the anomalous dimension \eqref{cftgammas} should be encoded in the diagrams that remain in the difference of the two-point amplitudes with $\Delta_+$ and $\Delta_-$ boundary conditions on the bulk scalar, which is shown in figure \ref{diffbc}. Since the change of boundary condition is just on the bulk scalar, only the diagrams involving a scalar in the loop, which are displayed on the first line of figure \ref{tot1loop} (diagrams (a), (b) and (c)), may generate non-trivial contributions in figure \ref{diffbc}. The diagrams on the first line of the latter were computed in \S \tcb{\ref{subsubsec::altquantads4}}, which arise from bubble diagrams (a) and (b) in figure \ref{tot1loop}. The total of which, given by the modulus of equation \eqref{totdiffbccont}, does not reproduce the anomalous dimension \eqref{cftgammas}. The discrepancy is quite large: The CFT result \eqref{cftgammas} asymptotes to a constant value for large $s$:
\begin{equation}
    \gamma_s \: \rightarrow \: \frac{8}{3\pi^2 N},
\end{equation}
while the total contribution \eqref{totdiffbccont} from the bubble diagrams seems to grow linearly with $s$ -- as shown in figure \ref{fig::Resum}. The remaining diagram (d) in figure \ref{diffbc}, which arises from the $\bep(14,10)\put(0,0){\line(1,0){14}}\put(7,4.25){\circle{8}}\eep$ tadpole diagram (c) in figure \ref{tot1loop} generated by the $s$-$s$-$0$-$0$ contact interactions, should thus give a significant non-trivial contribution of the equal but opposite magnitude as that from the total of diagrams (a), (b) and (c) in figure \ref{diffbc}.

\subsubsection{Discussion}
\label{subsubsec::discussion}

\subsubsection*{Sum over spin}

In computing the one-loop contributions to the type A higher-spin gauge theory two-point amplitude in the preceding section, we performed the sum over spin \emph{after} regularising the divergent two-point boundary conformal integrals \eqref{m1loop}. This is the standard prescription for computing Feynman diagrams in a field theory, where each diagram is evaluated separately and the amplitude is obtained from their total sum. However, since in higher-spin gauge theories an infinite number of diagrams must be summed for fixed external legs at each order in $1/N$ -- owing to the infinite spectrum of higher-spin gauge fields -- it is interesting to ask whether the infinite sum over spin and regularised integration over the boundary may be commuted.

\begin{figure}[h]
\captionsetup{width=0.8\textwidth}
\centering
\includegraphics[scale=0.4]{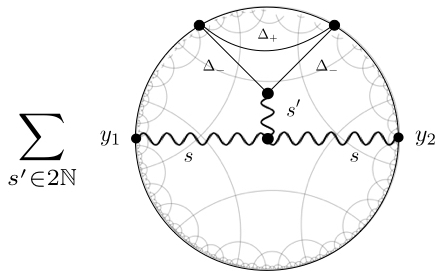}
\caption{Re-summation of tadpole diagrams with a single-cut of the scalar loop. The infinite sum over spin $s^\prime$ and the divergent integration over the boundary seem not to commute.}
\label{fig::taddiff}
\end{figure}

This point can be explored and is most illuminated by considering the contributions from $\bep(14,10)\put(0,0){\line(1,0){14}}\put(6,0){\line(0,1){4}}\put(6,7){\circle{6}}\eep$-type tadpole diagrams, which in \S \tcb{\ref{subsec::oneptspintad}} were argued to vanish individually. In performing the boundary integration \emph{before} summing over spin, such diagrams thus do not contribute to one-loop two-point amplitude. For simplicity, in the following let us restrict to the single-cut tadpole diagrams that would appear in the difference of the one-loop two-point amplitudes for the $\Delta_+$ and $\Delta_-$, shown in figure \ref{fig::taddiff}. These diagrams were not considered in \S \tcb{\ref{subsubsec::compwdcft}}, where they would appear in figure \ref{diffbc}, because there the sum over spin was being taken after performing the boundary integration and they thus did not contribute. To investigate instead summing over spin \emph{prior} to performing the boundary integration, it is useful to note that each individual such diagram in the sum over spin $s^\prime$ can be expressed as\footnote{\label{foo::bobu}The integration weighted by the $\Delta_+$ scalar bulk-to-boundary propagator in equation \eqref{sincutsprime} enforces the change of boundary condition on one of the external scalars from $\Delta_-$ to $\Delta_+$ \cite{Leonhardt:2003sn}, i.e. (see \S \tcb{\ref{subsec::shadprop}}): 
\begin{equation}\label{shadnonshadbubou}
    K_{\Delta_+,J}\left(x;y\right)=-\left(\Delta_+-\Delta_-\right) \int_{\partial \text{AdS}}d^d{\bar y}\, K_{\Delta_-,J}\left(x;{\bar y}\right) \cdot K_{\Delta_+,J}\left(y;{\bar y}\right).
\end{equation}} 
\begin{multline}\label{sincutsprime}
{\cal M}^{\Delta_+,\Delta_-}_{\text{tadpole},s^\prime}\left(y_1,y_2\right)\\ =  \frac{1}{2}\left(\Delta_+-\Delta_-\right)^2 \int_{\partial \text{AdS}}d^dy_3d^dy_4\,{\cal M}^{\text{tree-level exch.}}_{s,s|s^\prime|0^-,0^-}\left(y_1,y_2,y_3,y_4\right)K_{\Delta_+,0}\left(y_3,y_4\right).
\end{multline}
where ${\cal M}^{\text{tree-level exch.}}_{s,s|s^\prime|0^-,0^-}$ is the spin $s^\prime$ exchange diagram in the type A minimal theory with $\Delta_-$ boundary condition on both scalars, which was computed in \cite{Sleight:2017fpc}.\footnote{See also the preceding \cite{Bekaert:2014cea,Bekaert:2015tva} for the $s=0$ case, and also \cite{Francia:2007qt,Francia:2008hd}.} For the part of exchange diagrams corresponding to the genuine exchange of the single-particle (s.p.) state (i.e. as opposed to contact contributions associated to double-trace blocks) which is encoded in the traceless and transverse part of the bulk-to-bulk propagator \eqref{ttbubu}, the sum over exchanged spin is given by a higher-spin block \cite{Alday:2016njk,Sleight:2017pcz}:\footnote{Restricting to the single-particle contribution is the AdS analogue of restricting to single pole in Mandelstam variables in flat space exchange diagrams.} 
\begin{equation}
 {\cal H}_{\left(s,s|d-2|0^-,0^-\right)} = \sum_{s^\prime \in 2\mathbb{N}}   {\cal M}^{\text{tree-level exch.}}_{s,s|s^\prime|0^-,0^-}\Big|_{\text{s.p.}},
\end{equation}
which re-sums the contribution from the infinite tower of exchanged massless higher-spin particles. It is given explicitly by: 
\begin{multline}\label{hsblock}
    \mathcal{H}_{(s,s|d-2|0^-,0^-)}=\frac{c_{ss00}}{N\,(y_{12}^2)^{d-2}(y_{34}^2)^{d-2}}\left[\left(\frac{u}{v}\right)^{\frac{d-2}{2}}\ \left((2\,{\sf q}_{12})^{-\tfrac{d-4}4}\Gamma(\tfrac{d-2}2)J_{\tfrac{d-4}2}(\sqrt{2\,{\sf q}_{21}})\right){\sf Y}_{1,24}^s{\sf Y}_{2,31}^s\right]\\+\frac{c_{ss00}}{N\,(y_{12}^2)^{d-2}(y_{34}^2)^{d-2}}\left[u^{\frac{d-2}{2}}\left((2{\bar {\sf q}}_{12})^{-\tfrac{d-4}4}\Gamma(\tfrac{d-2}2)J_{\tfrac{d-4}2}(\sqrt{2\bar{{\sf q}}_{12}})\right)\mathsf{Y}_{1,23}^s\mathsf{Y}_{2,43}^s\right]\,.
\end{multline}
where
\begin{align}
{\sf q}_{12}&={\sf H}_{21}\pl_{{\sf Y}_{1,24}}\pl_{{\sf Y}_{2,31}}\,,&\bar{{\sf q}}_{12}&={\sf H}_{12}\pl_{{\sf Y}_{1,23}}\pl_{{\sf Y}_{2,41}}\,,
\end{align}
and with normalisation:
\begin{equation}
    c_{ss00}=\frac{\sqrt{\pi } 2^{-\Delta -s+4} \Gamma (s+1) \Gamma \left(s+\frac{\Delta }{2}\right) \Gamma (s+\Delta -1)}{N \Gamma \left(\frac{\Delta }{2}\right)^2 \Gamma \left(s+\frac{\Delta }{2}-\frac{1}{2}\right)}\,,
\end{equation}
corresponding to unit normalisation of the two point functions. The cross ratios in the $(12)$ channel are defined as:
\begin{align}
u=\frac{y_{12}^2y_{34}^2}{y_{13}^2y_{24}^2}\,, \qquad v=\frac{y_{14}^2y_{23}^2}{y_{13}^2y_{24}^2}\,.
\end{align}

The higher-spin block \eqref{hsblock} allows us to compute the contribution (dropping contact terms in exchange amplitudes) from the single-cut diagrams \eqref{sincutsprime} arising from $\bep(14,10)\put(0,0){\line(1,0){14}}\put(6,0){\line(0,1){4}}\put(6,7){\circle{6}}\eep$ tapoles by performing the sum over spin \emph{prior to} evaluating the boundary conformal integral. This is given by: 

{\small \begin{align}\nonumber
 {\cal M}^{\Delta_+,\Delta_-}_{\text{tadpole}}\left(y_1,y_2\right)&=\frac{1}{2}\left(\Delta_+-\Delta_-\right)^2 \int_{\partial \text{AdS}}d^dy_3d^dy_4\,\sum_{s^\prime \in 2 \mathbb{N}}{\cal M}^{\text{tree-level exch.}}_{s,s|s^\prime|0^-,0^-}\left(y_1,y_2,y_3,y_4\right)\Big|_{\text{s.p.}}K_{\Delta_+,0}\left(y_3,y_4\right)\\ \nonumber
 & = \frac{1}{2}\left(\Delta_+-\Delta_-\right)^2 \int_{\partial \text{AdS}}d^dy_3d^dy_4\, \mathcal{H}_{(s,s|d-2|0^-,0^-)}\left(y_1,y_2,y_3,y_4\right)K_{\Delta_+,0}\left(y_3,y_4\right) \\ 
 &= \frac12\left(\Delta_+-\Delta_-\right)^2 C_{\Delta_+,0}C_{\Delta_-,0}C_{s+d-2,s} \left[-\frac{2\pi ^d d(d-2)}{N\,\Gamma \left(\frac{d+2}{2}\right)^2}\right]\frac{\log(y_{12}^2)}{(y_{12}^2)^{d-2}}\,{\sf H}_{21}^s,\\
 &\equiv -\,C_{s+d-2,s} \gamma^{\Delta_+,\Delta_-}_{\text{tadpole}}\,\frac{\log(y_{12}^2)}{(y_{12}^2)^{d-2}}\,{\sf H}_{21}^s\label{tottadcut}
\end{align}}

\noindent where in the second-last equality we restricted to the log term that encodes the contribution to the anomalous dimension, as shown in the last equality, and which we note is non-vanishing. Upon recalling that:
\begin{equation}\label{norm}
    \left(\Delta_+-\Delta_-\right)^2 C_{\Delta_+,0}C_{\Delta_-,0}=\frac{1}{2} (d-4) \pi ^{-d-1} \sin \left(\frac{\pi  d}{2}\right) \Gamma (d-2)\,,
\end{equation}
for $d=3$, corresponding to AdS$_4$ in the bulk, this yields:
\begin{equation}
\gamma^{\Delta_+,\Delta_-}_{\text{tadpole}}=\frac{8}{3\pi^2 N},
\end{equation}
which is a non-zero and spin-independent contribution to the anomalous dimension. This is to be contrasted with the vanishing contribution obtained in \S \tcb{\ref{subsubsec::altquantads4}} instead by first performing the integration over the boundary, which seems to suggest that the sum over spin and boundary integration does not commute in higher-spin gauge theories.

While it may seem non-standard in field theory to first perform the sum over spin, which is more reminiscent of working directly with some analogue of string fields as opposed to expanding in spin, we note that it does the job of recovering the CFT anomalous dimension 
\eqref{cftgammas}: This is straightforward to see by noting that, by first summing over spin, the difference of one-loop two-point amplitudes for $\Delta_+$ and $\Delta_-$ boundary conditions considered in \S \tcb{\ref{subsubsec::compwdcft}} is given by:
\begin{multline}\label{diifbc2pt1}
{\cal M}^{\Delta_+,\text{total 1-loop}}_{s}\left(y_1,y_2\right)-{\cal M}^{\Delta_-,\text{total 1-loop}}_{s}\left(y_1,y_2\right)
 = -{\cal M}^{\Delta_+,\Delta_-}_{\Delta_+,\Delta_-}\left(y_1,y_2\right)\\+\frac{1}{2}\left(\Delta_+-\Delta_-\right)^2 \int_{\partial \text{AdS}}d^dy_3d^dy_4\,{\cal M}^{\text{tree-level 4pt}}_{s,s,0^-,0^-}\left(y_1,y_2,y_3,y_4\right)K_{\Delta_+,0}\left(y_3,y_4\right),
\end{multline}
where ${\cal M}^{\Delta_+,\Delta_-}_{\Delta_+,\Delta_-}$ is the double-cut diagram computed in \S \tcb{\ref{subsubsec::altquantads4}} and ${\cal M}^{\text{tree-level 4pt}}_{s,s,0^-,0^-}$ is the full connected tree-level four-point amplitude in the type A higher-spin gauge theory with two spin-$s$ external gauge fields and two external scalars with $\Delta_-$ boundary condition. Amplitudes in higher-spin gauge theories on AdS$_4$ are uniquely fixed by the global higher-spin symmetry \cite{Sleight:2017pcz}. In particular, in terms of {\sf s}-, {\sf t}- and {\sf u}-channel higher-spin blocks \eqref{hsblock} we have: 
\begin{align}\label{4pttreelev}
{\cal M}^{\text{tree-level 4pt}}_{s,s,0^-,0^-}\left(y_1,y_2,y_3,y_4\right)=&\frac{1}{2}\left[\mathcal{H}_{(s,s|d-2|0^-,0^-)}\left(y_1,y_3,y_2,y_4\right)\right.\\ \nonumber &+\left.\mathcal{H}_{(s,0^-|d-2|s,0^-)}\left(y_1,y_4,y_3,y_2\right)+\mathcal{H}_{(s,0^-|d-2|0^-,s)}\left(y_1,y_4,y_3,y_2\right)\right],
\end{align}
which neatly re-sums the contributions from the infinite tower of gauge fields in the spectrum. Performing now the boundary integration, we have
\begin{align}\label{inthsbt}
\frac{1}{2}\left(\Delta_+-\Delta_-\right)^2&\int d^dy_3 \,d^d y_4\,\mathcal{H}_{(s,0^-|d-2|s,0^-)}\left(y_1,y_3,y_2,y_4\right)\,K_{\Delta_+,0}\left(y_3,y_4\right)\Big|_{\log} \\\nonumber&=\frac{1}{2N}\left(\Delta_+-\Delta_-\right)^2 C_{\Delta_+,0}C_{\Delta_-,0}C_{s+d-2,s}\\\nonumber&\times\left[\frac{32 \pi ^{d-2}}{(d+2 s-4) (d+2 s-2) \Gamma \left(\frac{d}{2}-1\right)^2}-\frac{\pi ^d d(d-2)}{\Gamma \left(\frac{d+2}{2}\right)^2}\right]\frac{\log(y_{12}^2)}{(y_{12}^2)^{d-2}}\,{\sf H}_{21}^s\,,
\end{align}
and (which by symmetry in $y_3$ and $y_3$ is identical to \eqref{inthsbt}):
\begin{align}
\frac{1}{2}\left(\Delta_+-\Delta_-\right)^2&\int d^dy_3 \,d^d y_4\,\mathcal{H}_{(s,0^-|d-2|0^-,s)}\left(y_1,y_4,y_3,y_2\right)\,K_{\Delta_+,0}\left(y_3,y_4\right)\Big|_{\log} \\\nonumber&=\frac{1}{2N}\left(\Delta_+-\Delta_-\right)^2 C_{\Delta_+,0}C_{\Delta_-,0}C_{s+d-2,s}\\\nonumber&\times\left[\frac{32 \pi ^{d-2}}{(d+2 s-4) (d+2 s-2) \Gamma \left(\frac{d}{2}-1\right)^2}-\frac{\pi ^d d(d-2)}{\Gamma \left(\frac{d+2}{2}\right)^2}\right]\frac{\log(y_{12}^2)}{(y_{12}^2)^{d-2}}\,{\sf H}_{21}^s\,.
\end{align}
Combined with \eqref{tottadcut}, \eqref{norm}, and the result \eqref{dcutanom} for the double-cut ${\cal M}^{\Delta_+,\Delta_-}_{\Delta_+,\Delta_-}$, from \eqref{diifbc2pt1} upon factoring out the normalisation $C_{s+d-2,s}$ we obtain
\begin{equation} 
   \gamma_s=\frac{2^{d} (d-4) \sin \left(\frac{\pi  d}{2}\right) \Gamma \left(\frac{d-1}{2}\right) (d \,s!\, \Gamma (d-1)-2(s-1) (d+s-2) \Gamma (d+s-3))}{\pi ^{3/2} d (d+2 s-4) (d+2 s-2) \Gamma \left(\frac{d}{2}\right) \Gamma (d+s-3)N}\,, 
\end{equation}
which matches the result of \cite{Lang:1992zw,Hikida:2016wqj}, and in particular for $d=3$ reduces to the CFT result \eqref{cftgammas} for the anomalous dimensions in the $O(N)$ model:
\begin{equation}\label{cftres2}
    \gamma_s=\frac{16 (s-2)}{3 \pi ^2 (2 s-1)N}\,.
\end{equation}
Let us stress that, in first performing the sum over spin, once it is assumed that the duality with the $\Delta_-$ boundary condition holds, the recovery of the anomalous dimension \eqref{cftres2} from \eqref{diifbc2pt1} is trivial \cite{Giombi:2011ya}. A non-trivial question would be whether the same result can be recovered by treating higher-spin gauge theories as standard field theories, which entails using the approach taken in \S \tcb{\ref{subsubsec::altquantads4}} that instead sums over spin after performing the boundary integration.\footnote{If this turns out to be the case, a further question would be how this can be reconciled with the apparent non-commutativity of the sum over spin with the boundary integration observed earlier in this section.} Since we have seen that the contribution from  bubble diagrams \eqref{totdiffbccont} is insufficient, addressing this question requires to take into account $\bep(14,10)\put(0,0){\line(1,0){14}}\put(7,4.25){\circle{8}}\eep$-type tadpole diagrams, which we leave for future work. We would also like to stress that in using twist-blocks we are able to project out all double-trace contribution from the current exchange. This subtraction should be generated in the field theory computation by the quartic contact term and may justify the different behaviour of \eqref{cftres2} with respect to the behaviour in figure~\ref{fig::Resum}.

Let us note that also in performing first the sum over spin we can see that $\bep(14,10)\put(0,0){\line(1,0){14}}\put(7,4.25){\circle{8}}\eep$-type tadpole diagrams should give a non-trivial contribution to the anomalous dimension. The total contribution from the single-cut diagrams arising from $s-(s^\prime 0)-s$ bubbles in the difference of one-loop two-point amplitudes \eqref{diifbc2pt1} is given (modulo contact terms) by \eqref{inthsbt}, i.e.:
\begin{align}\label{inthsbt2}
{\cal M}^{\Delta_+,\Delta_-}_s&=\frac{1}{2}\left(\Delta_+-\Delta_-\right)^2\int d^dy_3 \,d^d y_4\,\mathcal{H}_{(s,0^-|d-2|s,0^-)}\left(y_1,y_3,y_2,y_4\right)\,K_{\Delta_+,0}\left(y_3,y_4\right)\Big|_{\log} \\\nonumber&=\frac{1}{2N}\left(\Delta_+-\Delta_-\right)^2 C_{\Delta_+,0}C_{\Delta_-,0}C_{s+d-2,s}\\\nonumber&\times\left[\frac{32 \pi ^{d-2}}{(d+2 s-4) (d+2 s-2) \Gamma \left(\frac{d}{2}-1\right)^2}-\frac{\pi ^d d(d-2)}{\Gamma \left(\frac{d+2}{2}\right)^2}\right]\frac{\log(y_{12}^2)}{(y_{12}^2)^{d-2}}\,{\sf H}_{21}^s\,,
\end{align}
which, either alone or together with the tadpole contributions \eqref{tottadcut} does not recover the contribution generated by the second line of \eqref{diifbc2pt1}.\footnote{In fact, the non-trivial contribution from $\bep(14,10)\put(0,0){\line(1,0){14}}\put(7,4.25){\circle{8}}\eep$-type tadpoles appears to arise from the $1/\Box$-type non-locality of quartic contact interactions in higher-spin gauge theories on AdS$_{d+1}$ \cite{Sleight:2017pcz}, which smears out the contact interaction to produce precisely the higher-spin blocks in the second line of \eqref{diifbc2pt1} needed to recover the anomalous dimension. Notice that the expression of the the four-point amplitude \eqref{4pttreelev} purely in terms of higher-spin blocks indicates that any genuine contact contributions (i.e. not of the $1/\Box$-type) cancel among each other to give a vanishing overall contribution to the anomalous dimension.}

\subsection*{Acknowledgements}

C. S. and M. T. thank Massimo Bianchi for useful discussions and the University of Rome Tor Vergata for the warm hospitality. We also thank Andres Collinucci for much appreciated technical support and Dean Carmi, Lorenzo Di Pietro and Shota Komatsu for discussions. The research of M. T. is partially supported by the Fund for Scientific Research-FNRS Belgium, grant FC 6369, the Russian Science Foundation grant 14-42-00047 in association with Lebedev Physical Institute and by the INFN within the program ``New Developments in AdS$_3$/CFT$_2$ Holography''. The research of C. S. was partially supported by the INFN and ACRI's (Associazione di Fondazioni e di Casse di Risparmio S.p.a.) Young Investigator Training Program, as part of the Galileo Galilei Institute for Theoretical Physics (GGI) workshop ``New Developments in AdS$_3$/CFT$_2$ Holography''. The work of S.G. was supported in part by the US NSF under Grant No. PHY-1620542.

\appendix

\section{Appendix of conformal integrals}
\label{app::confint}

In this appendix we outline the evaluation of various boundary conformal integrals utilised in this work. 

\subsection{Fourier Transform}
We recall the standard result:
\begin{align}
   \frac1{(2\pi)^{d/2}} \int \frac{d^dq}{\left[q^2\right]^{\Delta}}e^{iq\cdot p}=\frac1{(2\pi)^{d/2}}\frac1{\Gamma(\Delta)}\int_0^\infty\frac{dt}{t}\,t^{\Delta} \int d^dq\,e^{iq\cdot p-t\,q^2}=\frac1{2^{d/2}}\frac{\Gamma(\frac{d}{2}-\Delta)}{\Gamma(\Delta)}\left(\frac{4}{p^2}\right)^{\frac{d}{2}-\Delta}\,,
\end{align}
which we will use repeatedly in the following.

\subsection{Two-point and comments on regularisation}
\label{app::two2ptreg}

The two-point conformal integral
\begin{align}
    I_{\text{2pt}}\left(y_1,y_2\right) & = \int \frac{d^dy}{\left[\left(y_1-y\right)^2\right]^{a_1}\left[\left(y_2-y\right)^2\right]^{a_2}}, \qquad a_1+a_2=d,
\end{align}
appears universally in the computation of AdS two-point loop amplitudes. The regularisation of the latter integral generically produces two type of terms: one proportional to $\left(y^2_{12}\right)^{-\frac{d}{2}}$ and a term proportional to $\log(y_{12}^2)$, which is the fingerprint of the generation of anomalous dimensions. By conformal invariance all divergent diagrams, regardless they are bubble or tadpoles $\bep(14,10)\put(0,0){\line(1,0){14}}\put(7,4.25){\circle{8}}\eep$, are proportional to the above 2 pt integral. It can be evaluated by taking the Fourier transform

\begin{multline}
    \frac{1}{\left(2\pi\right)^{\frac{d}{2}}} \int d^d y_1\, I_{\text{2pt}}\left(y_1,0\right) e^{-i y_1 \cdot p} \\ = \frac{1}{\Gamma\left(a_1\right)\Gamma\left(a_2\right)} \int^{\infty}_0 \frac{dt_1 dt_2}{t_1 t_2} t^{a_1}_1 t^{a_2}_2 \frac{1}{\left(2\pi\right)^{\frac{d}{2}}}\left(\int d^d y_1\, e^{-t_1 y^2_1-i y_1 \cdot p}\right) \left(\int d^d y\, e^{-t_2 y^2-i y \cdot p}\right),
\end{multline}
where in the equality we sent $y_1 \rightarrow y_1+y$ and employed the Schwinger parameterisation 
\begin{equation}
    \frac{1}{\left(x^2\right)^a} = \frac{1}{\Gamma\left(a\right)} \int^\infty_0 \frac{dt}{t} t^a e^{-t x^2}.
\end{equation}
Evaluating the Gaussian integrals and performing the change of variables $t \rightarrow 1/t$, one finds 
\begin{align}
    \frac{1}{\left(2\pi\right)^{\frac{d}{2}}} \int d^d y_1\, I_{\text{2pt}}\left(y_1,0\right) e^{-i y_1 \cdot p} & = \left(\frac{\pi}{2}\right)^{\frac{d}{2}}\frac{1}{\Gamma\left(a_1\right)\Gamma\left(a_2\right)} \int^{\infty}_0 \frac{dt_1 dt_2}{t_1 t_2} t^{\frac{d}{2}-a_1}_1 t^{\frac{d}{2}-a_2}_2 e^{-\left(t_1+t_2\right) \frac{p^2}{4}} \\
    & = \left(\frac{\pi}{2}\right)^{\frac{d}{2}}\frac{\Gamma\left(\frac{d}{2}-a_1\right)\Gamma\left(\frac{d}{2}-a_2\right)}{\Gamma\left(a_1\right)\Gamma\left(a_2\right)} \left(\frac{4}{p^2}\right)^{d-a_1-a_2},
\end{align}
where in the second equality we used the integral representation of the Gamma function. Taking the inverse Fourier transform obtains the final expression
\begin{equation}
    \boxed{I_{\text{2pt}}\left(y_1,y_2\right)  = \pi^{\frac{d}{2}}\frac{\Gamma\left(\frac{d}{2}-a_1\right)\Gamma\left(\frac{d}{2}-a_2\right)}{\Gamma\left(a_1\right)\Gamma\left(a_2\right)} \frac{\Gamma\left(a_1+a_2-\frac{d}{2}\right)}{\Gamma\left(d-a_1-a_2\right)} \left(y^2_{12}\right)^{\frac{d}{2}-a_1-a_2}}\,,
\end{equation}
and, in particular, for $a_1+a_2=d$ employing the dimensional regularisation in eq.~\eqref{pseudo-dim} we have
\begin{subequations}
\begin{align}
    I_{\text{2pt}}\left(y_1,y_2\right)  & = \frac{2 \pi ^{d/2} (y_{12}^2)^{-\frac{d}{2}} \left(\log (\pi  (y_{12}^2))-\psi ^{(0)}\left(\frac{d}{2}\right)\right)}{\Gamma \left(\frac{d}{2}\right)}\qquad \qquad & a_1 = a_2 \\
    & = 0 \qquad & a_1 \ne a_2.
\end{align}
\end{subequations}
It is also interesting to study more generally the analytic structure of the above integral as a function of $d$, $a_1$ and $a_2$ which can be done in various ways. Considering a simple parameterisation of the type $a_1=\tfrac{d}2+\epsilon_1x$ and $a_2=\tfrac{d}{2}+\epsilon_2 x$ and expanding in $x$ one arrives at:
\begin{equation}
    I_{\text{2pt}}\left(y_1,y_2\right)\sim\frac{\pi ^{d/2} (y_{12}^2)^{-\frac{d}{2}} (\epsilon_1+\epsilon_2)^2 \log ((y_{12}^2))}{\epsilon_1 \epsilon_2 \Gamma \left(\frac{d}{2}\right)}-\frac{\pi ^{d/2} (y_{12}^2)^{-\frac{d}{2}} (\epsilon_1+\epsilon_2)}{x \epsilon_1 \epsilon_2 \Gamma \left(\frac{d}{2}\right)}\,.
\end{equation}
The variant of dimensional regularisation mentioned above (which is here referred to as a prescription to regulate a divergent integral) is instead achieved with the parameterisation:\footnote{To avoid any confusion it is useful to stress that a standard dimensional analytic continuation where one analytically continues the bulk Lagrangian to arbitrary dimensions does not define a regularisation of the theory in our case since this does not break the boundary conformal symmetry.}
\begin{align}\label{pseudo-dim}
d^\star&=d+\epsilon\,,& a_1&=\frac{d}{2}\,,& a_2&=\frac{d}{2}\,,
\end{align}
with $d^\star$ the dimension of the measure. This gives
\begin{multline}
    I_{\text{2pt}}\left(y_1,y_2\right)=\frac{\pi ^{\frac{d+\epsilon }{2}} \Gamma \left(\frac{\epsilon }{2}\right)^2 \Gamma \left(\frac{d-\epsilon }{2}\right)}{\Gamma \left(\frac{d}{2}\right)^2 \Gamma (\epsilon )}\,(y_{12}^2)^{\frac{\epsilon -d}{2}}\\\sim \frac{4 \pi ^{d/2} (y_{12}^2)^{-d/2}}{\epsilon\,  \Gamma \left(\frac{d}{2}\right)}+\frac{2 \pi ^{d/2} (y_{12}^2)^{-d/2} \left(\log (\pi  (y_{12}^2))-\psi ^{(0)}\left(\frac{d}{2}\right)\right)}{\Gamma \left(\frac{d}{2}\right)}\,.
\end{multline}

Another possible regularisation consists in taking the limit $a_1\to d/2$ at $a_2$ fixed and then take the limit $a_2\to d/2$. In this case one obtains:
\begin{equation}
   I_{\text{2pt}}\left(y_1,y_2\right)\sim-\frac{\pi ^{d/2} (y_{12}^2)^{-\frac{d}{2}}}{\epsilon_1 \Gamma \left(\frac{d}{2}\right)}-\frac{\pi ^{d/2} (y_{12}^2)^{-\frac{d}{2}}}{\epsilon_2 \Gamma \left(\frac{d}{2}\right)}+\frac{2 \pi ^{d/2} (y_{12}^2)^{-\frac{d}{2}} \log ((y_{12}^2))}{\Gamma \left(\frac{d}{2}\right)}\,,
\end{equation}
giving a $\log$ coefficient $\frac{2\pi^{d/2}}{\Gamma(d/2)}$ which is the same as for dimensional regularisation but in a different subtraction scheme, since no wave function renormalisation is generated. Other choices of $\epsilon_1=k\, \epsilon_2$ should not be admissible as they give different coefficients for the $\log$.

In this work we stick to the above generalised dimensional regularisation as this allows to keep $a_1=a_2=\frac{d}{2}$ in the regularisation process. This regularisation also matches known expectations in the large-$N$ expansion on the boundary side. Furthermore, it might be interesting to notice that all divergent conformal integrals we have encountered can be reduced to the same 2pt divergent conformal integral. Therefore, once a consistent regularisation scheme is identified for $I_{2\text{pt}}$, one should be able to consistently regulate all divergent conformal integrals.

\subsection{Three-point}

The three-point conformal integral
\begin{equation}
I_{\text{3pt}}\left(y_1,y_2,y_3\right)  = \int \frac{d^dy}{\left[\left(y_1-y\right)^2\right]^{a_1}\left[\left(y_2-y\right)^2\right]^{a_2}\left[\left(y_3-y\right)^2\right]^{a_3}}, \quad a_1+a_2+a_3=d
\end{equation}
arising in the computation of bubble diagrams can be evaluated using Schwinger parameterisation:
\begin{equation}
    I_{\text{3pt}}\left(y_1,y_2,y_3\right) = \int  \frac{d^dy}{\Gamma\left(a_1\right)\Gamma\left(a_2\right)\Gamma\left(a_3\right)} \int^{\infty}_0 \frac{dt_1dt_2dt_3}{t_1t_2t_3} t^{a_1}_1 t^{a_2}_2 t^{a_3}_3 e^{-\sum_i t_i \left(y_i-y\right)^2}.
\end{equation}
Writing
\begin{equation}\label{gaussianrel}
    \sum_i t_i \left(y_i-y\right)^2 = T \left(y - \frac{1}{T} \sum_i t_i y_i\right)^2 + \frac{1}{T} \sum_{i<j} t_it_j y^2_{ij}, \quad T = \sum_i t_i,
\end{equation}
 we can evaluate the integral in $y$ to give
\begin{equation}\label{Tdep}
    I_{\text{3pt}}\left(y_1,y_2,y_3\right) = \frac{\pi^{\frac{d}{2}}}{\Gamma\left(a_1\right)\Gamma\left(a_2\right)\Gamma\left(a_3\right)} \int^{\infty}_0 \frac{dt_1dt_2dt_3}{t_1t_2t_3} t^{a_1}_1 t^{a_2}_2 t^{a_3}_3 T^{-d/2} e^{-\frac{1}{T} \sum_{i<j} t_it_j y^2_{ij}}.
\end{equation}
The crucial observation of Symanzik \cite{Symanzik1972} was that, when $a_1+a_2+a_3=d$, \eqref{Tdep} is unchanged if we take instead $T = \sum_i \kappa_i t_i$ for any $\kappa_i \ge 0$.\footnote{This can be seen by making the change of variables $t_i = \sigma \alpha_i$ with $\alpha_i$ constrained by $\sum_i \kappa_i \alpha_i = 1$. For the integration measure we have
\begin{equation}
  \frac{dt_1dt_2dt_3}{t_1t_2t_3} t^{a_1}_1 t^{a_2}_2 t^{a_3}_3 = \frac{d\alpha_1d\alpha_2d\alpha_3}{\alpha_1\alpha_2\alpha_3} \alpha^{a_1}_1 \alpha^{a_2}_2 \alpha^{a_3}_3 \delta\left(1-\sum_i \kappa_i \alpha_i \right) d\sigma \sigma^{d-1}. 
\end{equation}
In performing the integration over $\sigma$ the explicit dependence on $T$ disappears.} We can thus simply take, for instance, $T = t_3$ which gives the following final expression upon using the integral representation of the gamma function
\begin{equation}
    \boxed{I_{\text{3pt}}\left(y_1,y_2,y_3\right) =  \frac{\pi^{d/2}}{\Gamma\left(a_1\right)\Gamma\left(a_2\right)\Gamma\left(a_3\right)} \frac{\Gamma\left(\frac{d}{2}-a_1\right)\Gamma\left(\frac{d}{2}-a_2\right)\Gamma\left(\frac{d}{2}-a_3\right)}{\left(y^{2}_{12}\right)^{\frac{d}{2}-a_3}\left(y^{2}_{13}\right)^{\frac{d}{2}-a_2}\left(y^{2}_{32}\right)^{\frac{d}{2}-a_1}}}.
\end{equation}

\subsection{$n$-point}\label{sec::Nc}
The 3pt conformal integral discussed in the previous section admits a straightforward extension to $n$-points:
\begin{align}
    I_{n\text{-pt}}&\equiv\int \frac{d^dy}{\prod_{i=1}^n\left[\left(y_i-y\right)^2\right]^{a_i}},& \sum_ia_i&=d\,,
\end{align}
via the Symanzik trick and employing the Cahen-Mellin identity:
\begin{equation}
    e^{-z}=\frac1{2\pi i}\int_{c-i\infty}^{c+i\infty}ds\,\Gamma(-s)\,z^s\,,
\end{equation}
valid for $c<0$ and $|\text{arg}(z)|<\tfrac{\pi}2$. The procedure is to first perform the Gaussian integration after employing the Schwinger parametrisation as in the 3pt case and use Cahen-Mellin formula in such a way to perform all Schwinger parameter integrations. The final result is given by Symanzik $\star$ formula and reads:
\begin{equation}
I_{n-pt}(y_i)=\frac{\pi^{d/2}}{\prod_{i}\Gamma(a_i)}\oint d\delta_{ij}\prod_{i<j}\Gamma(\delta_{ij})(y_{ij})^{-\delta_{ij}}\,,
\end{equation}
where the contour integration measure is defined as (see also \cite{Paulos:2012nu})
\begin{equation}
    \oint d\delta_{ij}\equiv \frac{2}{(2\pi i)^{\tfrac{n(n-3)}2}}\int_{c-i\infty}^{c+i\infty}\prod_{i<j}d\delta_{ij}\prod_{j\neq i}\delta\left(a_i-\sum_j\delta_{ij}\right)\,,
\end{equation}
where the constant $c$ is selected to ensure that all poles of gamma functions are on the left or right of the integration paths.

\subsection{Bubble Integral and alternative regularisations}
\label{app::biar}

In this section we study a different regularisation of the bubble conformal integrals which do not rely on analytically continuing the boundary dimension but instead a deformation of the bulk Harmonic functions appearing in the bulk-to-bulk propagators. In the spirit of large-$N$ conformal field theories one can indeed regularise all boundary conformal integrals deforming asymptotic behaviour of one of the bulk-to-boundary propagators in the split representation \eqref{spinharmfact} of the harmonic functions as:
\begin{equation}
    \Omega_{\nu,J}=\frac{\nu^2}{\pi}\int_{\pl \text{AdS}} dP\,C_{\tfrac{d}2+i\nu,J}\,C_{\tfrac{d}2-i\nu,J}\widehat{K}_{\tfrac{d}2+i\nu-\epsilon,J}\widehat{K}_{\tfrac{d}2-i\nu,J}
\end{equation}
where
\begin{equation}
    \widehat{K}_{\Delta,J}\left(X,U;P,Z\right)=
    \left(U \cdot Z-\frac{U \cdot P Z \cdot X}{P \cdot X}\right)^s\frac{1}{\left(-2P \cdot X\right)^{\Delta}}
\end{equation}
is the bulk-to-boundary propagator without normalisation factor.

With such deformed harmonic functions the basic scalar bubble conformal integral is not conformal:
\begin{align}
    \int d^dy\,d^d\bar{y}[[\mathcal{O}_{\Delta}(y_1)\mathcal{O}_{\tfrac{d}2+i\nu-\epsilon }(y)\mathcal{O}_{\tfrac{d}2+i\bar{\nu}-\epsilon}(\bar{y})]][[\mathcal{O}_{\tfrac{d}2-i\nu}(\bar{y})\mathcal{O}_{\tfrac{d}2-i\bar{\nu}}(y)\mathcal{O}_{\Delta}(y_2)]]\,.
\end{align}
One can still perform the integral rewriting it in Mellin space using the identity:
{\allowdisplaybreaks
\begin{align}
    \int d^dy_{\sf x}d^dy_{\sf y}\,&\frac{1}{\left(y_{1{\sf x}}^2\right)^{\alpha_1}\left(y_{2{\sf x}}^2\right)^{\alpha_2}\left(y_{\sf xy}^2\right)^{\gamma}(y_{1{\sf y}}^2)^{\beta_1}(y_{2{\sf y}}^2)^{\beta_2}}=\frac{1}{(y_{12}^2)^{d-\alpha_1-\alpha_2-\beta_1-\beta_2-\gamma}}\\\nonumber
    &\times\frac{\pi^d}{\Gamma(\alpha_1)\Gamma(\alpha_2)\Gamma(\gamma)\Gamma(d-\alpha_1-\alpha_2-\gamma)}\\\nonumber
    &\times\int_{-i\infty}^{+i\infty}\frac{ds\,dt}{(2\pi i)^2} \tfrac{\Gamma (-s) \Gamma (-t)  \Gamma \left(\frac{d}{2}+s+t-\gamma \right)  \Gamma (d+s-\alpha_2-\beta_2-\gamma ) \Gamma (d+t-\alpha_1-\beta_1-\gamma ) \Gamma (d+s+t-\alpha_1-\alpha_2-\gamma )}{  \Gamma (2 d+s+t-\alpha_1-\alpha_2-\beta_1-\beta_2-2 \gamma )}\\\nonumber
    &\times\tfrac{\Gamma \left(-\frac{d}{2}-s+\alpha_2+\gamma \right)\Gamma \left(-\frac{d}{2}-t+\alpha_1+\gamma \right)\Gamma \left(-\frac{3 d}{2}-s-t+\alpha_1+\alpha_2+\beta_1+\beta_2+2 \gamma \right)}{\Gamma \left(-\frac{d}{2}-s+\alpha_2+\beta_2+\gamma \right) \Gamma \left(-\frac{d}{2}-t+\alpha_1+\beta_1+\gamma \right)}\,.
\end{align}}
The limit $\epsilon\to 0$ can be performed as usual for Mellin integrals starting from a region where each $\Gamma$-function argument is positive and analytically continuing while keeping track of contour crossings. In our case the only contribution proportional to $\log (y_{12}^2)$ comes from the residue at $s=0$ and $t=0$ where for $\epsilon\to 0$ the integration contour is pinched. The result reads:
\begin{multline}
\int d^dy\,d^d\bar{y}[[\mathcal{O}_{\Delta}(y_1)\mathcal{O}_{\tfrac{d}2+i\nu-\epsilon }(y)\mathcal{O}_{\tfrac{d}2+i\bar{\nu}-\epsilon}(\bar{y})]][[\mathcal{O}_{\tfrac{d}2-i\nu}(\bar{y})\mathcal{O}_{\tfrac{d}2-i\bar{\nu}}(y)\mathcal{O}_{\Delta}(y_2)]]=\\
    \frac{2 \pi ^d \Gamma \left(\Delta -\frac{d}{2}\right) \Gamma \left(\frac{d}{2}-\Delta +\frac{\Delta -i ({\nu}-\bar{\nu})}{2}\right) \Gamma \left(\frac{d}{2}-\Delta +\frac{\Delta +i ({\nu}-\bar{\nu})}{2}\right)}{\Gamma \left(\frac{d}{2}\right) \Gamma (d-\Delta ) \Gamma \left(\frac{\Delta -i ({\nu}-\bar{\nu})}{2}\right) \Gamma \left(\frac{\Delta +i ({\nu}-\bar{\nu})}{2}\right)}\frac{\log (y_{12}^2)}{(y_{12}^2)^\Delta}+\ldots\,,
\end{multline}
where the $\ldots$ give terms not proportional to a $\log$ and the $\log$-term matches the result obtained by analytically continuing the boundary space-time dimension in \eqref{phi3specfun}. While the $\log$-term does not depend on the regularisation the $\ldots$ depend explicitly on the regularisation and in this case are expressed in terms of a Mellin-Barnes integral which contributes to the 2-pt function normalisation.

\subsection{Decomposition of bubble integrals}
\label{subsec::expansion}

In this appendix we explain how to decompose the conformal integrals \eqref{subsec::confint}: 
\begin{multline}\label{genconfintapp}
     \mathfrak{K}^{({\bf n,m})}_{s_1,s_2;s_{\sf x},s_{\sf y}}(\nu,{\bar \nu}\,;y_1,y_2)=\int d^dy_{\sf x}d^dy_{\sf y}\,[[{\cal O}_{\Delta_1,s_1}(y_1,z_1){\cal O}_{\Delta_{\sf x},s_{\sf x}}(y_{\sf x},\hat{\pl}_{z_{\sf x}}) {\cal O}_{\Delta_{\sf y},s_{\sf y}}(y_{\sf y},\hat{\pl}_{z_{\sf y}})]]^{(\text{{\bf n}})}\\\times[[{\cal O}_{d-\Delta_{\sf y},s_{\sf y}}(y_{\sf y},z_{\sf y}) {\cal O}_{d-\Delta_{\sf x},s_{\sf x}}(y_{\sf x},z_{\sf x}){\cal O}_{\Delta_2,s_2}(y_2,z_2)]]^{(\text{{\bf m}})}\,,
\end{multline}
which arise from spinning two-point bubble diagrams in terms of basic conformal integrals of the form: 
\begin{equation}\label{basicconfiapp}
    \mathfrak{I}^{a_1,a_2,b_1,b_2}_{\alpha_1,\alpha_2,\gamma,\beta_1,\beta_2}\equiv\int d^dy_{\sf x}d^dy_{\sf y}\,\frac{(z_1\cdot y_{1{\sf x}})^{a_1}(z_2\cdot y_{2{\sf x}})^{a_2}(z_1\cdot y_{1{\sf y}})^{b_1}(z_2\cdot y_{2{\sf y}})^{b_2}}{\left(y_{1{\sf x}}^2\right)^{\alpha_1}\left(y_{2{\sf x}}^2\right)^{\alpha_2}\left(y_{\sf xy}^2\right)^{\gamma}(y_{1{\sf y}}^2)^{\beta_1}(y_{2{\sf y}}^2)^{\beta_2}}\,,
\end{equation}
where conformal invariance requires:
\begin{align}
\alpha_1-a_1+\alpha_2-a_2+\gamma&=d\,,& \beta_1-b_1+\beta_2-b_2+\gamma&=d\,.
\end{align}
 By using the series expansion around $z=0$
\begin{equation}
    J_{\alpha}\left(z\right)=\sum^\infty_{k=0} \frac{\left(-1\right)^k}{k!\Gamma\left(k+\alpha+1\right)} \left(\frac{z}{2}\right)^{2k+\alpha},
\end{equation}
of the Bessel functions present in the three-point conformal structures \eqref{nicebasis}, the integrand of \eqref{genconfintapp} can be reduced to a linear sum of monomials of the form:
\begin{multline}\label{monomial6}
   \mathfrak{Q}_{{\bf p},{\bf \bar p}}=\left[ {\sf Y}_{1,{\sf y}{\sf x}}^{s_1-p_{\sf x}-p_{\sf y}}{\sf \bar{Y}}_{{\sf x},1{\sf y}}^{s_{\sf x}-p_{\sf y}-p_1}{\sf \bar{Y}}_{{\sf y},{\sf x}1}^{s_{\sf y}-p_1-p_{\sf x}}{\sf \bar{H}}_{{\sf y}{\sf x}}^{p_1}{\sf \bar{H}}_{1{\sf y}}^{p_{\sf x}}{\sf \bar{H}}_{{\sf x}1}^{p_{\sf y}}\right]
   \left[ {\sf Y}_{2,{\sf y}{\sf x}}^{s_2-\bar{p}_{\sf x}-\bar{p}_{\sf y}}{\sf Y}_{{\sf x},2{\sf y}}^{s_{\sf x}-\bar{p}_{\sf y}-{p}_2}{\sf Y}_{{\sf y},{\sf x}2}^{s_{\sf y}-{p}_2-\bar{p}_{\sf x}}{\sf H}_{{\sf y}{\sf x}}^{{p}_2}{\sf H}_{2{\sf y}}^{p_{\sf x}}{\sf H}_{{\sf x}2}^{p_{\sf y}}\right]\\
   \times \frac1{(y_{1{\sf x}}^2)^{\delta_{1{\sf x}}}(y_{\sf xy}^2)^{\delta_{{\sf xy}}}(y^2_{1{\sf y}})^{\delta_{{\sf y}1}}}\frac1{(y^2_{2{\sf y}})^{\bar{\delta}_{2{\sf y}}}(y^2_{\sf xy})^{\bar{\delta}_{{\sf yx}}}(y^2_{2{\sf x}})^{\bar{\delta}_{{\sf x}2}}},
\end{multline}
where 
{\small \begin{subequations}
\begin{equation}
\delta_{{\sf x}{\sf y}}=\frac12(\tau_{\sf x}+\tau_{\sf y}-\tau_1)\,,\quad \delta_{1{\sf x}}=\frac12(\tau_{1}+\tau_{\sf x}-\tau_{\sf y})\,,\quad \delta_{1{\sf y}}=\frac12(\tau_{1}+\tau_{\sf y}-\tau_{\sf x})\,,
\end{equation}
\begin{equation}
{\bar \delta}_{{\sf x}{\sf y}}=d-\Delta_{\sf x}-\Delta_{\sf y}+\frac12(\tau_{\sf x}+\tau_{\sf y}-\tau_2)\,,\quad {\bar \delta}_{2{\sf x}}=\Delta_{\sf y}-\Delta_{\sf x}+\frac12(\tau_{2}+\tau_{\sf x}-\tau_{\sf y})\,,\quad {\bar \delta}_{1{\sf y}}=\Delta_{\sf x}-\Delta_{\sf y}+\frac12(\tau_{2}+\tau_{\sf y}-\tau_{\sf x})\,,
\end{equation}
\end{subequations}}
with twists $\tau_i=\Delta_i-s_i$. The conformal building blocks in this case read explicitly:

{\small \begin{subequations}
\begin{align}
& {\sf Y}_{1,{\sf y}{\sf x}}= \frac{z_1\cdot y_{{\sf y}1}}{y_{{\sf y}1}^2}-\frac{z_1\cdot y_{{\sf x}1}}{y_{{\sf x}1}^2}, && {\sf Y}_{2,{\sf y}{\sf x}}= \frac{z_2\cdot y_{{\sf y}2}}{y_{{\sf y}2}^2}-\frac{z_2\cdot y_{{\sf x}2}}{y_{{\sf x}2}^2},\\
& {\sf \bar{Y}}_{{\sf x},1{\sf y}}=\frac{{\hat \partial}_{z_{{\sf x}}}\cdot y_{1{\sf x}}}{y_{1{\sf x}}^2}-\frac{{\hat \partial}_{z_{{\sf x}}}\cdot y_{{\sf yx}}}{y_{{\sf y x}}^2}, && {\sf Y}_{{\sf x},2{\sf y}}=\frac{z_{{\sf x}}\cdot y_{2{\sf y}}}{y_{2{\sf y}}^2}-\frac{z_{{\sf x}}\cdot y_{{\sf y x}}}{y_{{\sf y x}}^2},\\
& {\sf \bar{Y}}_{{\sf y},{\sf x}1}=\frac{{\hat \partial}_{z_{{\sf y}}}\cdot y_{{\sf x y}}}{y_{{\sf x y}}^2}-\frac{{\hat \partial}_{z_{{\sf y}}}\cdot y_{1{\sf y}}}{y_{1{\sf y}}^2}, && {{\sf Y}}_{{\sf y},{\sf x}2}=\frac{z_{{\sf y}}\cdot y_{{\sf x y}}}{y_{{\sf x y}}^2}-\frac{z_{{\sf y}}\cdot y_{2{\sf y}}}{y_{2{\sf y}}^2},\\
& {\sf \bar{H}}_{\sf yx}=\frac{1}{y_{{\sf xy}}^2}\left({\hat \partial}_{z_{{\sf x}}}\cdot {\hat \partial}_{z_{{\sf y}}}+\frac{2 {\hat \partial}_{z_{{\sf x}}}\cdot y_{{\sf xy}}\,{\hat \partial}_{z_{{\sf y}}}\cdot y_{{\sf yx}}}{y_{{\sf xy}}^2}\right), && {\sf H}_{\sf yx}=\frac{1}{y_{{\sf xy}}^2}\left(z_{{\sf x}}\cdot z_{{\sf y}}+\frac{2 z_{{\sf x}}\cdot y_{{\sf xy}}\,z_{{\sf y}}\cdot y_{{\sf yx}}}{y_{{\sf xy}}^2}\right),\\
& {\sf \bar{H}}_{1{\sf y}}=\frac{1}{y_{{\sf y}1}^2}\left({\hat \partial}_{z_{{\sf y}}}\cdot z_{1}+\frac{2 {\hat \partial}_{z_{{\sf y}}}\cdot y_{{\sf y}1}\,z_{1}\cdot y_{1{\sf y}}}{y_{{\sf xy}}^2}\right), && {{\sf H}}_{2{\sf y}}=\frac{1}{y_{{\sf y}2}^2}\left(z_{{\sf y}}\cdot z_{2}+\frac{2 z_{{\sf y}}\cdot y_{{\sf y}2}\,z_{2}\cdot y_{2{\sf y}}}{y_{{\sf xy}}^2}\right),\\
& {\sf \bar{H}}_{{\sf x}1}=\frac{1}{y_{1{\sf x}}^2}\left(z_{1}\cdot {\hat \partial}_{z_{{\sf x}}}+\frac{2 z_{1}\cdot y_{1{\sf x}}\,{\hat \partial}_{z_{{\sf x}}}\cdot y_{{\sf x}1}}{y_{1{\sf x}}^2}\right), && {{\sf H}}_{{\sf x}2}= \frac{1}{y_{2{\sf x}}^2}\left(z_{2}\cdot z_{{\sf x}}+\frac{2 z_{2}\cdot y_{2{\sf x}}\,z_{{\sf x}}\cdot y_{{\sf x}2}}{y_{2{\sf x}}^2}\right).
\end{align}
\end{subequations}}

The main step is to evaluate the Thomas derivatives ${\hat \partial}_{z_{{\sf x}}}$ and ${\hat \partial}_{z_{{\sf y}}}$ in \eqref{monomial6}. To this end, it's useful to introduce the combinations:
\begin{subequations}
\begin{align}
& \xi_{\sf x} \cdot {\hat \partial}_{z_{{\sf x}}} = {\sf \bar{Y}}_{{\sf x},1{\sf y}}+\lambda_{\sf y} {\sf \bar{H}}_{{\sf x}1}, &&  {\bar \xi}_{\sf x} \cdot z_{\sf x}= {{\sf Y}}_{{\sf x},2{\sf y}}+{\bar \lambda}_{\sf y} {{\sf H}}_{{\sf x}2}, \\
& \xi_{\sf y} \cdot {\hat \partial}_{z_{{\sf y}}}= {\sf \bar{Y}}_{{\sf y},{\sf x}1}+\lambda_{\sf x} {\sf \bar{H}}_{1{\sf y}}, &&  {\bar \xi}_{\sf y} \cdot z_{\sf x}= { {\sf Y}}_{{\sf y},{\sf x}2}+{\bar \lambda}_{\sf x} {{\sf H}}_{2{\sf y}},
\end{align}
\end{subequations}
and the differential operators: 
\begin{align}
\mathcal{O}_{{\sf \bar{H}}_{\sf yx}}&=\frac1{y_{\sf xy}^2}\left(\pl_{\xi_{\sf x}}\cdot\pl_{\xi_{\sf y}}-\frac{2}{y_{\sf xy}^2}\,y_{\sf xy}\cdot\pl_{\xi_{\sf x}} y_{\sf xy}\cdot\pl_{\xi_{\sf y}}\right)\,, \\
\mathcal{O}_{{\sf H}_{\sf yx}}&=\frac1{y_{\sf xy}^2}\left(\pl_{\bar{\xi}_{\sf x}}\cdot\pl_{\bar{\xi}_{\sf y}}-\frac{2}{y_{\sf xy}^2}\,y_{\sf xy}\cdot\pl_{\bar{\xi}_{\sf x}} y_{\sf xy}\cdot\pl_{\bar{\xi}_{\sf y}}\right)\,,
\end{align}
which have the property: $\mathcal{O}_{{\sf \bar{H}}_{\sf yx}}\left(\xi_{\sf x}\xi_{\sf y}\right)={\sf \bar{H}}_{\sf yx}$ and $\mathcal{O}_{{\sf H}_{\sf yx}}\left({\bar \xi}_{\sf x}{\bar \xi}_{\sf y}\right)={\sf H}_{\sf yx}$. This allows us to define the following generating function:
\begin{align}\label{genfuQ}
    \mathfrak{Q}\left(\lambda,{\bar \lambda}\right)& =\frac1{(s_{\sf x}-p_1+1)_{p_1}(s_{\sf y}-p_1+1)_{p_1}(s_{\sf x}-p_2+1)_{p_2}(s_{\sf y}-p_2+1)_{p_2}}\\ \nonumber & \hspace*{1cm}\times {\sf Y}_{1,{\sf yx}}^{s_1-p_{\sf x}-p_{\sf y}} {\sf Y}_{2,{\sf yx}}^{s_2-\bar{p}_{\sf x}-\bar{p}_{\sf y}} \mathcal{O}_{{\sf \bar{H}}_{\sf yx}}^{p_1}\mathcal{O}_{{\sf H}_{\sf yx}}^{p_2}\left[(\xi_{\sf x}\cdot \hat{\partial}_{z_{\sf x}})^{s_{\sf x}}(\bar{\xi}_{\sf x}\cdot {z_{\sf x}})^{s_{\sf x}}\right]\left[(\xi_{\sf y}\cdot \hat{\partial}_{z_{\sf y}})^{s_{\sf y}}(\bar{\xi}_{\sf y}\cdot {z_{\sf y}})^{s_{\sf y}}\right],
\end{align}
from which \eqref{monomial6} can be recovered via
\begin{equation}
    \mathfrak{Q}_{{\bf p},{\bf \bar p}}= \frac{(s_{\sf y}-p_1-p_{\sf x})!(s_{\sf y}-p_2-\bar{p}_{\sf x})!(s_{\sf x}-p_1-p_{\sf y})!(s_{\sf x}-p_2-\bar{p}_{\sf y})!}{(s_{\sf x}-p_1)!(s_{\sf y}-p_1)!(s_{\sf x}-p_2)!(s_{\sf y}-p_2)!}\pl_{\xi_{\sf x}}^{p_{\sf y}}\pl_{\xi_{\sf y}}^{p_{\sf x}}\pl_{{\bar \xi}_{\sf x}}^{\bar{p}_{\sf y}}\pl_{{\bar \xi}_{\sf y}}^{\bar{p}_{\sf x}}\mathfrak{Q}\left(\lambda,{\bar \lambda}\right).
\end{equation}
Above and also in the following discussion, for convenience the presence of the factor in the second line of \eqref{monomial6} is left implicit.
The generating function \eqref{genfuQ} is convenient, for it allows to straightforwardly evaluate the Thomas derivatives by simply using that 
\begin{equation}
   ( a \cdot \hat{\partial}_{z})^k  \left( b \cdot z\right)^k = \frac{k!}{2^k\left(\frac{d}{2}-1\right)_k}\left(a^2b^2\right)^{k/2}C^{\left(\frac{d}{2}-1\right)}_{k}\left(\frac{a \cdot b}{\sqrt{a^2b^2}}\right),
\end{equation}
in terms of a Gegenbauer polynomial. This gives
\begin{multline}
    \mathfrak{Q}\left(\lambda,{\bar \lambda}\right)=\tfrac1{(s_{\sf x}-p_1+1)_{p_1}(s_{\sf y}-p_1+1)_{p_1}(s_{\sf x}-p_2+1)_{p_2}(s_{\sf y}-p_2+1)_{p_2}}\tfrac{s_{\sf x}!}{2^{s_{\sf x}}(\frac{d}{2}-1)_{s_{\sf x}}}\tfrac{s_{\sf y}!}{2^{s_{\sf y}}(\frac{d}{2}-1)_{s_{\sf y}}}\\\times {\sf Y}_{1,{\sf yx}}^{s_1-p_{\sf x}-p_{\sf y}} {\sf Y}_{2,{\sf yx}}^{s_2-\bar{p}_{\sf x}-\bar{p}_{\sf y}} \mathcal{O}_{{\sf \bar{H}}_{\sf yx}}\mathcal{O}_{{\sf H}_{\sf yx}}\left\{[\xi_{\sf x}^2{\bar \xi}_{\sf x}^2]^{s_{\sf x}/2}C_{s_{\sf x}}^{(\frac{d}{2}-1)}\left(\tfrac{\xi_{\sf x}\cdot{\bar \xi}_{\sf x}}{[\xi_{\sf x}^2{\bar \xi}_{\sf x}^2]^{1/2}}\right)
   [\xi_{\sf y}^2{\bar \xi}_{\sf y}^2]^{s_{\sf y}/2}C_{s_{\sf y}}^{(\frac{d}{2}-1)}\left(\tfrac{\xi_{\sf y}\cdot{\bar \xi}_{\sf y}}{[\xi_{\sf y}^2{\bar \xi}_{\sf y}^2]^{1/2}}\right)\right\}.
\end{multline}
Upon expanding the Gegenbauer polynomials, one obtains
 
{\small\begin{multline}
   \mathfrak{Q}=\frac{2^{s_{\sf x}+s_{\sf y}}}{(s_{\sf x}-n_1+1)_{n_1}(s_{\sf y}-n_1+1)_{n_1}(s_{\sf x}-n_2+1)_{n_2}(s_{\sf y}-n_2+1)_{n_2}}\frac{s_{\sf x}!}{2^{s_{\sf x}}\Gamma(\frac{d}{2}-1+s_{\sf x})}\frac{s_{\sf y}!}{2^{s_{\sf y}}\Gamma(\frac{d}{2}-1+s_{\sf y})}  \\ \times 
   \sum^{\left\lfloor s_{\sf x}/2 \right\rfloor}_{k_1=0}\sum^{\left\lfloor s_{\sf y}/2 \right\rfloor}_{k_2=0}    
 (-1)^{k_1+k_2} \frac{\Gamma\left(s_{\sf x}-k_1+\frac{d}{2}-1\right)\Gamma\left(s_{\sf y}-k_2+\frac{d}{2}-1\right)}{2^{2\left(k_1+k_2\right)}k_1!\left(s_{\sf x}-2k_1\right)!k_2!\left(s_{\sf y}-2k_2\right)!}  
   \\\times {\sf Y}_{1,{\sf yx}}^{s_1-n_{\sf x}-n_{\sf y}} {\sf Y}_{2,{\sf yx}}^{s_2-\bar{n}_{\sf x}-\bar{n}_{\sf y}} \mathcal{O}_{{\sf \bar{H}}_{\sf yx}}^{n_1}\mathcal{O}_{{\sf H}_{\sf yx}}^{n_2}\left\{[\xi_{\sf x}^2{\bar \xi}_{\sf x}^2]^{2k_1-s_{\sf x}/2}
   [\xi_{\sf y}^2{\bar \xi}_{\sf y}^2]^{2k_2-s_{\sf y}/2}\left(\xi_{\sf x}\cdot{\bar \xi}_{\sf x}\right)^{s_{\sf x}-2k_1}\left(\xi_{\sf y}\cdot{\bar \xi}_{\sf y}\right)^{s_{\sf y}-2k_2}\right\},
\end{multline}}

which gives a nested sum of the conformal integrals \eqref{basicconfiapp} upon evaluating the $\mathcal{O}_{{\sf H}}$ and expanding the ${\sf Y}$'s, ${\sf H}$'s, $\xi$'s and ${\bar \xi}$'s, for which the following identities are useful:
\begin{align}
y_{ij}\cdot y_{kl}&=-y_{ik}\cdot y_{lk}+y_{jk}\cdot y_{lk}\,,\\
y_{ij}\cdot y_{kj}&=\frac12(y_{ij}^2+y_{kj}^2-y_{ik}^2)\,,\\
z_i\cdot y_{jk}&=z_i\cdot y_{ik}-z_i\cdot y_{ij}\,.
\end{align}

Particularly simple with respect to the general case is the situation in which one of the internal legs in the bubble is scalar. In this case indeed $n_1=n_2=0$ and the full conformal integral can be expressed by a Gegenbauer polynomial while the action of the differential operator trivialises.

\subsection{Shadow bulk-to-boundary propagator}
\label{subsec::shadprop}

In this section we prove the integral relationship \eqref{shadnonshadbubou} of footnote \ref{foo::bobu} between bulk-to-boundary propagators of different conformally invariant boundary conditions for the case $J=0$, as relevant for this work.

This is most straightforward working in ambient space. The RHS of \eqref{shadnonshadbubou} for $J=0$ reads:
\begin{multline}
  -\left(\Delta_+-\Delta_-\right)  \int_{\partial \text{AdS}} d{\bar P}\, K_{\Delta_+}\left(P;{\bar P}\right)K_{\Delta_-}\left(X;{\bar P}\right)\\ = -\left(\Delta_+-\Delta_-\right) C_{\Delta_+,0}C_{\Delta_-,0}  \int_{\partial \text{AdS}} d{\bar P}\,\frac{1}{\left(-2 P \cdot {\bar P}\right)^{\Delta_+}} \frac{1}{\left(-2 X \cdot {\bar P}\right)^{\Delta_-}}
\end{multline}
Using Feynman parameterisation:
\begin{multline}
    \int_{\partial \text{AdS}} d{\bar P}\,\frac{1}{\left(-2 P \cdot {\bar P}\right)^{\Delta_+}} \frac{1}{\left(-2 X \cdot {\bar P}\right)^{\Delta_-}} = \int_{\partial \text{AdS}} d{\bar P}\, \frac{\Gamma\left(d\right)}{\Gamma\left(\Delta_+\right)\Gamma\left(\Delta_-\right)} \int^\infty_0 d\lambda\,\frac{\lambda^{\Delta_+-1}}{\left(-2 {\bar P} \cdot Y\right)^d}, 
\end{multline}
where: $Y^A=X^A+\lambda P^A$, it is straightforward to perform the conformal integral in ${\bar P}$:
\begin{equation}
     \int_{\partial \text{AdS}} d{\bar P}\, \frac{1}{\left(-2 {\bar P} \cdot Y\right)^d} = \frac{\pi^{d/2}\Gamma\left(\frac{d}{2}\right)}{\Gamma\left(d\right)} \frac{1}{\left(-Y^2\right)^{d/2}}.
\end{equation}
The remaining integral in $\lambda$ is given by the Beta function, which yields:
\begin{equation}
      \int_{\partial \text{AdS}} d{\bar P}\,\frac{1}{\left(-2 P \cdot {\bar P}\right)^{\Delta_+}} \frac{1}{\left(-2 X \cdot {\bar P}\right)^{\Delta_-}}=\pi^{d/2} \frac{\Gamma\left(\frac{d}{2}-\Delta_+\right)}{\Gamma\left(\Delta_-\right)} \frac{1}{\left(-2 P\cdot X\right)^{\Delta_+}}.
\end{equation}
Using the explicit form \eqref{scbubounorm} of the propagator normalisation, this finally gives:
\begin{equation}
  -\left(\Delta_+-\Delta_-\right)  \int_{\partial \text{AdS}} d{\bar P}\, K_{\Delta_+}\left(P;{\bar P}\right)K_{\Delta_-}\left(X;{\bar P}\right)=K_{\Delta_+,0}\left(X;P\right).
\end{equation}

\section{Coincident Point Propagator}
\label{app:ccp}

In this appendix we show how the split representation relates to the standard expressions for the coincident point limit of the bulk-to-bulk propagator.
We will evaluate the following bulk integral:\footnote{For $s=0$ see \cite{Dobrev:1998md}.}
\begin{equation}
\mathcal{Z}_{\Delta,s}=\int_{\text{AdS}} \text{Tr}[G_{\Delta,s}(X,X)]\,.
\end{equation}
Without loss of generality we can restrict the attention to the TT part of the propagator which encodes the physical degrees of freedom. Using the split representation the above vacuum bubble therefore reads:
\begin{multline}
    \mathcal{Z}_{\Delta,s}=\int_{-\infty}^\infty d\nu\,\underbrace{\frac{\nu^2}{\pi}\frac{1}{\nu^2+(\Delta-\tfrac{d}{2})^2}\,C_{\tfrac{d}{2}+i\nu,s}C_{\tfrac{d}{2}-i\nu,s}}_{f_{(\Delta,s)}(\nu)}\\\times\int_{\partial \text{AdS}} dP\int_{\text{AdS}_{d+1}}dX\,\tfrac{(\pl_{W_1}\cdot D_{W_2})^s}{(s!)^2}\,\tfrac{\left\{\left[(-2P\cdot X)W_1+(2W_1\cdot P)X\right]\cdot \hat{\pl}_{Z}\right\}^s\left\{Z\cdot\left[(-2P\cdot X)W_2+(2W_2\cdot P)X\right]\right\}^s}{s!(-2P\cdot X)^d}\,,
\end{multline}
where $\tfrac{(\pl_{W_1}\cdot D_{W_2})^s}{(s!)^2\left(\tfrac{d-1}{2}\right)_s}$ defines the trace operation with respect to the tangent and light-like auxiliary variables $W_1$ and $W_2$ in terms of the AdS Thomas-D derivative:\footnote{It is convenient to use projected auxiliary variables such that $W_i^2=0$ and $W_i\cdot X=0$.}
\begin{subequations}
\begin{align}
D_{U^A} &= \left({\cal P} \cdot \partial_U\right)_{A}\\
{\hat D}_{W^A} &=  \partial_{W^A}-\frac{1}{d-1+2 W \cdot {\cal P} \cdot  \partial_W} W_{A}\left(\partial_{W} \cdot {\cal P} \cdot \partial_{W}\right),
\end{align}
\end{subequations}
Carrying the above derivative contractions and integrations using the identities:
\begin{align}
\frac1{s!}(A\cdot\hat{\pl}_Z)^s(Z\cdot B)^s&=\tfrac{s!}{2^s \left(\frac{d}{2}-1\right)_s}[A^2 B^2]^{s/2}G_{s}^{\left(\frac{d}{2}-1\right)}\left(\tfrac{A\cdot B}{\sqrt{A^2 B^2}}\right)\,,\\
\int_{\partial \text{AdS}} dP\,\frac{(W_1\cdot P)^2(W_2\cdot P)^s}{(-2P\cdot X)^{d+2s}}&=\frac{2^{1-d-3s}\pi^{\tfrac{d+1}2}s!}{\Gamma(s+\tfrac{d+1}2)}\,(W_1\cdot W_2)^s\,,\\
\frac{(\pl_{W_1}\cdot D_{W_2})^s}{(s!)^2}\,(W_1\cdot W_2)^s&=\frac{(d+2 s-1) (d+s-2)!}{(d-1)!\, s!}\,,
\end{align}
one arrives to the following equation:
\begin{equation}\label{Zres}
    \mathcal{Z}_{(\Delta,s)}=V_{\text{AdS}_{d+1}}\,\frac{2^{1-d} \pi ^{\frac{d+1}{2}}}{\Gamma \left(\frac{d+1}{2}\right)}\,g_s\int_{-\infty}^\infty d\nu f_{(\Delta,s)}(\nu)\,,
\end{equation}
where $V_{\text{AdS}_{d+1}}=\pi^{d/2}\Gamma\left(-\tfrac{d}2\right)$ is the AdS$_{d+1}$ regularised volume and one can recognise the spectral density:
\begin{equation}
    f_{(\Delta,s)}(\nu)=\frac{1}{4\pi^{d+2}}\frac{\nu ^2+\left(\frac{d-2}{2}+s\right)^2}{\nu ^2+\left(\Delta-\tfrac{d}{2}\right)^2} \nu\,  \sinh (\pi  \nu ) \Gamma \left(\frac{d}{2}-i \nu -1\right) \Gamma \left(\frac{d}{2}+i \nu -1\right)\,,
\end{equation}
the volume factor $V_{S^d}=\tfrac{2\pi^{(d+1)/2}}{\Gamma\left(\tfrac{d+1}2\right)}$ and we have expressed the result in terms of the number of degrees of freedom for a symmetric TT field
\begin{equation}
    g_s=\frac{(2s+d-2)(s+d-3)!}{(d-2)!\,s!}\,.
\end{equation}
As expected, equation \eqref{Zres} precisely matches the corresponding expression derived using $\zeta$-function techniques \cite{Camporesi:1994ga}:
\begin{equation}
    \mathcal{Z}_{(\Delta,s)}=\zeta_{(\Delta,s)}(1)\,.
\end{equation}

\subsection{Mellin-Barnes and sum over spins}

The spectral function integrals are naturally regulated as Mellin-Barnes integrals:
\begin{equation}
    \int_{-\infty}^\infty d\nu\,f_{(\Delta,s)}(\nu)\,z^{i\nu}\Big|_{z=1}\,.
\end{equation}
Such integrals can be straightforwardly evaluated as infinite series by closing the contour of integration in the appropriate convergence region and dropping the arc part of the contour. In the example above one can perform the spectral integral in full generality and for arbitrary dimensions:
\begin{multline}
    \lim_{z\to1}\int_{-\infty}^\infty d\nu f_{(\Delta,s)}(\nu)z^{i\nu}=\sum_{n=0}^{\infty}\frac{(d+2 n-2) (n-s) (d+n+s-2) \Gamma (d+n-2)\sin \left(\frac{\pi  d}{2}\right)}{4\,\pi ^{d+1} n! (\Delta +n-1) (d-\Delta +n-1)}\\-\frac{1}{4\pi^{d+1}} (\Delta +s-1)  (d-\Delta +s-1)\Gamma (\Delta -1)\Gamma (d-\Delta -1) \sin \left[\pi  \left(\Delta -\tfrac{d}{2}\right)\right]\,.
\end{multline}
The above series is divergent but with some effort it can be resummed in dimensional regularisation obtaining a remarkably simple answer:\footnote{We have checked that the expression below matches the expression obtained by $\zeta$-function regularisation in any even dimension. In odd dimension the two result differ but we expect that the main physical properties should remain unaffected.}
\begin{equation}
    \lim_{z\to1}\int_{-\infty}^\infty d\nu f_{(\Delta,s)}(\nu)z^{i\nu}=\frac{\sec \left(\frac{\pi  d}{2}\right) \csc (\pi  \Delta ) (\Delta +s-1) (d-\Delta +s-1)}{4\pi^{d-1} \Gamma (2-\Delta ) \Gamma (\Delta-d +2)}\,.
\end{equation}
Furthermore one can explicitely evaluate the sum over spins in dimensional regularisation using Gauss hypergeometric theorem. The sum over spins including ghosts gives:
\begin{align}
    \mathcal{Z}_{\text{HS}}&=\sum_{s=0}^\infty\left(\mathcal{Z}_{(d-2+s,s)}-\mathcal{Z}_{(d-1+s,s-1)}\right)=\frac{4 \pi  \csc (\pi  d) \Gamma (3-2 d)}{d\, \Gamma (3-d)^2}\,.
\end{align}
Remarkably the latter shows no pole in any CFT dimension $d>2$, signaling the cancellation of UV divergences upon summing over spins. Notice also that in the above expression we have included the regularised AdS volume.

\section{Graviton Bubble}
\label{sec::gravbub}

In this appendix we detail how to bring the $2-\left(20\right)-0$ bubble diagram involving the full de Donder gauge graviton propagator \eqref{gravdedon} into the form \eqref{gnericterm}. The diagram is given by four terms:
\begin{equation}\label{app2gravdec}
    {\cal M}^{\text{2pt-bubble}}={\cal M}^{\text{2pt-bubble}}_{1,0;1,0}+\tfrac{1}{2}\left(d-2\right){\cal M}^{\text{2pt-bubble}}_{1,0;0,1}+\tfrac{1}{2}\left(d-2\right){\cal M}^{\text{2pt-bubble}}_{0,1;1,0}+\tfrac{1}{4}\left(d-2\right)^2{\cal M}^{\text{2pt-bubble}}_{0,1;0,1},
\end{equation}
which each, via the spectral representation \eqref{gravdedon} of the full graviton propagator, decompose in terms of the three-point Witten diagrams \eqref{a3ptint} as:

\begin{enumerate}
    \item {\footnotesize
\begin{align}
{\cal M}^{\text{2pt-bubble}}_{1,0;1,0}\left(P_1,P_2\right) & = \frac{g^2}{\pi^2} \int^\infty_{-\infty} \nu^2 d\nu\, {\bar \nu}^2 d{\bar \nu}\, g^{(2)}_{0,0,0}\left(\nu\right)g^{(0)}_{0,0,0}\left({\bar \nu}\right) \int_{\partial \text{AdS}} dPd{\bar P}\\ & \hspace*{4cm}\times  \mathcal{A}^{1,0;0,0}_{\Delta_1,\frac{d}{2}+i{\bar \nu},\frac{d}{2}+i\nu}\left(P_1,P,{\bar P}\right) \cdot \mathcal{A}^{1,0;0,0}_{\Delta_2,\frac{d}{2}-i{\bar \nu},\frac{d}{2}-i\nu}\left(P_2,P,{\bar P}\right) \nonumber \\ \nonumber
& + \frac{g^2}{\pi^2} \int^\infty_{-\infty} \nu^2 d\nu\, {\bar \nu}^2 d{\bar \nu}\,g^{(2)}_{1,1,0}\left(\nu\right)g^{(0)}_{0,0,0}\left({\bar \nu}\right) \int_{\partial \text{AdS}} dPd{\bar P}\\ & \hspace*{4cm}\times  \mathcal{A}^{1,0;1,0}_{\Delta_1,\frac{d}{2}+i{\bar \nu},\frac{d}{2}+i\nu}\left(P_1,P,{\bar P}\right) \cdot \mathcal{A}^{1,0;1,0}_{\Delta_2,\frac{d}{2}-i{\bar \nu},\frac{d}{2}-i\nu}\left(P_2,P,{\bar P}\right) \nonumber\\
 \nonumber
& + \frac{g^2}{\pi^2} \int^\infty_{-\infty} \nu^2 d\nu\, {\bar \nu}^2 d{\bar \nu}\,g^{(2)}_{1,0,0}\left(\nu\right)g^{(0)}_{0,0,0}\left({\bar \nu}\right) \int_{\partial \text{AdS}} dPd{\bar P}\, \\ \nonumber & \hspace*{4cm}\times \mathcal{A}^{1,0;1,0}_{\Delta_1,\frac{d}{2}+i{\bar \nu},\frac{d}{2}+i\nu}\left(P_1,P,{\bar P}\right) \cdot \mathcal{A}^{1,0;0,2}_{\Delta_2,\frac{d}{2}-i{\bar \nu},\frac{d}{2}-i\nu}\left(P_2,P,{\bar P}\right) \\
 \nonumber
& + \frac{g^2}{\pi^2} \int^\infty_{-\infty} \nu^2 d\nu\, {\bar \nu}^2 d{\bar \nu}\,g^{(2)}_{1,0,0}\left(\nu\right)g^{(0)}_{0,0,0}\left({\bar \nu}\right) \int_{\partial \text{AdS}} dPd{\bar P}\, \\ \nonumber & \hspace*{4cm}\times \mathcal{A}^{1,0;0,2}_{\Delta_1,\frac{d}{2}+i{\bar \nu},\frac{d}{2}+i\nu}\left(P_1,P,{\bar P}\right) \cdot \mathcal{A}^{1,0;1,0}_{\Delta_2,\frac{d}{2}-i{\bar \nu},\frac{d}{2}-i\nu}\left(P_2,P,{\bar P}\right)
\\ \nonumber
& + \frac{g^2}{\pi^2} \int^\infty_{-\infty} \nu^2 d\nu\, {\bar \nu}^2 d{\bar \nu}\,g^{(2)}_{0,0,2}\left(\nu\right)g^{(0)}_{0,0,0}\left({\bar \nu}\right) \int_{\partial \text{AdS}} dPd{\bar P}\,\\ \nonumber & \hspace*{4cm}\times \mathcal{A}^{1,0;0,2}_{\Delta_1,\frac{d}{2}+i{\bar \nu},\frac{d}{2}+i\nu}\left(P_1,P,{\bar P}\right) \cdot \mathcal{A}^{1,0;0,2}_{\Delta_2,\frac{d}{2}-i{\bar \nu},\frac{d}{2}-i\nu}\left(P_2,P,{\bar P}\right),
\end{align}
}

\item 

{\footnotesize
\begin{align}
{\cal M}^{\text{2pt-bubble}}_{1,0;0,1}\left(P_1,P_2\right) & = \frac{g^2}{\pi^2} \int^\infty_{-\infty} \nu^2 d\nu\, {\bar \nu}^2 d{\bar \nu}\, g^{(2)}_{1,1,0}\left(\nu\right)g^{(0)}_{0,0,0}\left({\bar \nu}\right) \int_{\partial \text{AdS}} dPd{\bar P}\, \\ &  \hspace*{4cm}\times  \mathcal{A}^{1,0;1,0}_{\Delta_1,\frac{d}{2}+i{\bar \nu},\frac{d}{2}+i\nu}\left(P_1,P,{\bar P}\right) \cdot \mathcal{A}^{0,1;1,0}_{\Delta_2,\frac{d}{2}-i{\bar \nu},\frac{d}{2}-i\nu}\left(P_2,P,{\bar P}\right) \nonumber \\ \nonumber
& + \frac{g^2}{\pi^2} \int^\infty_{-\infty} \nu^2 d\nu\, {\bar \nu}^2 d{\bar \nu}\, g^{(2)}_{1,0,0}\left(\nu\right)g^{(0)}_{0,0,0}\left({\bar \nu}\right) \int_{\partial \text{AdS}} dPd{\bar P}\, \\ \nonumber & \hspace*{4cm}\times  \mathcal{A}^{1,0;0,2}_{\Delta_1,\frac{d}{2}+i{\bar \nu},\frac{d}{2}+i\nu}\left(P_1,P,{\bar P}\right) \cdot \mathcal{A}^{0,1;1,0}_{\Delta_2,\frac{d}{2}-i{\bar \nu},\frac{d}{2}-i\nu}\left(P_2,P,{\bar P}\right),
\end{align}
}

\item 

{\footnotesize
\begin{align}
{\cal M}^{\text{2pt-bubble}}_{0,1;1,0}\left(P_1,P_2\right) & = \frac{g^2}{\pi^2} \int^\infty_{-\infty} \nu^2 d\nu\, {\bar \nu}^2 d{\bar \nu}\, g^{(2)}_{1,1,0}\left(\nu\right)g^{(0)}_{0,0,0}\left({\bar \nu}\right) \int_{\partial \text{AdS}} dPd{\bar P}\, \\ &  \hspace*{4cm}\times  \mathcal{A}^{0,1;1,0}_{\Delta_1,\frac{d}{2}+i{\bar \nu},\frac{d}{2}+i\nu}\left(P_1,P,{\bar P}\right) \cdot \mathcal{A}^{1,0;1,0}_{\Delta_2,\frac{d}{2}-i{\bar \nu},\frac{d}{2}-i\nu}\left(P_2,P,{\bar P}\right) \nonumber \\ \nonumber
& + \frac{g^2}{\pi^2} \int^\infty_{-\infty} \nu^2 d\nu\, {\bar \nu}^2 d{\bar \nu}\, g^{(2)}_{1,0,0}\left(\nu\right)g^{(0)}_{0,0,0}\left({\bar \nu}\right) \int_{\partial \text{AdS}} dPd{\bar P}\, \\ \nonumber & \hspace*{4cm}\times  \mathcal{A}^{0,1;1,0}_{\Delta_1,\frac{d}{2}+i{\bar \nu},\frac{d}{2}+i\nu}\left(P_1,P,{\bar P}\right) \cdot \mathcal{A}^{1,0;0,2}_{\Delta_2,\frac{d}{2}-i{\bar \nu},\frac{d}{2}-i\nu}\left(P_2,P,{\bar P}\right),
\end{align}
}

\item {\footnotesize
\begin{align}
{\cal M}^{\text{2pt-bubble}}_{0,1;0,1}\left(P_1,P_2\right) & = \frac{g^2}{\pi^2} \int^\infty_{-\infty} \nu^2 d\nu\, {\bar \nu}^2 d{\bar \nu}\, g^{(2)}_{1,1,0}\left(\nu\right)g^{(0)}_{0,0,0}\left({\bar \nu}\right) \int_{\partial \text{AdS}} dPd{\bar P}\, \\ &  \hspace*{4cm}\times  \mathcal{A}^{0,1;1,0}_{\Delta_1,\frac{d}{2}+i{\bar \nu},\frac{d}{2}+i\nu}\left(P_1,P,{\bar P}\right) \cdot \mathcal{A}^{0,1;1,0}_{\Delta_2,\frac{d}{2}-i{\bar \nu},\frac{d}{2}-i\nu}\left(P_2,P,{\bar P}\right), \nonumber 
\end{align}
}
\end{enumerate}
The three-point Witten diagrams \eqref{a3ptint} can be straightforwardly evaluated in the present case, in particular since the three-point conformal structure generated is unique. We have:
\begin{enumerate}
    \item
\begin{multline}
\mathcal{A}^{1,0;0,0}_{\Delta_1,\Delta_2,\Delta_3}\left(y_1,y_2,y_3\right)\\={\sf B}\left(0,0,2;{\bf 0};\Delta_1,\Delta_2,\Delta_3-2\right) \left[\left[{\cal O}_{\Delta_1,0}\left(y_1\right){\cal O}_{\Delta_2,0}\left(y_2\right){\cal O}_{\Delta_3,2}\left(y_3,z\right)\right]\right]^{(\bf 0)},
\end{multline}
\item \begin{subequations}
\begin{align}
\mathcal{A}^{1,0;1,0}_{\Delta_1,\Delta_2,\Delta_3}\left(y_1,y_2,y_3\right)&=f^{1,0;1,0}_{\Delta_1,\Delta_2,\Delta_3} \left[\left[{\cal O}_{\Delta_1,0}\left(y_1\right){\cal O}_{\Delta_2,0}\left(y_2\right){\cal O}_{\Delta_3,0}\left(y_3\right)\right]\right]^{(\bf 0)}\\
f^{1,0;1,0}_{\Delta_1,\Delta_2,\Delta_3}&=2\left(\Delta_2+1-\tfrac{d}{2}\right)_2 \frac{C_{\Delta_2+2,0}}{C_{\Delta_2,0}}
{\sf B}\left(0,0,0;{\bf 0};\Delta_1,\Delta_2,\Delta_3\right),
\end{align}
\end{subequations}
\item \begin{subequations}
\begin{align}
\mathcal{A}^{0,1;0,2}_{\Delta_1,\Delta_2,\Delta_3}\left(y_1,y_2,y_3\right)&=f^{0,1;0,2}_{\Delta_1,\Delta_2,\Delta_3} \left[\left[{\cal O}_{\Delta_1,0}\left(y_1\right){\cal O}_{\Delta_2,0}\left(y_2\right){\cal O}_{\Delta_3,0}\left(y_3\right)\right]\right]^{(\bf 0)},
\end{align}
{\small \begin{multline}
f^{0,1;0,2}_{\Delta_1,\Delta_2,\Delta_3}=2\left[ \Delta_2 (\Delta_2+1) \Delta_3^2 \right. \\
\left. + \tfrac{1}{4} (\Delta_1-\Delta_2-\Delta_3) (-d+\Delta_1+\Delta_2+\Delta_3) (d (-\Delta_1+\Delta_2+\Delta_3+2) +(\Delta_1+\Delta_2-\Delta_3) (\Delta_1-\Delta_2+\Delta_3))\right]\\ \times 
{\sf B}\left(0,0,0;{\bf 0};\Delta_1,\Delta_2,\Delta_3\right),
\end{multline}}
\end{subequations}
\item \begin{subequations}
 \begin{align}
 \mathcal{A}^{0,1;1,0}_{\Delta_1,\Delta_2,\Delta_3}\left(y_1,y_2,y_3\right) & =f^{0,1;1,0}_{\Delta_1,\Delta_2,\Delta_3}
\left[\left[{\cal O}_{\Delta_1,0}\left(y_1\right){\cal O}_{\Delta_2,0}\left(y_2\right){\cal O}_{\Delta_3,0}\left(y_3\right)\right]\right]^{(\bf 0)}\\
f^{0,1;1,0}_{\Delta_1,\Delta_2,\Delta_3}&=2\left(d+1\right)
{\sf B}\left(0,0,0;{\bf 0};\Delta_1,\Delta_2,\Delta_3\right).
\end{align}
\end{subequations}
\end{enumerate}
Putting everything together in \eqref{app2gravdec} gives:
\begin{multline}
{\cal M}^{\text{2pt-bubble}}\left(y_1,y_2\right) = \int^\infty_{-\infty} d\nu d{\bar \nu} {\cal F}^{\text{2pt-bubble}}_{TT}\left(\nu,{\bar \nu }\right)\mathfrak{K}^{({\bf 0},{\bf 0})}_{2;0,0}\left(\nu,{\bar \nu};y_1,y_2\right) \\ + \int^\infty_{-\infty} d\nu d{\bar \nu} {\cal F}^{\text{2pt-bubble}}_{\text{contact}}\left(\nu,{\bar \nu }\right)\mathfrak{K}^{({\bf 0},{\bf 0})}_{0;0,0}\left(\nu,{\bar \nu};y_1,y_2\right),
\end{multline}
with the usual traceless and transverse contribution \eqref{ttanomspec}:
\begin{multline}
{\cal F}^{\text{2pt-bubble}}_{TT}\left(\nu,{\bar \nu }\right) =g^2 \frac{\nu^2}{\pi \left[\nu^2+\left(\frac{d}{2}\right)^2\right]}\frac{{\bar \nu}^2}{\pi \left[{\bar \nu}^2+\left(\Delta-\frac{d}{2}\right)^2\right]} \\ \times {\sf B}\left(0,0,2;{\bf 0};\Delta_1,\tfrac{d}{2}+i{\bar \nu},\tfrac{d}{2}+i{\nu}-2\right){\sf B}\left(0,0,2;{\bf 0};\Delta_2,\tfrac{d}{2}-i{\bar \nu},\tfrac{d}{2}-i{\nu}-2\right),
\end{multline}
and purely contact contribution:
\begin{align}
&{\cal F}^{\text{2pt-bubble}}_{\text{contact}}\left(\nu,{\bar \nu }\right) \\ \nonumber
&\hspace*{0.5cm}
 =g^2 \frac{\nu^2  {\bar \nu}^2}{\pi^2} g^{(0)}_{0,0,0}\left({\bar \nu}\right)\left[ g^{(2)}_{1,1,0}\left(\nu\right)f^{1,0;1,0}_{\Delta_1,\frac{d}{2}+i{\bar \nu},\frac{d}{2}+i\nu}  f^{1,0;1,0}_{\Delta_2,\frac{d}{2}-i{\bar \nu},\frac{d}{2}-i\nu} +  g^{(2)}_{1,0,0}\left(\nu\right)f^{1,0;1,0}_{\Delta_1,\frac{d}{2}+i{\bar \nu},\frac{d}{2}+i\nu}f^{1,0;0,2}_{\Delta_2,\frac{d}{2}-i{\bar \nu},\frac{d}{2}-i\nu} \right.\\ \nonumber
&\hspace*{0.5cm} +  g^{(2)}_{1,0,0}\left(\nu\right) f^{1,0;0,2}_{\Delta_1,\frac{d}{2}+i{\bar \nu},\frac{d}{2}+i\nu}f^{1,0;1,0}_{\Delta_2,\frac{d}{2}-i{\bar \nu},\frac{d}{2}-i\nu}   + g^{(2)}_{0,0,2}\left(\nu\right)f^{1,0;0,2}_{\Delta_1,\frac{d}{2}+i{\bar \nu},\frac{d}{2}+i\nu}f^{1,0;0,2}_{\Delta_2,\frac{d}{2}-i{\bar \nu},\frac{d}{2}-i\nu} \\ \nonumber
& \hspace*{0.5cm} + \frac{1}{2}\left(d-2\right)\left( g^{(2)}_{1,1,0}\left(\nu\right)f^{1,0;1,0}_{\Delta_1,\frac{d}{2}+i{\bar \nu},\frac{d}{2}+i\nu}f^{0,1;1,0}_{\Delta_2,\frac{d}{2}-i{\bar \nu},\frac{d}{2}-i\nu}+ g^{(2)}_{1,0,0}\left(\nu\right)f^{1,0;0,2}_{\Delta_1,\frac{d}{2}+i{\bar \nu},\frac{d}{2}+i\nu}f^{0,1;1,0}_{\Delta_2,\frac{d}{2}-i{\bar \nu},\frac{d}{2}-i\nu} \right. \\ \nonumber
&  \hspace*{0.5cm}+  g^{(2)}_{1,1,0}\left(\nu\right) f^{0,1;1,0}_{\Delta_1,\frac{d}{2}+i{\bar \nu},\frac{d}{2}+i\nu} f^{1,0;1,0}_{\Delta_2,\frac{d}{2}-i{\bar \nu},\frac{d}{2}-i\nu} +\left.g^{(2)}_{1,0,0}\left(\nu\right) f^{0,1;1,0}_{\Delta_1,\frac{d}{2}+i{\bar \nu},\frac{d}{2}+i\nu}f^{1,0;0,2}_{\Delta_2,\frac{d}{2}-i{\bar \nu},\frac{d}{2}-i\nu}\right) \\ \nonumber
&  \hspace*{0.5cm} + \left. \frac{1}{4}\left(d-2\right)^2 g^{(2)}_{1,1,0}\left(\nu\right)f^{0,1;1,0}_{\Delta_1,\frac{d}{2}+i{\bar \nu},\frac{d}{2}+i\nu}f^{0,1;1,0}_{\Delta_2,\frac{d}{2}-i{\bar \nu},\frac{d}{2}-i\nu}\right],
\end{align}
which arises from considering the full propagator \eqref{gravdedon} as opposed to just its traceless and transverse part \eqref{ttbubu}.

\section{Full single-cut bubble diagrams}
\label{secLLfullsinglecut}

In this appendix we present some examples of the single-cut bubble diagrams considered in \S \tcb{\ref{subsubsec::altquantads4}} using the full bulk-to-bulk propagator -- i.e. including all contact terms. We work with Fronsdal higher-spin fields $\varphi_s$ in the de Donder gauge:
\begin{equation}
   \left[\left( \nabla \cdot \partial\right)-\frac{1}{2} \left(u \cdot \nabla \right) \left(\partial_u \cdot \partial_u\right)\right]\varphi_s\left(x,u\right)=0.
\end{equation}
It is useful to express the double-traceless Fronsdal field in terms of its traceless components:
\begin{equation}\label{irredecomp}
\varphi_s\left(x,u\right) = {\tilde \varphi}_{s}\left(x,u\right)+\frac{u^2}{2\left(d-3+2s\right)}  \varphi^{\prime}_s\left(x,u\right),
\end{equation}
where
\begin{equation}
\left(\partial_u \cdot \partial_u\right) \varphi_s\left(x,u\right) = \varphi^{\prime}_s\left(x,u\right), \qquad \left(\partial_u \cdot \partial_u\right){\tilde \varphi}_{s}\left(x,u\right)=\left(\partial_u \cdot \partial_u\right)\varphi^{\prime}_s\left(x,u\right)=0.
\end{equation}
The $s-s^\prime-0$ cubic coupling in de Donder gauge then reads \cite{Sleight:2016dba,Sleight:2017cax}:\footnote{See also \cite{Francia:2016weg} for other recent developments on off-shell interactions of higher-spin gauge fields, among which includes interactions the Maxwell-like formulation \cite{Francia:2011qa,Campoleoni:2012th}.}
\begin{multline}\label{dedondver}
    \mathcal{V}_{s,s^\prime,0}=g_{s,0,s^\prime}\left[{\cal Y}_1^{s}{\cal Y}_3^{s^\prime}{\tilde \varphi}_s{\tilde \varphi}_{s^\prime}\phi-(s-s^\prime)\binom{s}{2}\,\frac{d-4+s+s^\prime}{d-3+2s}\,{\cal Y}_1^{s-2}{\cal Y}_2^{s^\prime}{ \varphi}_s^\prime{\tilde \varphi}_{s^\prime}\phi\right.\\\left.-(s^\prime-s)\binom{s^\prime}{2}\,\frac{d-4+s+s^\prime}{d-3+2s^\prime}\,{\cal Y}_1^{s}{\cal Y}_3^{s^\prime-2}{\tilde \varphi}_s{ \varphi}^\prime_{s^\prime}\phi\right]\,,\qquad s\geq s^\prime\,.
\end{multline}
Notice that above we have only displayed the terms at most linear in the traces of the Fronsdal fields, since terms involving two traces do not contribute to bubble diagrams with one scalar propagating in the loop. Furthermore, in order to avoid double counting of vertices we assume $s\geq s^\prime$. One can then see that if the exchanged spin inside the loop is greater than the external spin, the contact contribution generated by the trace terms in the vertex changes sign with respect to the diagrams where the internal spin is lower than the external one.

For this computation we will use the following result for Witten diagrams involving traceless symmetrised gradients of harmonic functions:
\begin{align}
\int_{\text{AdS}_{d+1}}dX\,\mathcal{Y}_1^2\mathcal{Y}_3^2&\,K_{d,2}\,K_{\Delta_2,0}\,(w_3\cdot\nabla_3)^2K_{\Delta_3,0}\\\nonumber&=-\frac{2 (\Delta_3-1) \Delta_3 (\Delta_2-\Delta_3-3) (\Delta_2-\Delta_3+2) \left(\Delta_2^2+\Delta_2-(\Delta_3-4) (\Delta_3+1)\right)}{(\Delta_2-\Delta_3-1) (\Delta_2-\Delta_3+1) (\Delta_2-\Delta_3+5) (\Delta_2+\Delta_3+2)}\\\nonumber&\times {\sf B}(2,0,2;{\bf{0}};d-2,\Delta_2,\Delta_3-2)\left[\left[{\cal O}_{d,2}\left(P_1,Z\right){\cal O}_{\Delta_2,0}\left(P_2\right){\cal O}_{\Delta_3,0}\left(P_3\right)\right]\right]^{\left({\bf 0}\right)}\,,\\
\int_{\text{AdS}_{d+1}}dX\,\mathcal{Y}_1^4\mathcal{Y}_3^2&\,K_{d+2,4}\,K_{\Delta_2,0}\,(w_3\cdot\nabla_3)^2K_{\Delta_3,0}\\\nonumber&\hspace{-50pt}=-\tfrac{2 (\Delta_3-1) \Delta_3 (\Delta_2-\Delta_3-7) \left((5-2 \Delta_2 (\Delta_2+5)) \Delta_3^2+6 (\Delta_2 (\Delta_2+5)+2) \Delta_3+(\Delta_2-1) \Delta_2 (\Delta_2+5) (\Delta_2+6)+\Delta_3^4-6 \Delta_3^3\right)}{(\Delta_2-\Delta_3-1) (\Delta_2-\Delta_3+1) (\Delta_2-\Delta_3+9) (\Delta_2+\Delta_3-1) (\Delta_2+\Delta_3+6)}\\\nonumber&\times {\sf B}(4,0,2;{\bf{0}};d-2,\Delta_2,\Delta_3-2)\left[\left[{\cal O}_{d+2,4}\left(P_1,Z\right){\cal O}_{\Delta_2,0}\left(P_2\right){\cal O}_{\Delta_3,0}\left(P_3\right)\right]\right]^{\left({\bf 0}\right)}.
\end{align}

\subsection{2-(20)-2}

In this case the coupling (like for all $s-s-0$ couplings which are of the $R^2$ form) is traceless with respect to the $s^\prime=2$ leg. 
Following the same approach as in \S \tcb{\ref{subsec::gravloops}}, including all terms in the graviton propagator \eqref{gravdedon} we obtain the following spectral integral for the single-cut in $d=3$:
\begin{multline}
    \gamma_{2,2}=-g_{2,0,2}^2\int_{-\infty}^{\infty} d\nu_1\,\left(\frac{\nu_1^{13}}{51840 \pi ^3}+\frac{23 \nu_1^{11}}{103680 \pi ^3}-\frac{5993 \nu_1^9}{829440 \pi ^3}-\frac{24491 \nu_1^7}{165888 \pi ^3}-\frac{12295649 \nu_1^5}{13271040 \pi ^3}\right.\\\left.-\frac{56596249 \nu_1^3}{26542080 \pi ^3}-\frac{51048983 \nu_1}{212336640 \pi ^3}-\frac{1024 \nu_1}{135 \pi ^3 \left(4 \nu_1^2+33\right)}\right) \tanh (\pi  \nu_1) \text{sech}(\pi  \nu_1)\,.
\end{multline}
which, apart from the rightmost term on the second line, can be evaluated analytically using the techniques developed in this work. The part of the integral which we are able to evaluate analytically gives
\begin{equation}
    \gamma^{\text{an.}}_{2,2}=\frac{1757}{4320 \pi ^2}\sim 0.0412086\,g_{2,0,2}^2\,,
\end{equation}
while the total result is given numerically by:
\begin{equation}
    \gamma_{2,2}^{\text{full}}=0.0432286\,g_{2,0,2}^2\,.
\end{equation}
It is interesting to compare the above result with the $TT$ contribution \eqref{typabubminus2}. The latter is:
\begin{equation}
    \gamma_{2,2}^{\Delta_+ \Delta_-}\sim \frac{253}{480 \pi ^2}\,g_{2,0,2}^2\,,
\end{equation}
and differs from the full result by $|\gamma_{2,2}^{\text{full}}-\gamma_{2,2}^{\text{TT}}|\sim 0.0101761\,g_{2,0,2}^2$.

\subsection{4-(20)-4}

In this case using the full graviton propagator \eqref{gravdedon} we have
\begin{multline}
    \gamma_{2,4}=-g_{2,4,0}^2\int_{-\infty}^{\infty} d\nu_1\,\nu_1 \left(4 \nu_1^2+1\right) \left(4 \nu_1^2+25\right) \left(4 \nu_1^2+49\right) \left(4 \nu_1^2+81\right) \left(4 \nu_1^2+121\right) \left(4 \nu_1^2+169\right)\\\times \left(4 \nu_1^2+225\right)\frac{ \left(256 \nu_1^8-20224 \nu_1^6-778144 \nu_1^4-8790256 \nu_1^2-28691327\right)}{3262849744896000 \pi ^3 \left(4 \nu_1^2+33\right)}\,,
\end{multline}
which can be evaluated analytically apart from the term $\frac{1274544128 \nu_1}{2701125 \pi ^3 \left(4 \nu_1^2+33\right)}$. The part of the integral which we are able to evaluate analytically gives
\begin{equation}
    \gamma^{\text{an.}}_{2,2}=\frac{3938687}{105840 \pi ^2}\sim 3.77053\,g_{4,2,0}^2\,,
\end{equation}
while the total result is given numerically by:
\begin{equation}
    \gamma_{4,2}^{\text{full}}=3.74762\,g_{4,2,0}^2\,.
\end{equation}
The $TT$ contribution \eqref{typabubminus2} in this case is
\begin{equation}
    \gamma_{4,2}^{\Delta_+ \Delta_-}\sim \frac{87491}{2352 \pi ^2}\,g_{4,2,0}^2\,,
\end{equation}
which differs from the full result by $|\gamma_{4,2}^{\text{full}}-\gamma_{4,2}^{\text{TT}}|\sim 0.0213821\,g_{4,2,0}^2$.

\bibliography{refs}
\bibliographystyle{JHEP}
\end{document}